\documentclass[12pt,notitlepage,fleqn]{article}
\usepackage{graphicx}
\usepackage{amsmath}
\usepackage{indentfirst}

\newtheorem{problem}{Hypothesis}

\begin{document}

\begin{center}
{\LARGE A Possible Structure Model of the Vacuum}

The Body Center Cubic Model of the Vacuum Material

{\normalsize Jiao Lin Xu}

{\small The Center for Simulational Physics, The Department of Physics and
Astronomy}

{\small University of Georgia, Athens, GA 30602, USA.}

E- mail: {\small \ Jxu@Hal.Physast.uga.edu}

\bigskip

\textbf{Abstract}
\end{center}

{\small This paper has deduced the baryon spectrum using only 2 flavored
quarks u and d (each of them has three colored members). From Dirac's sea
concept and the quark confinement idea, we conjecture and then assume that
the quarks (in the vacuum state) compose colorless particles (uud and udd,
hereafter will be called the} \textbf{Lee Particles}{\small ). Moreover, the
Lee Particles lead in a natural way to the construction }-{\small \ a body
center cubic lattice in the vacuum. Consequently, there exists} {\small a
strong interaction periodic field with body center cubic periodic
symmetries\ and an electromagnetic interaction periodic field (from the
charged Lee Particles uud) with simple cubic periodic symmetries in the
vacuum. In terms of the energy band theory, using }a {\small point baryon
approximation (i.e., omitting the structure of the Lee Particles), and from
the symmetries of the strong interaction periodic field, we have deduced the
intrinsic quantum numbers (I, S, C, b, Q) of all baryons and a unified mass
formula from which all baryon masses can be found. This theoretical baryon
spectrum is in accordance with the experimental results in both intrinsic
quantum numbers and masses. These results show that various baryons are the
energy band excited states of the Lee Particles. We also predict some new
baryons: }$\Lambda ^{0}(2559),${\small \ }$\Lambda ^{0}(4279),${\small \ }$%
\Xi _{C}(3169),${\small \ }$\Omega ^{-}(3619)${\small , }$\Lambda
_{C}^{+}(6659)${\small , }$\Lambda _{b}^{0}(10159)${\small ... Furthermore,
from the electromagnetic interaction periodic field of the vacuum material,
we predict a super heavy electron spectrum }$E(166${\small \ Gev}$)${\small %
, }$E(332${\small \ Gev}$)${\small , }$E(498${\small \ Gev}$)${\small ... We
suggest that experiments to find the super heavy electron spectrum and long
lifetime baryon }$\Lambda ^{0}(2559)${\small \ should be done.}

\section{Introduction}

The Quark Model \cite{QuarkModel} has already explained the baryon spectrum
in terms of the quarks. It successfully gives intrinsic quantum numbers (the
isospin $I$, the strange number $S$, the charmed number $C$, the bottom
number $b$, the electric charge $Q$) of all baryons. However, (1) it has not
given a satisfactory mass spectrum of baryons in a united mass formula \cite
{RELATION of MASS}; (2) it needs too many elementary particles (6 flavors $%
\times $ 3 colors $\times $ 2 (quark and antiquark) = 36 quarks) \cite
{QUARKS} \cite{Three Color}; (3) the intrinsic quantum numbers of the quarks
are ``entered by hand'' \cite{Vacuum engineering} \cite{Hand in}; (4) on the
one hand it assumes \cite{QuarkModel} that all quarks (u, d, s, c, b, t) are
independent elementary particles, but on the other hand it assumes that the
higher energy quarks can decay into lower energy quarks \cite{QUARK DECAY},
the two ``hands'' do not cooperate with each other; (5) all free quark
searches since 1977 have had negative results \cite{Free QUARK}. This paper
will try to find a possible solution to the above problems of the Quark
Model.

According to the confinement idea \cite{CONFINEMENT}, quarks are confined
inside hadrons. The question is, are quarks also confined inside hadrons
when they are in the vacuum state? The fact that free quarks have not yet
been found may imply that the quarks are confined in the vacuum state. We
also think that \textbf{the (Quark Model's) assumption that all quarks (u,
d, s, c, b, t) are independent elementary particles has already blocked the
correct way to find a united mass formula of the baryons. }Considering the
fact \cite{particle} that all higher energy baryons decay into lower energy
baryons and finally decay into nucleons, we guess that (1) there are only
two kinds of the elementary quarks ($u$ and $d$) in the vacuum; (2) they
have already formed two colorless quark groups uud and udd (we call the
colorless quark groups uud and udd \textbf{the Lee Particles }\cite
{Leeparticle} \cite{asymmetry} \cite{li pARTICLE}) in the vacuum; (3) the
above facts (all higher energy baryons decay into lower energy baryons and
finally decay into nucleons)\ may indicate that \textbf{all baryons may be
the same kind of particles (the Lee Particles) in different states with
different energy and quantum numbers}; (4) the strange number S, the charmed
number C, and the bottom number b reflect some symmetry properties of the
different states of the Lee Particles rather than coming from new quarks (s,
c, b). Now we need a mechanism which can generate the baryon
spectrum--various states (with different properties and masses) of the Lee
Particles.

Twenty years ago, T. D. Lee had already noticed that \cite{Vacuum
engineering} : ``the standard model... needs $\sim $ 20 parameters: e, G, $%
\theta _{w},$various masses for the three generations of leptons and quarks
and the four weak decay angles $\theta _{1},\theta _{2},\theta _{3,}$and $%
\delta $.'' ``Therefore, while we may have achieved a rather effective
description of physical processes up to about 100 Gev, the theory we have
should more appropriately be viewed as essentially phenomenological. After
all, who has ever heard of a fundamental theory that requires twenty - some
parameters?'' He then suggested some possible directions that ``may change
our present frame of thinking'': ``1. Size of leptons and quarks.'' ``2.
Possibility of vacuum engineering.'' ``3. Improvement on conventional
quantum mechanics.'' Furthermore, in the section which discussed the
possibility of vacuum engineering, T. D. Lee pointed out: ``we believe our
vacuum, though Lorentz invariant, to be quite complicated. Like any other
physical medium, it can carry long-range-order parameters and it may also
undergo phase transitions... If we can create vacuum excitations or vacuum
phase transitions, then any of the constants in our present theory, $\theta
_{w},\theta _{c},m_{u},m_{d,}...$ $,$ can be subject to change.''

Recently Frank Wilczek, the J. Robert Oppenheimer Professor at the Institute
for Advanced Study in Princeton, further elaborated Lee's idea \cite{wilczek}%
: ``empty space--the vacuum--is in reality a richly structured, though
highly symmetrical, medium. Dirac's sea was an early indication of this
feature, which is deeply embedded in quantum field theory and the Standard
Model. Because the vacuum is a complicated material governed by locality and
symmetry, one can learn how to analyze it by studying other such
materials--that is, condensed matter.'' Professor Wilczek not only pointed
out one of the most important and most urgent research directions of modern
physics--studying the structure of the vacuum, but also provided a very
practical and efficient way for the studying--learning from studying
condensed matter.

Applying the Lee-Wilczek idea, this paper conjectures a structure of the
vacuum (body center cubic symmetry), which will be used as the mechanism to
generate the baryon\ spectrum (various excited states of the Lee Particles).
This paper will deduce the intrinsic quantum numbers (including S, C, and b)
and the masses of all baryons in terms of a phenomenological model (the BCC
model), using only the u quarks and the d quarks. The BCC model will show
that although baryons ($\Delta ,$ $N,$ $\Lambda ,$ $\Sigma ,$ $\Xi ,$ $%
\Omega ,$ $\Lambda _{C,}$ $\Xi _{C},$ $\Sigma _{C},$ and $\Lambda _{C}$) are
so different from one another in I, S, C, b, Q, and M, they are the same
kind of particles (the Lee Particles) which are in different energy band
states. The theoretical results of the BCC model on the intrinsic numbers
and masses of baryons are in good agreement with the experimental results.
The BCC model also predicts a super heavy electron spectrum. The super heavy
electron spectrum may be a touchstone of the model of the vacuum medium.%
\textbf{\ }

According to Dirac's sea concept \cite{diarcsea}, there are electron Dirac
sea, $\mu $ lepton Dirac sea, $\tau $ lepton Dirac sea, $u$ quark Dirac sea, 
$d$ quark Dirac sea, $s$ quark Dirac sea, $c$ quark Dirac sea, $b$ quark
Dirac sea...in the vacuum. All of these Dirac seas are in the same space, at
any location, that is, at any physical space point. According to the Quark
Model and the quantum chromodynamics \cite{Chromodynamics}, there are
super-strong color attractive interactions among the quarks, causing three
quarks of different colors to be confined together and form a colorless
baryon ($p$, $n$, $\Lambda $, $\Sigma $, $\Xi $, $\Omega ..$). These
baryons, electrons, leptons, etc. will interact with one another and form
the perfect vacuum material. However, some kinds of particles do not play an
important role in forming the vacuum material. First, the main force which
makes and holds the structure of the vacuum material must be the strong
interactions, not the weak-electromagnetic, or the gravitational
interactions. Hence, in considering the structure of the vacuum material, we
leave out the Dirac seas of those particles which do not have strong
interactions ($e,$ $\mu $, $\tau $). Secondly, it is unlikely that the super
stable vacuum material is composed of unstable blocks, hence we also omit
the unstable particles (such as: $\Lambda $, $\Sigma $, $\Xi $, $\Omega $,
...). Finally, there are only two kinds of possible particles left: the
vacuum state protons (uud-Lee Particle) and the vacuum state neutrons
(udd-Lee Particle). It is well known that there are strong attractive forces
between the protons and the neutrons inside a nucleus. Similarly, there
should also exist the strong attractive forces between the Lee Particles
(uud) and the Lee Particles (udd) which will make and hold the densest
structure of the vacuum state Lee Particles.

According to solid state physics \cite{Solidstates}, if two kinds of
particles (with radius $R_{1}<R_{2}$) satisfy the condition $%
1>R_{1}/R_{2}>0.73$, the densest structure is the body center cubic crystal 
\cite{bodycenter}. We know: first, the Lee Particle (uud) and the Lee
Particle (udd) are not completely the same, thus $R_{1}\neq R_{2}$; second,
they are similar to each other (they have the same $S=C=b=0$, and their
first excited states have essentially the same masses: $M_{uud}=938.3$ Mev, $%
M_{udd}=939.6$ Mev), thus $R_{1}\approx R_{2}$. Hence, if $R_{1}<R_{2}$ (or $%
R_{2}<R_{1}$), we have $1>R_{1}/R_{2}>0.73$ (or $1>R_{2}/R_{1}>0.73$).
Therefore, we conjecture that the vacuum state Lee Particles construct the
densest structure which is the body center cubic lattice (in this paper it
will be regarded as \textbf{the BCC model}). At the same time, the vacuum
Lee Particles (uud) alone forms a simple cubic lattice. In addition, there
should be a vacuum electron (with the electric charge $-1$) in each body
center cubic cell to preserve electric neutrality, because there is a vacuum
Lee Particle (uud) (with the positive electric charge $+1$) in the cubic
cell.

Similar to a crystal, which has a periodic field, there are also periodic
fields in the vacuum. Particularly, there will be two kinds of periodic
fields: the periodic field with the body center cubic symmetries is a strong
interaction field; the other with the simple cubic symmetries of vacuum
state protons is an electromagnetic interaction field.

For the strong interaction periodic field, from energy band theory \cite
{eband} and the phenomenological fundamental hypotheses of the BCC model, we
can deduce all intrinsic quantum numbers of all baryons (the isospin $I$,
the strange number $S$, the charmed number $C$, the bottom number $b$, the
electric charge $Q$) which are consistent with the experimental results \cite
{particle}. Likewise, we can calculate the masses of all baryons which are
in very good agreement with the experimental results \cite{particle}, using
a united mass formula.

Similarly, from the simple cubic electromagnetic periodic field, we get a
super heavy electron spectrum. Super heavy electrons are so heavy that even
the lightest one is about $177$ times heavier than a proton. So far none of
the super heavy electrons has been confirmed by the experiments completely.
However, the interesting thing is that a heavy particle of approximately $%
177 $ times heavier than a proton has already been discovered \cite{TOPQUARK}%
.

We should note that an ideal crystal does not scatter particles. Because the
Hamiltonian of a particle in the ideal crystal is independent of time, the
particle is in a stationary state. If there is no perturbation, the particle
will remain in the state forever. Thus, as a property of the body center
cubic symmetric structure, the vacuum material should not scatter particles
as well. The vacuum material is like an ultra-superconductor. Just as
electrons can move inside superconductors without encountering electric
resistance, any particle can move inside the ultra-superconductor (vacuum
material) not only without encountering electric resistance, but also
without encountering mechanical resistance.

The structure of the vacuum material may help us understand many fundamental
problems. For example, the periodic field of the vacuum material may lead to
an explanation for the particle-wave duality in quantum mechanics; the
vacuum material may be the source of asymmetry \cite{asymmetry}; a broken
hole of the perfect vacuum material may help us explain the black hole...

This paper is organized as follows: The fundamental hypotheses are presented
in \emph{Section II}. The motion equation is solved with a free particle
approximation \cite{freeparticle} in \emph{Section III.} The recognition of
baryons is accomplished in \emph{Section IV}. A comparison of the results of
the BCC model and the experimental results is listed in \emph{Section V}.
The predictions of the model (including the super-heavy electron spectrum),
recommendations for experimental tests, and discussions are stated in \emph{%
SectionVI}. The conclusions are in \emph{SectionVII}.

\section{Fundamental Hypotheses}

The BCC model attempts to explain the spectrum of baryons in terms of the
outside environment. For simplicity, the intrinsic structure (three quarks)
of baryons will be ignored temporarily while the outside influences of the
vacuum is being considered. In other words, the baryons are treated as
elementary particles without intrinsic structure in the phenomenological BCC
model. We would like to call this simplification \textbf{the} \textbf{point
baryon approximation}. We will call \textbf{the} \textbf{point
approximations of the colorless quark group uud and udd as the Lee Particles
(uud - charged Lee Particle, udd - neutral Lee Particle). }The approximation
is based on the the quark confinement theory \cite{CONFINEMENT} and the
experimental results \cite{Free QUARK} that a baryon always appears as a
whole particle. Indeed, we have never seen a part of a baryon alone in any
experiment. By ignoring the intrinsic structure of baryons, we can focus our
attention on the outside influence.

In order to explain our model accurately and concisely, we will start from
the phenomenological fundamental hypotheses in an axiomatic form.

\begin{problem}
There are only two kinds of fundamental quarks u and d in the quark family.
There exist super-strong color attractive interactions between the\ colored
quarks. Three quarks (uud or udd) compose a kind of colorless Fermi
particles in the vacuum state.
\end{problem}

\textbf{According to the Quark Model} \cite{QuarkModel}, above Fermi
particles (the Lee Particles) are unflavored ($S=C=b=0$) with spin $s=1/2$
and isotropic spin $I=1/2.$ From the quantum field theory \cite{quantumfield}%
, a Lee Particle serves only as the background for the physical vacuum (the
baryon number $B=0$) when it is in the vacuum state. However, once it is
excited from the vacuum, it becomes an observable particle ($B=1$, $S=C=b=0,$
$s=1/2,$ and $I=1/2$). Generally, the excited Lee Particles ($B=1,$ $S=C=b=0$%
) with the third component of the isospin $I_{z}=+1/2$ are protons, and with
the third component of the isospin $I_{z}=-1/2$ are neutrons.

\begin{problem}
There are strong attractive interactions between the Lee Particles, the
interactions will make and hold the densest structure of the Lee Particles -
the body center cubic Lee Particle Lattice in the vacuum. The lattice forms
a strong interaction periodic field with body center cubic symmetries in the
vacuum, where the periodic constant $a_{x}$ is much smaller than the
magnitude of the radii of the nuclei.
\end{problem}

\begin{problem}
Quantum mechanics applies to the ultra-microscopic world \cite{TDLEE}. Thus,
the energy band theory \cite{eband} is also valid in the ultra-microscopic
world.
\end{problem}

According to the energy band theory, an excited Lee Particle (from vacuum),
inside the body center cubic periodic field, will be in a state of the
energy bands. \textbf{The energy band excited states of the Lee Particles
will be various baryons}. According to the point baryon approximation, we
will treat the Lee Particle as a point particle without intrinsic structure
in this paper..

\begin{problem}
Due to the effect of the periodic field, fluctuations of an excited Lee
Particle state may exist. Thus, the fluctuations of energy $\varepsilon $%
{\LARGE \ }and intrinsic quantum numbers (such as the strange number $S$)
may exist also. The fluctuation of the Strange number, if exists, is always $%
\Delta S=\pm 1$ \cite{real value of S}. From the fluctuation of the Strange
number we will be able to deduce new quantum numbers, such as the Charmed
number $C$ and the Bottom number $b$.
\end{problem}

\begin{problem}
The energy band excited states (except the energy bands in the first
Brillouin zone) of the Lee Particles are unstable baryons (we call all
baryons except protons and neutrons unstable baryons). Their quantum numbers
and masses are determined as follows (note: the quantum numbers of the
ground energy bands in the first Brillouin zone are determind by \textbf{%
Hypothesis I):}
\end{problem}

\begin{enumerate}
\item  Baryon number $B$: according to \textbf{Hypothesis I}, all energy
band states have 
\begin{equation}
B=1.  \label{baryon}
\end{equation}

\item  The ground energy bands (in the firs Brillouin zone) are the free
excited states of the Lee Particles. They have I =1/2, S = C = b =0, from 
\textbf{Hypothesis I. In other words, they are }nucleons.

\item  Isospin number $I$: the maximum isospin $I_{m}$ is determined by the
energy band degeneracy $d$ \cite{eband}, where 
\begin{equation}
d=2I_{m}+1,  \label{isomax}
\end{equation}

and another possible isospin value is determined by 
\begin{equation}
I=I_{m}-1,\text{ \ \ }I\geq 0.  \label{isonext}
\end{equation}

\item  Strange number $S$: the Strange number $S$ is determined by the
rotary fold $R$ of the symmetry axis \cite{eband} with 
\begin{equation}
S=R-4,  \label{strange}
\end{equation}

where the number $4$ is the highest possible rotary fold number.

\item  Electric charge $Q$: after obtaining $B,$ $S$ and $I$, we can find
the charge $Q$ from the Gell-Mann-Nishijiman relationship \cite{GellMann}: 
\begin{equation}
Q=I_{z}+1/2(S+B).  \label{charge}
\end{equation}

\item  Charmed number $C$ and Bottom number $b$: Since the Lee Particles do
not have any partial charge and the unstable baryons are the energy band
excited states of the Lee Particles, the unstable baryons shall not have
partial charges. Thus, if a partial charge is resulted from (\ref{strange})
and (\ref{charge}), (\ref{strange}) will be changed into 
\begin{equation}
\bar{S}=R-4.  \label{strangebar}
\end{equation}

From \textbf{Hypothesis IV (}$\Delta S=\pm 1$\textbf{),} the real value of $%
S $ is 
\begin{equation}
S=\bar{S}+\Delta S=(R-4)\pm 1.  \label{strangeflu}
\end{equation}

The ``Strange number'' $S$ in (\ref{strangeflu}) is not completely the same
as the strange number in (\ref{strange}). In order to compare it with the
experimental results, we would like to give it a new name under certain
circumstances. Based on \textbf{Hypothesis IV}, the new names will be the 
\textbf{Charmed} number and the \textbf{Bottom} number: 
\begin{gather}
\text{if }S=+1\text{ which originates from the fluctuation }\Delta S=+1\text{%
, }  \notag \\
\text{then we call it the \textbf{Charmed} number }C\text{ }(C=+1)\text{;}
\label{charmed}
\end{gather}
\begin{gather}
\text{if }S=-1\text{ which originates from the fluctuation }\Delta S=+1\text{%
, }  \notag \\
\text{and if there is an energy fluctuation,}  \notag \\
\text{then we call it the \textbf{Bottom} number }b\text{ }(b=-1)\text{.}
\label{bottom}
\end{gather}

Thus, (\ref{charge}) needs to be generalized to 
\begin{equation}
Q=I_{z}+1/2(B+S_{G})=I_{z}+1/2(B+S+C+b),  \label{chargeflu}
\end{equation}

where we define the generalized strange number as 
\begin{equation}
S_{G}=S+C+b.  \label{strangegen}
\end{equation}

\item  Charmed strange baryon $\Xi _{C}$ and $\Omega _{C}$: if the energy
band degeneracy $d$ is larger than the rotary fold $R$, the degeneracy will
be divided. Sometimes degeneracies should be divided more than once. After
the first division, the sub-degeneracy energy bands have $S_{Sub}=\bar{S}%
+\Delta S$. For the second division of a degeneracy bands, we have: 
\begin{gather}
\text{if the second division has fluctuation }\Delta S=+1\text{, \ }  \notag
\\
\text{ then }S_{Sub}\text{ may be unchanged and we may have }  \notag \\
\text{ a Charmed number }C\text{ from }C=\Delta S=+1.  \label{C&S}
\end{gather}
Therefore, we can obtain charmed strange baryons $\Xi _{C}$ and $\Omega _{C}$%
.

\item  We assume that a baryon's static mass is the minimum energy of the
energy curved surface which represents the baryon.
\end{enumerate}

\section{The Energy Bands}

\subsection{The Motion Equation of the Lee Particle}

When a Lee Particle is excited from vacuum, it becomes an observable
physical particle. Since the Lee Particle is a Fermion, its motion equation
should be the Dirac equation. Taking into account that (according to the
renormalization theory \cite{renormal}) the bare mass of the Lee Particle is
much larger than the empirical values of the baryon masses, we use the
Schr\"{o}dinger equation instead of the Dirac equation (our results will
show that this is a very good approximation): 
\begin{equation}
\frac{\text{%
%TCIMACRO{\UNICODE{0x127}}%
%BeginExpansion
h\hskip-.2em\llap{\protect\rule[1.1ex]{.325em}{.1ex}}\hskip.2em%
%EndExpansion
}^{2}}{2m_{b}}\nabla ^{2}\Psi +(\varepsilon -V(\vec{r}))\Psi =0,
\label{Schrodinger}
\end{equation}
where $V(\vec{r})$ denotes the strong interaction periodic field\ with body
center cubic symmetries and $m_{b}$ is the bare mass of the Lee Particle.

\subsection{Finding the Energy Bands}

Using the energy band theory \cite{eband} and the free particle
approximation \cite{freeparticle} (taking $V(\vec{r})=V_{0}$ constant and
making the wave functions satisfy the body center cubic periodic
symmetries), we have 
\begin{equation}
\frac{\text{%
%TCIMACRO{\UNICODE{0x127}}%
%BeginExpansion
h\hskip-.2em\llap{\protect\rule[1.1ex]{.325em}{.1ex}}\hskip.2em%
%EndExpansion
}^{2}}{2m_{b}}\nabla ^{2}\Psi +(\varepsilon -V_{0})\Psi =0,  \label{motion}
\end{equation}
where $V_{0}$ is a constant potential. The solution of Eq.(\ref{motion}) is
a plane wave 
\begin{equation}
\Psi _{\vec{k}}(\vec{r})=\exp \{-i(2\pi /a_{x})[(n_{1}-\xi )x+(n_{2}-\eta
)y+(n_{3}-\zeta )z]\},  \label{wave}
\end{equation}
where the wave vector $\vec{k}=(2\pi /a_{x})(\xi ,\eta ,\zeta )$, $a_{x}$ is
the periodic constant, and $n_{1}$, $n_{2}$, $n_{3}$ are integers satisfying
the condition 
\begin{equation}
n_{1}+n_{2}+n_{3}=\pm \text{ even number or }0.  \label{condition}
\end{equation}
Condition (\ref{condition}) implies that the vector $\vec{n}%
=(n_{1},n_{2},n_{3})$ can only take certain values. For example, $\vec{n}$
can not take $(0,0,1)$ or $\left( 1,1,-1\right) $, but can take $(0,0,2)$
and $(1,-1,2)$.

The zeroth-order approximation of the energy \cite{freeparticle} is 
\begin{equation}
\varepsilon ^{(0)}(\vec{k},\vec{n})=V_{0}+\alpha E(\vec{k},\vec{n}),
\label{mass}
\end{equation}
\begin{equation}
\alpha =h^{2}/2m_{b}a_{x}^{2},  \label{C-ALPHAR}
\end{equation}
\begin{equation}
E(\vec{k},\vec{n})=(n_{1}-\xi )^{2}+(n_{2}-\eta )^{2}+(n_{3}-\zeta )^{2}.
\label{energy}
\end{equation}

Now we will demonstrate how to find the energy bands.

The first Brillouin zone \cite{Brillouin} of the body center cubic lattice
is shown in Fig. 1. In Fig. 1 (depicted from \cite{eband} (Fig. 1) and \cite
{Brillouin}(Fig. 8.10)), the $(\xi ,\eta ,\zeta )$ coordinates of the
symmetry points are: 
\begin{gather}
\Gamma =(\text{0, 0, 0}),\text{ }H=(\text{0, 0, 1}),\text{ }P=(\text{1/2,
1/2, 1/2}),  \notag \\
N=(\text{1/2, 1/2, 0}),\text{ }M=(\text{1, 0, 0}),
\end{gather}
and the $(\xi ,\eta ,\zeta )$ coordinates of the symmetry axes are: 
\begin{eqnarray}
\Delta &=&(\text{0, 0, }\zeta ),\text{\ }0<\zeta <1;\text{ \ \ \ \ \ \ }%
\Lambda =(\xi \text{, }\xi \text{, }\xi ),\text{ }0<\xi <1/2;  \notag \\
\Sigma &=&(\xi \text{, }\xi \text{, 0}),\text{ }0<\xi <1/2;\text{ \ \ \ }D=(%
\text{1/2, 1/2, }\xi ),\text{ }0<\xi <1/2;  \notag \\
G &=&(\xi \text{, 1-}\xi \text{, 0}),\text{ }1/2<\xi <1;\text{ \ }F=(\xi 
\text{, }\xi \text{, 1-}\xi ),\text{ }0<\xi <1/2.
\end{eqnarray}

For any valid value of the vector $\vec{n}$, substituting the $(\xi ,\eta
,\zeta )$ coordinates of the symmetry points or the symmetry axes into Eq.(%
\ref{energy}) and Eq.(\ref{wave}), we can get the $E(\vec{k},\vec{n})$
values and the wave functions at the symmetry points and on the symmetry
axes. In order to show how to calculate the energy bands, we give the
calculation of some low energy bands in the symmetry axis $\Delta $ as an
example (the results are illustrated in Fig. 2(a)).

First, from (\ref{energy}) and (\ref{wave}) we find the formulas for the $E(%
\vec{k},\vec{n})$ values and the wave functions at the end points $\Gamma $
and $H$ of the symmetry axis $\Delta $, as well as on the symmetry axis $%
\Delta $ itself: 
\begin{equation}
E_{\Gamma }=n_{1}^{2}+n_{2}^{2}+n_{3}^{2},  \label{gammae}
\end{equation}
\begin{equation}
\Psi _{\Gamma }=\exp \{-i(2\pi /a_{x})[n_{1}x+n_{2}y+n_{3}z]\}.
\label{gammaksi}
\end{equation}
\begin{equation}
E_{H}=n_{1}^{2}+n_{2}^{2}+(n_{3}-1)^{2},  \label{he}
\end{equation}
\begin{equation}
\Psi _{H}=\exp \{-i(2\pi /a_{x})[n_{1}x+n_{2}y+(n_{3}-1)z]\}.  \label{hksi}
\end{equation}
\begin{equation}
E_{\Delta }=n_{1}^{2}+n_{2}^{2}+(n_{3}-\zeta )^{2},  \label{deltae}
\end{equation}
\begin{equation}
\Psi _{\Delta }=\exp \{-i(2\pi /a_{x})[n_{1}x+n_{2}y+(n_{3}-\zeta )z]\}.
\label{deltaksi}
\end{equation}

Then, using (\ref{gammae})--(\ref{deltaksi}), beginning from the lowest
possible energy, we can obtain the corresponding integer vectors $\vec{n}%
=(n_{1},n_{2},n_{3})$ (satisfying (\ref{condition})) and the wave functions:

\begin{enumerate}
\item  The lowest $E(\vec{k},\vec{n})$ is at $(\xi ,\eta ,\zeta )=0$ (the
point $\Gamma $) and with only one value of $\vec{n}=(0,0,0)$ (see (\ref
{gammae}) and (\ref{gammaksi})): 
\begin{equation}
\vec{n}=(\text{0, 0, 0})\text{ \ \ \ \ \ }E_{\Gamma }=0\text{ \ \ \ \ \ }%
\Psi _{\Gamma }=1.  \label{GROUND E-W}
\end{equation}

\item  Starting from $E_{\Gamma }=0$, along the axis $\Delta $, there is one
energy band (the lowest energy band $E_{\Delta }=\zeta ^{2}$, with $%
n_{1}=n_{2}=n_{3}=0$ (see (\ref{deltae}) and (\ref{deltaksi})) ended at the
point $E_{H}=1$: 
\begin{gather}
\vec{n}=(\text{0, 0, 0})\text{, \ \ \ \ \ }E_{\Gamma }=0\rightarrow
E_{\Delta }=\zeta ^{2}\rightarrow E_{H}=1\text{, }  \notag \\
\Psi _{\Delta }=\exp [i(2\pi /a_{x})(\zeta z)]\text{. \ \ \ \ \ \ \ \ \ }
\end{gather}

\item  At the end point $H$ of the energy band $E_{\Gamma }=0\rightarrow
E_{\Delta }=\zeta ^{2}\rightarrow E_{H}=1$, the energy $E_{H}=1$. Also at
the point $H$, $E_{H}=1$ when $n=(\pm 1,0,1)$, $(0,\pm 1,1)$, and $(0,0,2)$
(see (\ref{he}) and (\ref{hksi})): 
\begin{equation}
E_{H}=1\text{, \ \ }\Psi _{H}=[e^{[i(2\pi /a_{x})(\pm x)]},e^{[i(2\pi
/a_{x})(\pm y)]},e^{[i(2\pi /a_{x})(\pm z)]}]\text{.}
\end{equation}

\item  Starting from $E_{H}=1$, along the axis $\Delta $, there are three
energy bands ended at the points $E_{\Gamma }=0$, $E_{\Gamma }=2$, and $%
E_{\Gamma }=4$, respectively: 
\begin{gather}
\vec{n}=(\text{0,0,0})\text{, \ \ \ }E_{H}=1\rightarrow E_{\Delta }=\zeta
^{2}\rightarrow E_{\Gamma }=0,\text{ }  \notag \\
\Psi _{\Delta }=\exp [i(2\pi /a_{x})(\zeta z)]\text{. \ \ \ \ \ \ \ \ \ }
\end{gather}
\begin{gather}
\vec{n}=(\text{0,0,2})\text{, \ \ }E_{H}=1\rightarrow E_{\Delta }=\text{(2-}%
\zeta \text{)}^{2}\rightarrow E_{\Gamma }=4\text{,}  \notag \\
\text{ }\Psi _{\Delta }=\exp {[i(2\pi /a_{x})(2-\zeta )z)]}\text{. \ \ \ \ \
\ \ \ }
\end{gather}
\begin{gather}
\vec{n}=(\pm \text{1,0,1})(\text{0,}\pm \text{1,1})\text{, \ \ }%
E_{H}=1\rightarrow E_{\Delta }=\text{1+(1-}\zeta \text{)}^{2}\rightarrow
E_{\Gamma }=2\text{,}  \notag \\
\text{ }\Psi _{\Delta }=e^{{\{-i(2\pi /a_{x})[\pm x+(1-\zeta )z]\}}},e^{{%
\{-i(2\pi /a_{x})[\pm y+(1-\zeta )z]\}}}\text{. \ \ \ \ \ }
\end{gather}
The energy band ($E_{H}=1\rightarrow E_{\Delta }=$1+(1-$\zeta $)$%
^{2}\rightarrow E_{\Gamma }=2)$ is a 4 fold degeneracy band.

\item  The energy bands with $4$ sets of values $\vec{n}$ $\ (\vec{n}=(\pm $%
1,0,1$),$ $($0,$\pm $1,1$))$ ended at $E_{\Gamma }=2$. From (\ref{gammae}), $%
E_{\Gamma }=2$ also when $\vec{n}$ takes other $8$ sets of values: $\vec{n}%
=(1,\pm 1,0)$, $(-1,\pm 1,0)$, and $(\pm 1,0,-1)$, $(0,\pm 1,-1)$. Putting
the $12$ sets of $\vec{n}$ values into Eq. (\ref{gammaksi}), we can obtain $%
12$ plane wave functions: 
\begin{equation}
E_{\Gamma }=2\text{, }\Psi _{\Gamma }=[e^{i(2\pi /a_{x})(\pm x\pm
y)},e^{i(2\pi /a_{x})(\pm y\pm z)},e^{i(2\pi /a_{x})(\pm z\pm x)}]\text{.}
\end{equation}

\item  Starting from $E_{\Gamma }=2$, along the axis $\Delta $, there are
three 4 fold degeneracy energy bands ended at the points $E_{H}=1$, $E_{H}=3$%
, and $E_{H}=5$, respectively: 
\begin{equation}
\vec{n}=(\pm \text{1,0,1})(\text{0,}\pm \text{1,1})\text{,\ }E_{\Gamma
}=2\rightarrow E_{\Delta }=\text{1+(1-}\zeta \text{)}^{2}\rightarrow E_{H}=1%
\text{,}
\end{equation}
\begin{equation}
\vec{n}=(\text{1,}\pm \text{1,0})(\text{-1,}\pm \text{1,0})\text{,\ }%
E_{\Gamma }=2\rightarrow E_{\Delta }=\text{2+}\zeta ^{2}\rightarrow E_{H}=3%
\text{,}
\end{equation}
\begin{equation}
\vec{n}=(\pm \text{1,0,-1})(\text{0,}\pm \text{1,-1})\text{,\ }E_{\Gamma
}=2\rightarrow E_{\Delta }=\text{1+(}\zeta \text{+1)}^{2}\rightarrow E_{H}=5%
\text{.}
\end{equation}
\end{enumerate}

Continuing the process, we can find all the energy bands and the
corresponding wave functions. The wave functions are not needed for the
zeroth order approximation, so we only show the energy bands in Fig. 2-5.
There are six small figures in Fig. 2-4. Each of them shows the energy bands
in one of the six axes in Fig. 1. Each small figure is a schematic one where
the straight lines that show the energy bands shold be parabolic curves. The
numbers above the lines are the values of $\vec{n}$ = ($n_{1}$, $n_{2}$, $%
n_{3}$). The numbers under the lines are the fold numbers of the energy
bands with the same energy (the zeroth order approximation). The numbers
beside both ends of an energy band (the intersection of the energy band line
and the vertical lines) represent the highest and lowest E($\vec{k}$,$\vec{n}
$) values (see Eq. (\ref{energy})) of the band. Putting the values of the E($%
\vec{k}$,$\vec{n}$) into Eq. (\ref{mass}), we get the zeroth order energy
approximation values (in Mev).

\section{The Recognition of the Baryons}

It is worth while to emphasize that \textbf{there are significant
differences between nucleons (proton, neutron) and unstable baryons (}$%
\Delta ,$\textbf{\ }$N,$\textbf{\ }$\Lambda ,$\textbf{\ }$\Sigma ,$\textbf{\ 
}$\Xi ,$\textbf{\ }$\Omega ,$\textbf{\ }$\Lambda _{C,}$\textbf{\ }$\Xi _{C},$%
\textbf{\ }$\Sigma _{C},$\textbf{\ }$\Lambda _{b}...$\textbf{) in the BCC
model}. According to \textbf{Hypothesis I}, the excited Lee Particles with
the third component of the isospin $I_{z}=+1/2$ are protons, and the excited
Lee Particles with the third component of the isospin $I_{z}=-1/2$ are
neutrons. On the other hand, the unstable baryons are the energy band
excited states of the Lee Particles. Their intrinsic quantum numbers and
masses can be found using \textbf{Hypothesis V} and (\ref{mass})\textbf{.}
However, they are excited from the same particles-the Lee Particles. The
nucleons are the ground bands. Therefore, we can determine $V_{0}$ in
formula (\ref{mass})$,$ using the static masses (static energy) $M_{nucleon}$
of the nucleons. The static energy ($M_{nucleon}=939$ Mev (\cite{particle}))
of the nucleons should be the lowest energy ($V_{0})$ of the energy bands in
(\ref{mass}). Thus, at the ground states, we have 
\begin{equation}
\varepsilon ^{(0)}=V_{0}=M_{nucleon}=939\text{ Mev.}  \label{vzero}
\end{equation}
Fitting the theoretical mass spectrum to the empirical mass spectrum of the
baryons, we can also determine 
\begin{equation}
\alpha =h^{2}/2m_{b}a_{x}^{2}=360\text{ Mev}  \label{alpha}
\end{equation}
in (\ref{mass}). Thus, we have 
\begin{equation}
\varepsilon ^{(0)}(\vec{k},\vec{n})=V_{0}+\alpha E(\vec{k},\vec{n})=939+360E(%
\vec{k},\vec{n})\text{ \ \ (Mev).}  \label{massconst}
\end{equation}

If there were no strong interaction periodic field, we would only see
protons and neutrons as excited free Lee Particles ($N(939)$). We could not
see any other baryons because they would not exist. Due to the periodic
field, although protons and neutrons are still the essential state of the
excited Lee Particles, there is a slight chance that the Lee Particles are
excited to \textbf{the symmetry points} (see Fig. 1) of the periodic field.
Once at the symmetry points, the Lee Particles will show special symmetric
properties. Due to the periodic field, the parabolic energy curve of the
free Lee Particle will be changed to energy bands (see Eq. (\ref{mass}) and
Fig. 2-5). Also, there will be energy \textbf{gaps} on the surfaces of the
Brillouin zones which originate from the periodic field. \textbf{The
symmetry points} and \textbf{the energy gaps} will give the excited Lee
Particles some special properties and longer lives, which are different from
those of the free nucleons ($N(939)$) with the same energy. Because of these
properties, physicists naturally regard them as new baryons that are
different from protons and neutrons.

Using \textbf{Hypothesis V} and the energy bands (Fig. 2-5), we can find the
quantum numbers and masses of all excited energy bands. Then, from the
quantum numbers and the masses we can recognize the\textbf{\ unstable baryons%
}.

About the quantum numbers, considering the energy bands, we have:

\begin{enumerate}
\item  All energy band states have the baryon number $B=1$ from (\ref{baryon}%
).

\item  All energy bands with $\vec{n}$ = ($0$,$0$,$0$) (in the first
Brillouin zone) have $S=C=b=0$, spin s =1/2, and I=1/2 from \textbf{%
Hypothesis I}. They represent the baryon N(939) from (\ref{vzero}). 
\begin{equation}
\begin{tabular}{lllll}
$\vec{n}$ = ($0$,$0$,$0$) & S=C=b=0 & I =1/2 & s =1/2 & N(939).
\end{tabular}
\label{QN of N(939)}
\end{equation}

\item  Except for the first Brillouin zone, the isospin $I$ is determined by
(\ref{isomax}) and (\ref{isonext}). In some cases the degeneracy $d$ should
be divided into sub-degeneracies before using the formulas. Specifically, if
the degeneracy $d$ is larger than the rotary fold $R$ of the symmetry axis: 
\begin{equation}
d>R\text{,}  \label{degeneracy}
\end{equation}
then we assume that the degeneracy will be divided into $\gamma $
sub-degeneracies, where 
\begin{equation}
\gamma =d/R\text{.}  \label{subdegen}
\end{equation}
For the three axes which pass through the center point $\Gamma $ (the axis $%
\Delta (\Gamma -H)$, the axis $\Lambda (\Gamma -P)$, the axis $\Sigma
(\Gamma -N)$), the energy bands in the same degeneracy group have symmetric $%
\vec{n}$ ($\vec{n}=(n_{1},n_{2},n_{3})$) values (see Fig. 2(a), 2(b) and
3(a)). Hence, if the sub-degeneracy $d_{sub}\leq R$, it will not be divided
further. About ``symmetric'' $\vec{n}$ , we give a definition: \textbf{A
group of }$\vec{n}$\textbf{\ = (}$n_{1},n_{2},n_{3}$\textbf{) values is said
to be symmetric if any two }$\vec{n}$\textbf{\ values in the group can
transform to each other by various permutation (change component order) and
changing the sign ``}$\pm "$\textbf{\ (multiplied by ``-1'' ) of the
components (one, two, or all three). }For example, $(-2,-1,3)$ and $(-3,2,1)$
are symmetric, but $(-3,0,2)$ and $(-3,0,1)$ are asymmetric.

\qquad For other three symmetry axes which are on the surface of the first
Brillouin zone (the axis $D(P-N)$, the axis $F(P-H)$, the axis $G(M-N)$),
the energy bands in the same degeneracy group may have asymmetric $\vec{n}$
values (see Fig. 3(b), 4(a) and 4(b)). This may indicate that they belong to
different Brillouin zones. In such cases, even if a sub-degeneracy $%
d_{sub}\leq R$, it may still be divided further. Finding the criteria for
dividing the degeneracy depends on the structures of the Brillouin zones,
the irreducible representations of the single and double point groups \cite
{DoubleGroup}, and requires a higher order approximation. Hence, it is
beyond the scope of this paper. In order to simplify, we assume (some
phenomenological rules) that (a) the degeneracy of the energy bands which
are in the first and second Brillouin zones will be divided. (b) if $\Delta
\varepsilon \neq 0$ (see (\ref{eformula})), an asymmetric sub-degeneracy
should be divided at the end point $N$ (the lowest symmetry point, only has $%
8$ symmetric operations \cite{SymmetryOp}); may or may not be divided at end
point $P$ ($24$ symmetric operations \cite{SymmetryOp}) with a possibility
of $\ 50\%$ ; but should not be divided at the end points $H$ and $M$ (the
highest symmetry points, $48$ symmetric operations \cite{SymmetryOp}): 
\begin{eqnarray}
&&\text{if }\Delta \varepsilon \neq 0\text{, divided at end point }N\text{;}
\notag \\
&&\text{not divided at end points }H\text{ and }M\text{;}  \notag \\
&&\text{divided at end point }P\text{ with possibility of 50\%.}
\label{dividing}
\end{eqnarray}
However, if $\Delta \varepsilon =0$, the asymmetric sub-degeneracy should
not be divided. After finding the sub-degeneracy $d_{sub}$, we can use (\ref
{isomax}) ($d_{sub}$ = 2I+1) to find the isospin $I$.

\item  Except for the first Brillouin zone, the strange number $S$ is
determined by (\ref{strange}), where the number $4$ is the highest rotary
fold number ($R$) of the highest rotary symmetry axis. To be specific, from
Eq. (\ref{strange}) and Fig. 1, we get 
\begin{equation}
\Delta (\Gamma -H)\text{ is a }4\text{-fold rotation axis, }R=4\rightarrow
S=0;  \label{deltas}
\end{equation}
\begin{equation}
\Lambda (\Gamma -P)\text{ is a }3\text{-fold rotation axis, }R=3\rightarrow
S=-1;  \label{lambdas}
\end{equation}
\begin{equation}
\Sigma (\Gamma -N)\text{ is a }2\text{-fold rotation axis, }R=2\rightarrow
S=-2.  \label{sigmas}
\end{equation}
For the other three symmetry axes $D(P-N)$, $F(P-H)$, and $G(M-N),$ which
are on the surface of the first Brillouin zone (see Fig. 1), we determine
the strange numbers as follows: 
\begin{equation}
D(P-N)\text{ is parallel to axis }\Delta \text{, }S_{D}=S_{\Delta }=0;
\label{ds}
\end{equation}
\begin{equation}
F\text{ is parallel to an axis equivalent to }\Lambda \text{, }%
S_{F}=S_{\Lambda }=-1;  \label{fs}
\end{equation}
\begin{equation}
G\text{ is parallel to an axis equivalent to }\Sigma \text{, }%
S_{G}=S_{\Sigma }=-2\text{.}  \label{gs}
\end{equation}

\item  Except for the first Brillouin zone, other quantum numbers will also
be determind by \textbf{Hypothesis V}.

\item  When the energy is low, the baryons from different axes in the same
Brillouin zone, and with the same S, C, b, Q, I, I$_{Z}$, and $%
\overrightarrow{n}$ values, are ragarded as the same baryon. The mass of the
baryon is the lowest value of their energy curved surface. For example, in
the second Brillouin zone, the energy bands (on diffrent symmetry axes) with
n = (1, 1, 0) are the same band (the same energy curved surface in the three
dimensional phase space). They have the same quantum numbers ($%
S=-1,C=b=Q=I=I_{Z}=0$ ). The lowest energy is 1119 - the baryon $\Lambda
(1119)$.
\end{enumerate}

Starting from the axis $\Delta (\Gamma -H),$ we recognize the \textbf{%
unstable baryons}.

\subsection{The Axis $\Delta (\Gamma -H)$}

The axis $\Delta (\Gamma -H)$ is a $4$ fold rotary symmetry axis, $R=4$.
From (\ref{deltas}), we get the strange number $S=0$. For low energy levels,
there are $8$ and $4$ fold degenerate energy bands and single bands on the
axis. Since the axis has $R=4$, from (\ref{degeneracy}) and (\ref{subdegen}%
), the energy bands of degeneracy $8$ will be divided into two 4 fold
degenerate bands.

\subsubsection{The four fold degenerate bands on the axis $\Delta (\Gamma
-H) $}

For the 4 fold degenerate bands (see Fig. 2(a) and Fig. 5(a)), using (\ref
{isomax}), we get the isospin $I_{m}=3/2$, and using (\ref{charge}), we have 
$Q=2$, $1$, $0$, $-1$. Comparing them with the experimental results \cite
{particle} that the baryon families $\Delta (\Delta ^{++},\Delta ^{+},\Delta
^{0},\Delta ^{-})$ have $S=0$, $I=3/2$, $Q=2$, $1$, $0$, $-1$, we discover
that each four fold degenerate band represents a baryon family $\Delta $.
Using (\ref{isonext}), we get\ another $I=3/2-1=1/2$, and from (\ref{charge}%
), we get $Q=1$, $0$. From the facts \cite{particle} that the baryon
families $N(N^{+},N^{0})$ have $S=0$, $I=1/2$, and $Q=1$, $0$, we know that
there is another baryon family $N$ corresponding to each $\Delta $ family.
Using Fig. 2(a) and Fig. 5(a), we can get $E_{\Gamma },$ $E_{H},$ and $\vec{n%
}$ values. Then, putting the values of the $E_{\Gamma }$ and $E_{H}$ into
the energy formula (\ref{massconst}), we can find the values of the energy $%
\varepsilon ^{(0)}$. Finally, we have 
\begin{equation}
\begin{array}{lllll}
E_{H}=1 & \vec{n}=(\text{101,-101,011,0-11}) & \varepsilon ^{(0)}=1299 & 
\Delta (1299); & N(1299) \\ 
E_{\Gamma }=2\text{ } & \vec{n}=(\text{110,1-10,-110,-1-10}) & \varepsilon
^{(0)}=1659 & \Delta (1659); & N(1659) \\ 
E_{\Gamma }=2 & \vec{n}=(\text{10-1,-10-1,01-1,0-1-1}) & \varepsilon
^{(0)}=1659 & \Delta (1659); & N(1659) \\ 
E_{H}=3 & \vec{n}=(\text{112,1-12,-112,-1-12}) & \varepsilon ^{(0)}=2019 & 
\Delta (2019); & N(2019) \\ 
E_{\Gamma }=4 & \vec{n}=(\text{200,-200,020,0-20}) & \varepsilon ^{(0)}=2379
& \Delta (2379); & N(2379) \\ 
E_{H}=5 & \vec{n}=(\text{121,1-21,-121,--1-21}, & \varepsilon ^{(0)}=2739 & 
\Delta (2739); & N(2739) \\ 
& \text{ \ \ \ \ \ \ \ 211,2-11,-211,-2-11}) & \varepsilon ^{(0)}=2739 & 
\Delta (2739); & N(2739) \\ 
E_{H}=5 & \vec{n}=(\text{202,-202,022,0-22}) & \varepsilon ^{(0)}=2739 & 
\Delta (2739); & N(2739) \\ 
E_{H}=5 & \vec{n}=(\text{013,0-13,103,-103}) & \varepsilon ^{(0)}=2739 & 
\Delta (2739); & N(2739) \\ 
\ldots &  &  &  & 
\end{array}
\label{DELTA_4}
\end{equation}

\subsubsection{The single bands on the axis $\Delta (\Gamma -H)$}

From Fig. 2(a) and Fig. 5(b), we can see that there exist single bands on
the axis $\Delta $. From (\ref{isomax}), we have $I=0$. Using (\ref{strange}%
) and (\ref{charge}), we get $S=0$ and $Q=0+1/2(S+B)=1/2$ (a partial
charge). According to \textbf{Hypothesis V. 6}, we should use (\ref
{strangeflu}) instead of (\ref{strange}). Therefore, we have 
\begin{equation}
S_{\text{Single}}=\bar{S}_{\Delta }\pm \Delta S=0\pm 1\text{,}
\label{single}
\end{equation}
where $\Delta S=\pm 1$ from \textbf{Hypothesis IV}. The best way to
guarantee the validity of Eq.(\ref{strangebar}) in any small region is to
assume that $\Delta S$ takes $+1$ and $-1$ alternately from the lowest
energy band to higher ones. In fact, the $\vec{n}$ values are really
alternately taking positive and negative values: $E_{H}=1,$ $\vec{n}=($0, 0,
2$);$ $E_{\Gamma }=4,$ $\vec{n}=($0, 0, -2$);$ $E_{H}=9,$ $\vec{n}=($0, 0, 4$%
);$ $E_{\Gamma }=16,$ $\vec{n}=($0, 0, -4$);$ $E_{H}=25,$ $\vec{n}=($0, 0, 6$%
);$ $E_{\Gamma }=36,$ $\vec{n}=($0, 0, -6$)$ .... Using the fact, we can
find a phenomenological formula. If we define a function $Sign(\vec{n})$

\begin{equation}
Sign(\vec{n})=\frac{n_{1}+n_{2}+n_{3}}{\left| n_{1}\right| +\left|
n_{2}\right| +\left| n_{3}\right| }\text{ ,}  \label{Sign}
\end{equation}
then the \textbf{phenomenological formula is}

\begin{equation}
\Delta S=-(1+S_{axis})Sign(\vec{n}).  \label{Sign+-0}
\end{equation}
For the single states on the axis $\Delta ,$ we have $S_{axis}=0.$ Thue,
from (\ref{Sign+-0}), we get 
\begin{equation}
\Delta S=-(1+S_{axis})Sign(\vec{n})=-Sign(\vec{n}).  \tag{54-A}
\end{equation}
And for the single states on the axis $\Sigma ,$ since $S_{axis}=-2,$ we
have 
\begin{equation}
\Delta S=-(1+S_{axis})Sign(\vec{n})=Sign(\vec{n}).  \tag{54-B}
\end{equation}

Before recognizing the baryons, we need to discuss the fluctuation of energy.

The fluctuation of the strange number will be accompanied by an energy
change (\textbf{Hypothesis IV}). We assume that the change of the energy is
proportional to $(\Delta S),$ a number $K\equiv 4-R$ ($R$ is the rotary
number of the axis)$,$ and a number $J$ representing the energy level with a 
\textbf{phenomenological formula:} 
\begin{equation}
\Delta \varepsilon =\left\{ 
\begin{tabular}{l}
$(-1)^{K}100[(J-1)\times K-\delta (K)]\Delta S\text{ \ \ }J=1\text{, }2\text{%
, ...}$ \\ 
$\text{ \ \ \ \ \ \ \ \ }0\text{ \ \ \ \ \ \ \ \ \ \ \ \ \ \ \ \ \ \ \ \ \ \
\ \ \ \ \ \ \ \ \ \ \ \ \ \ \ \ \ \ \ \ \ }J=0\text{\ ,}$%
\end{tabular}
\right\}  \label{eformula}
\end{equation}
where $\delta (K)$ is a Dirac function ($\delta (K)=1$ when $K=0,$ and $%
\delta (K)=0$ when $K\neq 0$.), and $J$ is the energy level number ($J=0,1$, 
$2$, $3,$...) with asymmetric $\vec{n}$ values (or with partial electric
charge from (\ref{strange}) for single energy bands) at the lowest point of
the energy band. If the two end points of the axis have the same symmetries
(such as the points $\Gamma $ and $H$ of the axis $\Delta $), $J$ will take
0, 1, 2 ... from the lowest energy band to the higher ones, no matter at
which end point the lowest energy point is located (for example, see (\ref
{DELTA-ONE})). However, if the two end points of the axis have different
symmetries (such as the points $\Gamma $ and N of the axis $\Sigma $), $J$
will take 0, 1, 2 ... from the lowest energy band to the higher ones for
each of the two end points (for example, see (\ref{SIGMA_1})).

Applying (\ref{eformula}) to the symmetry axes, we have:

for the\ axis $\Delta $, $K=R-4=0$, 
\begin{equation}
\Delta \varepsilon =\left\{ 
\begin{tabular}{l}
$-100\times \Delta S\text{ \ J = 1, 2, ..., }$ \\ 
$\text{ \ \ \ \ \ \ \ \ }0\text{ \ \ \ \ \ \ \ \ J = 0\ \ \ \ \ \ \ \ \ ;}$%
\end{tabular}
\right\}  \label{flua}
\end{equation}
{}

for the\ axes $\Lambda $ and $F$, $K=4-R=1$,

\begin{equation}
\Delta \varepsilon =\left\{ 
\begin{tabular}{l}
$-100\times (J-1)\Delta S\text{ \ J = 1, 2, ..., }$ \\ 
$\text{ \ \ \ \ \ \ \ \ }0\text{ \ \ \ \ \ \ \ \ \ \ \ \ \ \ \ \ \ \ \ J =
0\ \ \ \ \ \ \ \ \ ;}$%
\end{tabular}
\right\}  \label{flub}
\end{equation}
\ \qquad \qquad \qquad \qquad\ \ \ \qquad \qquad \qquad

for the\ axes $\Sigma $, $G$, and $D$, $K=R-4=2$

\begin{equation}
\Delta \varepsilon =\left\{ 
\begin{tabular}{l}
$200\times (J-1)\Delta S\text{ \ J = 1, 2, ..., }$ \\ 
$\text{ \ \ \ \ \ \ \ \ }0\text{ \ \ \ \ \ \ \ \ \ \ \ \ \ \ \ \ J = 0\ \ \
\ \ \ \ \ \ ;}$%
\end{tabular}
\right\}  \label{fluc}
\end{equation}

Due to the fluctuation, the energy formula (\ref{massconst}) should be
changed to 
\begin{eqnarray}
\mathbf{\varepsilon } &=&\mathbf{\varepsilon }^{(0)}\mathbf{(\vec{k},\vec{n}%
)+\Delta \varepsilon }  \notag \\
&=&\mathbf{939+360E(\vec{k},\vec{n})+\Delta \varepsilon }\text{ \ .}
\label{UNITEDMASS}
\end{eqnarray}
\textbf{Formula (\ref{UNITEDMASS}) is the united mass formula which can give
masses of all the baryons.}\qquad \qquad \qquad \qquad \qquad \qquad \qquad
\qquad \qquad \qquad \qquad \qquad \qquad \qquad \qquad \qquad \qquad \qquad
\qquad \qquad \qquad \qquad \qquad \qquad \qquad \qquad \qquad \qquad \qquad
\qquad \qquad \qquad \qquad \qquad \qquad \qquad \qquad \qquad \qquad \qquad
\qquad \qquad \qquad \qquad \qquad \qquad \qquad \qquad \qquad \qquad \qquad
\qquad \qquad

After obtaining the energy fluctuation formula, we come back to the study of
the single bands on the axis $\Delta $.

First, at $E_{\Gamma }=0,J_{\Gamma }=0$, $\Delta \varepsilon =0$ from (\ref
{flua}), the lowest energy band with $\vec{n}=(0,0,0)$ represents the baryon 
$N(939)$ from\ (\ref{QN of N(939)}).

Then, we study the second lowest single energy band with $\vec{n}=(0,0,2)$
and $J=1$. The lowest $E$ of the band is at $E_{H}=1$. From (54-A), $\Delta
S=-Sign(\vec{n})$ $=-1$. Thus $S=\bar{S}+\Delta S=0+\Delta S=-1$, and $%
\Delta \varepsilon =100$ Mev from (\ref{flua})$\rightarrow $ the energy $%
\varepsilon =939+360E+\Delta \varepsilon =1399$ Mev from (\ref{UNITEDMASS}),
as well as $I=Q=0$ from (\ref{charge}). Therefore, it represents the baryon $%
\Lambda (1399)$.

For the third lowest band with $\vec{n}=(0,0,-2)$, the lowest $E$ of the
band is at $E_{\Gamma }=4$. Using (54-A), we get $\Delta S=+1$ . Thus, $%
\mathbf{S=\bar{S}}_{\Delta }\mathbf{+1=1}$\textbf{. }The energy $\varepsilon
=939+360\times 4+\Delta \varepsilon $ $=2379-100=2279$ from (\ref{UNITEDMASS}%
) and (\ref{flua}). Here $S=+1$ originates from the fluctuation $\Delta S=+1$
and there is an energy fluctuation of $\Delta \varepsilon =-100$. From 
\textbf{Hypothesis V. 6 (\ref{charmed})}, we know the energy band has a
charmed number $C=+1$. \textbf{It represents a new baryon with }$I=0$, $C=+1$%
, and $Q=+1$.\ Since it has a charmed number $C=+1$, \textbf{we will call it
the CHARMED baryon }$\Lambda _{C}^{+}(2279)$ \cite{charmed}. It is very
important to pay attention to \textbf{the Charmed baryon }$\Lambda
_{C}^{+}(2279)$\textbf{\ born here, on the single energy band, and from the
fluctuation } $\Delta S=+1$ and $\Delta \varepsilon =-100$ Mev.

Continuing the above procedure, from Fig. 5 (b), (\ref{UNITEDMASS}), (54-A)
and (\ref{flua}), we have (notice that the point H and the point $\Gamma $
of the axis have the same symmetries): 
\begin{equation}
\begin{array}{llllll}
E_{H}=1 & \vec{n}=(\text{002}) & \Delta S=-1 & J=1 & \Delta \varepsilon =+100
& \Lambda (1399) \\ 
E_{\Gamma }=4 & \vec{n}=(\text{00-2}) & \Delta S=+1 & J=2 & \Delta
\varepsilon =-100 & \Lambda _{C}^{+}(2279) \\ 
E_{H}=9 & \vec{n}=(\text{004}) & \Delta S=-1 & J=3 & \Delta \varepsilon =+100
& \Lambda (4279) \\ 
E_{\Gamma }=16 & \vec{n}=(\text{00-4}) & \Delta S=+1 & J=4 & \Delta
\varepsilon =-100 & \Lambda _{C}^{+}(6599) \\ 
E_{H}=25 & \vec{n}=(\text{006}) & \Delta S=-1 & J=5 & \Delta \varepsilon
=+100 & \Lambda (10039) \\ 
E_{\Gamma }=36 & \vec{n}=(\text{00-6}) & \Delta S=+1 & J=6 & \Delta
\varepsilon =-100 & \Lambda _{C}^{+}(13799) \\ 
\ldots &  &  &  &  & 
\end{array}
\label{DELTA-ONE}
\end{equation}

\subsection{The Axis $\Lambda (\Gamma -P)$}

The axis $\Lambda (\Gamma -P)$ is a $3$ fold rotary symmetry axis, $R=3$ and 
$S=-1$ from (\ref{lambdas}). From Fig. 2(b), we see that there is a single
energy band with $\vec{n}=(0,0,0)$, and all other bands are $3$ fold
degenerate energy bands ($d=3$) and $6$ fold degenerate bands ($d=6$).

At $E_{\Gamma }=0,$ $J_{\Gamma }=0$, $\Delta \varepsilon =0$ from (\ref{flub}%
), \textbf{the energy band with }$\vec{n}=(0,0,0)$ \textbf{represents the
baryon }$N\left( 939\right) $ from (\ref{QN of N(939)}) .

From (\ref{degeneracy}) and (\ref{subdegen}), the $6$ fold degenerate energy
bands will be divided into two energy bands with $3$ fold degeneracy. For
the 3 fold degenerate energy bands, using (\ref{isomax}) and (\ref{charge}),
we have $I_{m}=1$ and $Q=1,0,-1$. Comparing the intrinsic numbers $S$, $I$, $%
Q$ of the energy bands with those of the experimental baryon families $%
\Sigma (\Sigma ^{+},\Sigma ^{0},\Sigma ^{-})$, we know that each $3$ fold
degenerate energy band represents a baryon family $\Sigma (\Sigma
^{+},\Sigma ^{0},\Sigma ^{-})$. Furthermore, from (\ref{isonext}), we get
another possible value of the isospin $I=I_{m}-1=0$, and $Q=0$ from (\ref
{charge}). Consequently, there is also a baryon $\Lambda $ with $S=-1$, $I=0$
and $Q=0,$ corresponding to each $\Sigma $ family. Using Fig. 2(b), we get 
\begin{equation}
\begin{array}{llccc}
E_{P}=3/4 & \vec{n}=(\text{101,011,110}) & \varepsilon ^{(0)}=1209 & \Sigma
(1209)\text{;} & \Lambda (1209) \\ 
E_{\Gamma }=2 & \vec{n}=(\text{1-10,-110,01-1,} & \varepsilon ^{(0)}=1659 & 
\Sigma (1659)\text{;} & \Lambda (1659) \\ 
& \text{ \ \ \ \ \ \ \ 0-11,10-1,-101}) & \varepsilon ^{(0)}=1659 & \Sigma
(1659)\text{;} & \Lambda (1659) \\ 
E_{\Gamma }=2 & \vec{n}=(\text{-10-1,0-1-1,-1-10}) & \varepsilon ^{(0)}=1659
& \Sigma (1659)\text{;} & \Lambda (1659) \\ 
E_{P}=11/4 & \vec{n}=(\text{020,002,200}) & \varepsilon ^{(0)}=1929 & \Sigma
(1929)\text{;} & \Lambda (1929) \\ 
E_{P}=11/4 & \vec{n}=(\text{121,211,112}) & \varepsilon ^{(0)}=1929 & \Sigma
(1929)\text{;} & \Lambda (1929) \\ 
E_{\Gamma }=4 & \vec{n}=(\text{0-20,-200,00-2}) & \varepsilon ^{(0)}=2379 & 
\Sigma (2379)\text{;} & \Lambda (2379) \\ 
E_{P}=19/4 & \vec{n}=(\text{1-12,-112,21-1,} & \varepsilon ^{(0)}=2649 & 
\Sigma (2649)\text{;} & \Lambda (2649) \\ 
& \text{ \ \ \ \ \ \ \ 2-11,12-1,-121}) & \varepsilon ^{(0)}=2649 & \Sigma
(2649)\text{;} & \Lambda (2649) \\ 
E_{P}=19/4 & \vec{n}=(\text{202,022,220}) & \varepsilon ^{(0)}=2649 & \Sigma
(2649)\text{;} & \Lambda (2649) \\ 
E_{\Gamma }=6 & \vec{n}=(\text{-211,2-1-1,2-1-1}, & \varepsilon ^{(0)}=3099
& \Sigma (3099)\text{;} & \Lambda (3099) \\ 
& \text{\ \ \ \ \ \ \ 11-2,-12-11-21}) & \varepsilon ^{(0)}=3099 & \Sigma
(3099)\text{;} & \Lambda (3099) \\ 
& \vec{n}\text{=(-1-21,1-2-1,-11-2,} & \varepsilon ^{(0)}=3099 & \Sigma
(3099)\text{;} & \Lambda (3099) \\ 
& \text{\ \ \ \ \ \ 1-1-2,-21-1,-2-11}) & \varepsilon ^{(0)}=3099 & \Sigma
(3099)\text{;} & \Lambda (3099) \\ 
& \vec{n}\text{=(-1-2-1,-1-1-2,-2-1-1)} & \varepsilon ^{(0)}=3099 & \Sigma
(3099)\text{;} & \Lambda (3099) \\ 
\ldots &  &  &  & 
\end{array}
\label{LAMBDA_3}
\end{equation}

\subsection{The Axis $\Sigma (\Gamma -N)$}

The axis $\Sigma (\Gamma -N)$ is a $2$ fold symmetry axis, $R=2$ and $S=-2$
from (\ref{sigmas}). For low energy levels, there are $4$ fold degenerate
energy bands, $2$ fold degenerate energy bands, and single energy bands on
the axis (see Fig. 3(a)). According to (\ref{degeneracy}) and (\ref{subdegen}%
), each $4$ fold energy band will be divided into two $2$ fold degenerate
energy bands.

\subsubsection{The two fold degenerate energy bands on the axis $\Sigma
(\Gamma -N)$}

For the two fold degenerate energy bands (see Fig. 3(a)), $I=1/2$ from (\ref
{isomax}), and $Q=0,-1$ from (\ref{charge}). Comparing the intrinsic numbers 
$S$, $I$, $Q$ of the $2$ fold energy band with those of the experimental
baryon families $\Xi (\Xi ^{0}$, $\Xi ^{-})$, we know that each $2$ fold
degenerate energy band represents a baryon family $\Xi (\Xi ^{0}$, $\Xi
^{-}) $ because they have the same quantum numbers ( $I=1/2$, $S=-2$ and $%
Q=0,-1$): 
\begin{equation}
\begin{array}{llcc}
E_{\Gamma }=2 & \vec{n}=(\text{1-10,-110}) & \varepsilon ^{(0)}=1659 & \Xi
(1659) \\ 
E_{N}=5/2 & \vec{n}=(\text{200,020}) & \varepsilon ^{(0)}=1839 & \Xi (1839)
\\ 
E_{\Gamma }=4 & \vec{n}=(\text{002,00-2}) & \varepsilon ^{(0)}=2379 & \Xi
(2379) \\ 
& \vec{n}=(\text{-200,0-20}) & \varepsilon ^{(0)}=2379 & \Xi (2379) \\ 
E_{N}=9/2 & \vec{n}=(\text{112,11-2}) & \varepsilon ^{(0)}=2559 & \Xi (2559)
\\ 
E_{\Gamma }=6 & \vec{n}=(\text{-1-12,-1-1-2}) & \varepsilon ^{(0)}=3099 & 
\Xi (3099) \\ 
\ldots &  &  & 
\end{array}
\label{Segma-2}
\end{equation}

\subsubsection{The four fold degenerate energy bands on the axis $\Sigma
(\Gamma -N)$}

According to (\ref{subdegen}), each $4$ degenerate energy band (see Fig.
3(a)) on the symmetry axis $\Sigma $ will be divided into two $2$ fold
degenerate bands, which represent $2$ baryon families $\Xi (\Xi ^{0}$, $\Xi
^{-})$. Considering that the symmetry axis is a $2$ fold rotary symmetric
axis, we know that the division is reasonable: 
\begin{equation}
\begin{array}{llcc}
E_{N}=3/2 & \vec{n}=(\text{101,10-1,011,01-1}) & \varepsilon ^{(0)}=1479 & 2%
\text{ }\Xi (1479) \\ 
E_{\Gamma }=2 & \vec{n}=(\text{-101,-10-1,0-11,0-1-1}) & \varepsilon
^{(0)}=1659 & 2\text{ }\Xi (1659) \\ 
E_{N}=7/2 & \vec{n}=(\text{121,12-1,211,21-1}) & \varepsilon ^{(0)}=2199 & 2%
\text{ }\Xi (2199) \\ 
E_{N}=11/2 & \vec{n}=(\text{-121,-12-1,2-11,2-1-1}) & \varepsilon ^{(0)}=2919
& 2\text{ }\Xi (2919) \\ 
E_{\Gamma }=6 & \vec{n}=(\text{1-12,1-1-2,-112,-11-2}) & \varepsilon
^{(0)}=3099 & 2\text{ }\Xi (3099) \\ 
E_{\Gamma }=6 & \vec{n}=(\text{1-21,1-2-1,-211,-21-1}) & \varepsilon
^{(0)}=3099 & 2\text{ }\Xi (3099) \\ 
E_{\Gamma }=6 & \vec{n}=(\text{-1-21,-1-2-1,-2-11,-2-1-1}) & \varepsilon
^{(0)}=3099 & 2\text{ }\Xi (3099) \\ 
\ldots &  &  & 
\end{array}
\label{SEGMA-4}
\end{equation}

\subsubsection{The single energy bands on the axis $\Sigma (\Gamma -N)$}

From Fig. 3(a) and Fig. 5(c), we can see that single bands exist on\ the
axis $\Sigma $. For the single energy bands, we have $S=-2$ , $I=0$, and $%
Q=-1/2$ from (\ref{charge}). According to \textbf{Hypothesis V}. 6, we have
to use (\ref{strangeflu}) instead of (\ref{strange}): 
\begin{equation}
S_{\text{Single}}=\bar{S}_{\Sigma }\pm \Delta S=-2\pm 1\text{.}
\label{sigmasingle}
\end{equation}
Using (54-B), we can find the $\Delta S$ for the single bands.\ The strange
number will take $-1$ and $-3$ alternately from the lower to the higher
energy bands.

From (\ref{fluc}), we can find the energy fluctuations. Since the end points 
$\Gamma $ and N of the axis $\Sigma $ have different symmetries, J will take
0, 1, 2, ... from the lowest energy band to higher ones for each of the two
end points respectively. Particularly, at $E_{\Gamma }=0,$ $J_{\Gamma }=0,$ $%
\Delta \varepsilon =0,$ at $E_{\Gamma }=2,$ $J_{\Gamma }=1$, $\Delta
\varepsilon =0$, at $E_{\Gamma }=8,$ $J_{\Gamma }=2$, $\Delta \varepsilon
=-200$... Similarly, at $E_{N}=1/2$, $J_{N}=0$, $\Delta \varepsilon =0;$ at $%
E_{N}=9/2$, $J_{N}=1$, $\Delta \varepsilon =0;$ at $E_{N}=25/2$, $J_{N}=2$, $%
\Delta \varepsilon =200...$

At $E_{\Gamma }=0,$ $J_{\Gamma }=0$, $\Delta \varepsilon =0$ from (\ref{fluc}%
), \textbf{the energy band with }$\vec{n}=(0,0,0)$ \textbf{represents the
baryon }$N\left( 939\right) $ from (\ref{QN of N(939)}) .

At $E_{N}=1/2$, $J_{N}=0$, $\Delta \varepsilon =0$ from (\ref{fluc}), the
second lowest energy band with $\vec{n}=(1,1,0),$ $\Delta S=+1$ from (54-B) $%
\rightarrow S=\overline{S}+\Delta S=-1$. Using (\ref{UNITEDMASS}), the
energy of the energy band is $\varepsilon =1119$ Mev. Therefore, the energy
band represents the baryon $\Lambda (1119).$

At $E_{\Gamma }=2,$ $J_{\Gamma }=1$, $\Delta \varepsilon =0$ from (\ref{fluc}%
), the third lowest band with $\vec{n}=($-1, -1, 0$)$ should have $\Delta
S=-1$ from from (54-B). So that $S=-3$, $I=0$, and $Q=-1$. From the
experimental results that the baryon $\Omega ^{-}$ \cite{omiga} has $S=-3$, $%
I=0$, and $Q=-1$, we obtain that the energy band represents a baryon $\Omega
^{-}(1659)$.

At $E_{N}=9/2$, $J_{N}=1\rightarrow \Delta \varepsilon =0$ from (\ref{fluc})$%
,$ the fourth band ($\vec{n}=($2, 2, 0$)$) has $\Delta S=+1$ from (54-B), $%
S=-2+1=-1,$and $J_{N}=1,$ $\varepsilon =2559$ Mev. Since $\Delta \varepsilon
=0,$ from (\ref{bottom}), we infer that this energy band represents a baryon 
$\Lambda (2559)$.

At $E_{\Gamma }=8,$ $J_{\Gamma }=2,$ the fifth one ($\vec{n}$=(-2, -2, 0)$%
\rightarrow \Delta $S=-1 from (54-B)) has $S=-3$, $I=0$ and $Q=-1$, so it
represents a baryon $\Omega ^{-}(3619)$.

At $E_{N}=25/2,$ $J_{N}=2$, the sixth one ($\vec{n}=($3, 3, 0$))$ has $%
\Delta S=+1$ from (54-B)$\rightarrow $ $S=-2+1=-1$ from (\ref{strangeflu}),
and $\varepsilon =5439+200=5639$ from (\ref{UNITEDMASS}). According to 
\textbf{Hypothesis V}. 6 (\ref{bottom}), we know that the energy band has a
bottomed number $b=-1$. It represents a baryon with $I=0$, $b=-1$, $Q=0$,
and $\varepsilon =5639$. Since it has a bottomed number $b=-1$ , we call
this baryon as the \textbf{Bottom baryon }$\mathbf{\Lambda }_{b}\mathbf{%
(5639)}$ \cite{bottom}. It is very important to pay attention to the \textbf{%
experimentally confirmed bottom baryon }$\mathbf{\Lambda }_{b}\mathbf{(5639)}
$ \textbf{born here, on the single energy band, and from the fluctuation }$%
\Delta S=+1$ and $\Delta \varepsilon =200$~Mev. Using Fig. 5(c), we find the
baryons: 
\begin{equation}
\begin{array}{llllll}
E_{N}=1/2 & \vec{n}=(\text{110}) & \Delta S=+1 & J_{N}=0 & \Delta
\varepsilon =0 & \Lambda (1119) \\ 
E_{\Gamma }=2 & \vec{n}=(\text{-1-10}) & \Delta S=-1 & J_{\Gamma }=1 & 
\Delta \varepsilon =0 & \Omega ^{-}(1659) \\ 
E_{N}=9/2 & \vec{n}=(\text{220}) & \Delta S=+1 & J_{N}=1 & \Delta
\varepsilon =0 & \Lambda (2559) \\ 
E_{\Gamma }=8 & \vec{n}=(\text{-2-20}) & \Delta S=-1 & J_{\Gamma }=2 & 
\Delta \varepsilon =-200 & \Omega ^{-}(3619) \\ 
E_{N}=25/2 & \vec{n}=(\text{330}) & \Delta S=+1 & J_{N}=2 & \Delta
\varepsilon =200 & \Lambda _{b}(5639) \\ 
E_{\Gamma }=18 & \vec{n}=(\text{-3-30}) & \Delta S=-1 & J_{\Gamma }=3 & 
\Delta \varepsilon =-400 & \Omega ^{-}(7019) \\ 
E_{N}=49/2 & \vec{n}=(\text{440}) & \Delta S=+1 & J_{N}=3 & \Delta
\varepsilon =400 & \Lambda _{b}(10159) \\ 
\ldots &  &  &  &  & 
\end{array}
\label{SIGMA_1}
\end{equation}

\subsection{The Axis $D(P-N)$}

The axis $D(P-N)$ is a 2 fold rotary symmetry axis, $R=2$. It is parallel to
the axis $\Delta $ (see Fig. 1), thus $S=0$ from (\ref{ds}). For low energy
levels, there are $4$ fold degenerate energy bands and $2$ fold degenerate
energy bands on the axis (see Fig. 3(b)).

\subsubsection{The four fold energy bands on the axis $D(P-N)$}

From Fig. 3(b), we can see that there are $4$ fold degenerate energy bands
on the axis. Using (\ref{isomax}), we get $I=3/2$, and from (\ref{charge}),
we get $Q=2,1$, $0,$ $-1$. From the fact \cite{particle} that the baryon
families $\Delta (\Delta ^{++},\Delta ^{+},\Delta ^{0},\Delta ^{-})$ have $%
S=0$, $I=3/2$, and $Q=2,1$, $0$,-1 we know that each $4$ fold degeneracy
energy band represents a baryon family $\Delta $.\ However,\ each $4$ fold
degenerate energy band has 4 asymmetric $\vec{n}$ values. They can be
divided into two groups at the point N (with 50\% possibility at the point
P). Each of them has $2$ symmetric $\vec{n}$ values. Using (\ref{isomax}),
we get $I=1/2$, and from (\ref{charge}), we get $Q=1$, $0$. From the fact 
\cite{particle} that the baryon families $N(N^{+},N^{0})$ have $S=0$, $I=1/2$%
, and $Q=1$, $0$, we know that each $2$ fold sub-degeneracy energy band
represents a baryon family $N$. Hence, for the $4$ fold energy bands we have 
\begin{equation}
\begin{array}{llll}
E_{N}=5/2 & \vec{n}=(\text{1-10,-110,020,200}) & \varepsilon ^{(0)}=1839 & 
\\ 
\text{ \ }\Delta S=0 & \vec{n}=(\text{1-10,-110}) & S=0 & N(1839) \\ 
\text{ \ }\Delta S=0 & \vec{n}=(\text{020,200}) & S=0 & N(1839) \\ 
E_{P}=11/4 & \vec{n}=(\text{-101,0-11,211,121}) & \varepsilon ^{(0)}=1929 & 
\Delta (1929) \\ 
\text{ \ }\Delta S=0 & \vec{n}=(\text{-101,0-11}) & S=0 & N(1929) \\ 
\text{ \ }\Delta S=0 & \vec{n}=(\text{211,121}) & S=0 & N(1929) \\ 
E_{N}=7/2 & \vec{n}=(\text{12-1,21-1,-10-1,0-1-1}) & \varepsilon ^{(0)}=2199
&  \\ 
\text{ \ }\Delta S=0 & \vec{n}=(\text{12-1,21-1}) & S=0 & N(2199) \\ 
\text{ \ }\Delta S=0 & \vec{n}=(\text{-10-1,0-1-1}) & S=0 & N(2199) \\ 
E_{P}=19/4 & \vec{n}=(\text{-112,1-12,202,022}) & \varepsilon ^{(0)}=2649 & 
\Delta (2649) \\ 
\text{ \ }\Delta S=0 & \vec{n}=(\text{-112,1-12}) & S=0 & N(2649) \\ 
\text{ \ }\Delta S=0 & \vec{n}=(\text{202,022}) & S=0 & N(2649) \\ 
\ldots &  &  & 
\end{array}
\label{D-4}
\end{equation}

\subsubsection{The two fold energy bands on the axis $D(P-N)$}

There are symmetric and asymmetric $\vec{n}$ values in the $2$ fold energy
bands (see Fig. 3(b)). The $2$ fold energy bands with symmetric $\vec{n}$
values will not be divided because the degeneracy is not greater than the
rotary number ($d=R=2$). The bands have $S=0$, $I=1/2$, and $Q=+1$, $0$.
Since the experimental baryon family $N(N^{+}$,$N^{0})$ has the same $S$, $I$
and $Q$, we know that the two energy bands represent a baryon family $N$.
However, the case for $2$ fold energy bands with asymmetric $\vec{n}$ values
is not so simple.

At $E_{N}=1/2$, $J_{N}=0$ $\rightarrow \Delta \varepsilon =0$ from (\ref
{fluc}), there are two energy bands with asymmetric $\vec{n}=(000,110)$.
Since the two energy bands are in different Brillouin zones, they will be
divided into $2$ single bands. The energy band with $\vec{n}=(110)$ belongs
to the second Brillouin zone, and it represents the baryon $\Lambda (1119)$
with S = -1 from the first row of (\ref{SIGMA_1}). For another band with $%
\vec{n}=(000)$, it represents the baryon $N(939)$\textbf{\ }from (\ref{QN of
N(939)}).

At $E_{P}=11/4$, $J_{P}=1(J_{P}=0$ at $E_{P}=3/4)\rightarrow \Delta
\varepsilon =0$ from (\ref{fluc}), $\vec{n}=(002,112)-$ asymmetric $\vec{n}$
values. Using (\ref{dividing}), the two energy bands may be divided with a
possibility of $50\%$. However, the fluctuation of energy $\Delta
\varepsilon =0$, hence there is not enough energy fluctuation for the two
bands to be divided. Thus, the two energy bands may not be divided. They
represent a baryon family $N(1929)$.

There are two asymmetric 2 fold energy bands at $E_{N}=9/2$ ($\vec{n}%
=(220,-1-10)$ and $\vec{n}=(11-2,00-2)$). Using (\ref{fluc}), the energy
fluctuation for the first 2 energy bands ($\vec{n}=(220,-1-10)$, $J_{N}=1$)
is $\Delta \varepsilon =0$. Hence, they can not be divided. They represent a
baryon family $N(2559)$. However, the energy fluctuation for the second 2
energy bands ($\vec{n}=(11-2,00-2)$, $J_{N}=2$) is $\Delta \varepsilon
=200(2-1)\Delta S=200\Delta S$. Using (\ref{dividing}), the 2 fold bands
should be divided into 2 single energy bands. Thus, from (\ref{strangeflu}), 
$S=0+\Delta S=\pm 1$. According to (\ref{charmed}), one of the 2 single
energy bands ($\vec{n}=(00-2)$ from (\ref{DELTA-ONE}))\ has a charmed number 
$C=S=+1$ (with $I=0$ and $Q=+1$), and it represents a Charmed baryon $%
\Lambda _{C}^{+}(2759)$; the other has a strange number $S=-1$ (with $I=0$
and $Q=0$), and it represents a baryon $\Lambda (2359)$.

Therefore, we have

\begin{equation}
\begin{array}{llll}
E_{N}=1/2 & \vec{n}=(\text{000,110}) & \varepsilon ^{(0)}=1119 &  \\ 
\text{\ \ }J_{N}=0 & \vec{n}=(\text{000}) & \text{S=C=b=0} & N(939) \\ 
\text{ \ }\Delta \varepsilon =0 & \vec{n}=(\text{110}) & S=-1 & \Lambda
(1119) \\ 
E_{P}=3/4 & \vec{n}=(\text{101,011}) & \varepsilon ^{(0)}=1209 & N(1209) \\ 
E_{N}=3/2 & \vec{n}=(\text{10-1,01-1}) & \varepsilon ^{(0)}=1479 & N(1479)
\\ 
E_{P}=11/4 & \vec{n}=(\text{002,112}) & \varepsilon ^{(0)}=1929 &  \\ 
\text{ \ }J_{P}=1 & \text{ \ }\Delta S=0 & \Delta \varepsilon =0 & N(1929)
\\ 
E_{N}=9/2 &  & \varepsilon ^{(0)}=2559 &  \\ 
\text{ \ }J_{N}=1 & \vec{n}=(\text{220,-1-10}) & \Delta \varepsilon =0 & 
N(2559) \\ 
\text{ \ }J_{N}=2 & \vec{n}=(\text{11-2,00-2}) &  &  \\ 
\text{ \ \ \ \ }\vec{n}\text{=00-2} & \text{ \ }\Delta S=+1 & \Delta
\varepsilon =+200 & \Lambda _{C}^{+}(2759) \\ 
\text{ \ \ \ \ }\vec{n}\text{=11-2} & \text{ \ }\Delta S=-1 & \Delta
\varepsilon =-200 & \Lambda (2359) \\ 
E_{P}=19/4 & \vec{n}=(\text{-121,2-11}) & \varepsilon ^{(0)}=2649 & N(2649)
\\ 
E_{N}=11/2 & \vec{n}=(\text{2-1-1,-12-1}) & \varepsilon ^{(0)}=2919 & N(2919)
\\ 
\ldots &  &  & 
\end{array}
\label{d_two}
\end{equation}

\subsection{The Axis $F(P-H)$}

The axis $F(P-H)$ is a $3$ fold symmetry axis, from (\ref{fs}), $S=-1$. For
low energy levels, there are $6$ fold energy bands, $3$ fold energy bands,
and single energy bands on the axis (see Fig. 4(a)).

\subsubsection{The single energy bands on the axis $F(P-H)$}

For the single energy band, the strange number $S=-1$, and $I=Q=0$ from (\ref
{isomax}) and (\ref{charge}). Each single energy band represents a baryon $%
\Lambda $. From Fig. 4(a) we have 
\begin{equation}
\begin{array}{llllll}
E_{P}=3/4 & \vec{n}=(\text{110}) & \varepsilon ^{(0)}=1209 & S=-1 & I=Q=0 & 
\Lambda (1209) \\ 
E_{H}=3 & \vec{n}=(\text{-1-12}) & \varepsilon ^{(0)}=2019 & S=-1 & I=Q=0 & 
\Lambda (2019) \\ 
\ldots &  &  &  &  & 
\end{array}
\label{F-1}
\end{equation}

\subsubsection{The three fold energy bands on the axis $F(P-H)$}

For the 3 fold energy band (see Fig. 4(a)), $S=-1$, $I=1$ from (\ref{isomax}%
), and $Q=1,0,-1$ from (\ref{charge}). Comparing them with the experimental
result that the families $\Sigma (\Sigma ^{+}$, $\Sigma ^{0}$, $\Sigma ^{-})$
have the same intrinsic quantum numbers, we find that the 3 fold degenerate
energy bands represent the baryon families $\Sigma (\Sigma ^{+}$, $\Sigma
^{0}$, $\Sigma ^{-})$. For another possible isospin, we have $I=0$ from (\ref
{isonext}), and $Q=0$ from (\ref{charge}). This suggests the existence of
another baryon $\Lambda $. Using (\ref{dividing}), we know that the 3 fold
degenerate energy bands (all have asymmetric $\vec{n}$ values) will not be
divided at the point $H$, but may be divided at the point $P$ with $a$
possibility of $50\%$. The division at point $P$ will result in a single
band representing a baryon $\Lambda $, and a 2 fold baryon (with symmetric $%
\vec{n}$ values) representing a baryon family $N$ ($\Delta S=+1$) or $\Xi $($%
\Delta S=-1$).

At $E_{P}=3/4$, $J_{P}=0,$ the 3 energy bands with asymmetric $\vec{n}%
=(000,101,011)$ are in different Brillouin zones. They will be divided into
a single band with $\vec{n}=(000)$ (in the first Brillouin zone) and a two
fold energy band with $\vec{n}=(101,011)$ (in the second Brillouin zone). 
\textbf{The single energy band\ with }$\vec{n}$\textbf{\ }$\mathbf{=(0,0,0)}$
\textbf{represents }$N(939)$\textbf{\ from (\ref{QN of N(939)}).} The two
fold energy band with $\vec{n}=(101,011)$ represents $N(1209)$ from the
fourth line of (\ref{d_two}).

At $E_{P}=11/4$, $\vec{n}=(112,1-10,-110),$ $J_{P}=1\rightarrow \Delta
\varepsilon =0$ from (\ref{flub}). Since $\Delta \varepsilon =0,$ the three
energy bands will not be divided (see (\ref{dividing})). Thus, the energy
bands represent $\Sigma (1929)$ and $\Lambda (1929)$.

At $E_{P}=19/4$, $J_{P}=2\rightarrow \Delta \varepsilon =-100\Delta S$ from (%
\ref{flub})$,$ the three energy bands with $\vec{n}=($220,21-1,12-1$)$ may
be divided with $a$ possibility of $50\%$ (see (\ref{dividing})). Hence the
energy bands may represent $\Sigma (2649)$ and $\Lambda (2649)$ (if not
divided), or 2$\Lambda (2649)$ and $N(2549)$ or $\Xi (2749)$ (if divided).

We have

\begin{equation}
\begin{array}{llll}
E_{P}=3/4 & \vec{n}=(\text{000,011,101}) & \varepsilon ^{(0)}=1209 &  \\ 
\text{ \ }J_{P}=0 & \vec{n}=(\text{000}) & \text{S=C=b=0} & N(939) \\ 
\text{ \ }\Delta \varepsilon =0 & \vec{n}=(\text{011,101}) & \Delta S=+1 & 
N(1209) \\ 
E_{H}=1 & \vec{n}=(\text{002,-101,0-11}) & \Sigma (1299) & \Lambda (1299) \\ 
E_{P}=11/4 & \vec{n}=(\text{112,1-10,-110}) & \Sigma (1929) & \Lambda (1929)
\\ 
\text{ \ }J_{P}=1 & \Delta \varepsilon =0 &  &  \\ 
E_{H}=3 & \vec{n}=(\text{-1-10,112,1-12}) & \Sigma (2019) & \Lambda (2019)
\\ 
E_{P}=19/4 & \vec{n}=(\text{220,21-1,12-1}) & \Sigma (2649) & \Lambda (2649)
\\ 
\text{\ \ }J_{P}=2 & \vec{n}=(\text{220}) & \Delta S=0 & \Lambda (2649) \\ 
\text{ \ }\Delta \varepsilon =-100 & \vec{n}=(\text{21-1,12-1}) & \Delta S=+1
& N(2549) \\ 
&  & \Delta S=-1 & \Xi (2749) \\ 
... &  &  & 
\end{array}
\label{F-3}
\end{equation}

\subsubsection{The six fold energy bands ($d=6$) on the axis $F(P-H)$}

The $6$ fold energy bands on the axis $F$ are a special case. The $6$
asymmetric $\vec{n}$ values consist of three groups, each of them has $2$
symmetric $\vec{n}$ values. However, since the symmetry axis $F$ has a
rotary $R=3$, there are two ways to divide the energy bands: A) dividing the
energy bands according to (\ref{degeneracy}) and (\ref{subdegen}); B)
dividing the energy bands according to the symmetry of $\vec{n}$ values.

(A). From (\ref{degeneracy}) and (\ref{subdegen}), each $6$ fold energy band
will be divided into two $3$ fold energy bands first. Since each $3$ fold
sub-degeneracy band has asymmetric $\vec{n}$ values, according to (\ref
{dividing}), the $3$ energy bands may be divided (second division) further
at point $P$ with $a$ possibility of $50$\%, but not be at the point $H$. If
the two $3$ fold sub-degeneracy bands are not divided further, they will
represent two baryon families $\Sigma (\Sigma ^{+}$, $\Sigma ^{0}$, $\Sigma
^{-})$ ($S=-1$, $I=1$, $Q=1,0,-1$). However, if the second division occurs
at the point $P$, in order to keep (\ref{strangebar}), both of the two $3$
fold sub-degeneracy bands will be divided, resulting in two $2$ fold bands
(one with $\Delta S=+1$, the other with $\Delta S=-1$) and two single bands.
According to (\ref{C&S}), the $2$ fold energy band with $\Delta S=+1$ will
keep $S$ unchanged while increasing the Charmed number $C$ by $1$. Thus, the
two $2$ fold energy bands represent a baryon family $\Xi _{C}(\Xi _{C}^{+}$, 
$\Xi _{C}^{0})$ ($C=+1$, $S=-1$, $I=1/2$, $Q=1,$ $0$) and a baryon family $%
\Xi (\Xi ^{0}$, $\Xi ^{-})$ ($S=-2$, $I=1/2$, $Q=0,-1$), while the two
single bands represent two $\Lambda $ baryons ($S=-1$, $I=0$, $Q=0$).

At $E_{P}=11/4$, $\vec{n}=($01-1,10-1,121,211,020,200$)$, $J_{P}=1$ ($%
J_{P}=0 $ at $E_{P}=3/4$), $\Delta \varepsilon =0$ from (\ref{flub}). Since
there is not enough fluctuation energy to divide the two $3$ fold
sub-degeneracies (\ref{dividing}), they are not divided. Thus, the $6$ fold
band represents 
\begin{equation}
\Sigma (1929)_{\text{({\small 01-1,10-1,121})}}+\Sigma (1929)_{\text{{\small %
(211,020,200)}}}\text{ and }\Lambda (1929)_{121}+\Lambda (1929)_{211}
\end{equation}

At $E_{P}=19/4$, $\vec{n}=($202,022,-121,2-11,0-1-1,-10-1$)$, $J_{P}=2$.
According to (\ref{flub}), the energy fluctuation $\ \Delta \varepsilon
=-100\times \Delta S$. \ Thus, the two $3$ fold bands represent (with a
possibility of $50$\% to be divided at the point $P$) 
\begin{equation}
\Sigma (2649)\text{+}\Sigma (2649)\rightarrow \lbrack \Xi _{C}(2549)\text{+}%
\Lambda (2649)]\text{+}[\Xi (2749)\text{+}\Lambda (2649)]  \label{KERSA-C}
\end{equation}
It is very important to pay attention to the\textbf{\ baryon }$\Xi
_{C}(2549)_{(202,}$ $_{022)}$\cite{KESI_C} \textbf{born here, on the 6 fold
energy band, after the second division from the fluctuation } $\Delta S=+1$
and $\Delta \varepsilon =-100$ Mev.

At $E_{H}=5$, $\vec{n}=($0-20,-200,-211,1-21,013,103$)$. According to (\ref
{dividing}), the two $3$ fold sub-degeneracies are not divided. Thus, the $6$
fold band represents 
\begin{equation}
\Sigma (2739)+\Sigma (2739)\text{ and }\Lambda (2739)+\Lambda (2739)
\end{equation}

At $E_{H}=5$, $\vec{n}=($0-22,-202,-2-11,-1-21,0-13,-103$)$. Similarly, we
have 
\begin{equation}
\Sigma (2739)+\Sigma (2739)\text{ and }\Lambda (2739)+\Lambda (2739)
\end{equation}

At $E_{P}=27/4,\vec{n}=($-12-1,2-1-1,301,031,222,00-2$),J_{P}=3$. Similar to
the case of $E_{P}=19/4$, $\Delta \varepsilon =-200\times \Delta S$ and the $%
6$ fold band represents (with $a$ possibility of $50$\% to be divided at the
point $P$) 
\begin{equation}
2\text{ }\Sigma (3369)\rightarrow \lbrack \Xi _{C}(3169)+\Lambda
(3369)]+[\Xi (3569)+\Lambda (3369)]\text{ ...}  \label{F_6_KESIC}
\end{equation}
\qquad \qquad \qquad \qquad \newline

\ \ (B). According to the symmetry values of $\vec{n}$, each $6$ fold
degeneracy can be divided into a $2$ fold sub-degeneracy and a $4$ fold
sub-degeneracy. Since this kind of division will result in partial charges
from (\ref{strange}) and (\ref{charge}), we get $S=\bar{S}+\Delta S=-1\pm 1$
from (\ref{strangeflu}). To keep (\ref{strangebar}), the $2$ fold energy\
band will have $\Delta S=-1$, while the $4$ fold band will have $\Delta S=+1$%
. (Another possibility is that the $2$ fold energy\ band has $\Delta S=+1$,
while the $4$ fold band has $\Delta S=-1$ ($S=-2)$. However, the possibility
of obtaining a $4$ fold band with strange number $S=-2$ is very small.
Hence, it will be ignored here).

Since the symmetric axis is a $3$ rotary one, the $4$ fold energy band will
be divided further. In order to balance the $3$ rotary symmetry of the axis
and the $2$ fold symmetry of the $\vec{n}$ values, the second division
should \textbf{keep the 3 rotary symmetry} (\textbf{may break the }$\mathbf{2%
}$\textbf{\ fold symmetry of }$\vec{n}$ ) because the first division has
kept the $2$ fold symmetry of the $\vec{n}$ values.\ Thus, the $4$ fold
energy band ( $S=-1+1=0$, $\Delta $ family ) may be divided into ($\Lambda
_{C}^{+}+\Sigma $) or ($\Lambda +\Sigma _{C}$)\ ( if $\Delta \varepsilon
\neq 0$ ). But if $\Delta \varepsilon =0,$ the $\Delta $ family may not be
divided because the above division requires the fluctuation of energy.\ When
the energy increases, the fluctuation of energy ($\Delta \varepsilon $) also
increases, so the probability of further division increases as well.

At $E_{P}=11/4$, $\vec{n}=($01-1,10-1,121,211,020,200$)$, $J_{P}=1$, $\Delta
\varepsilon =0$ from (\ref{flub}).\ Since $\Delta \varepsilon =0$, the $4$
fold energy band $\Delta (1929)$ may not be divided. From Fig. 2(b),we can
see that there are 3 energy bands at the point P: (1) $E_{\Gamma }=2$ $%
\rightarrow E_{P}=11/4,$ $\vec{n}=($1-10, -110, 01-1, 0-11, \ 10-1, -101);
(2) E$_{P}=11/4\rightarrow E_{\Gamma }=4,$ $\vec{n}=(20$0, 020, 200$);(3)$ E$%
_{P}=11/4\rightarrow $ $E_{\Gamma }=6,$ \ $\vec{n}=(121,211,112)$. Two near
bands may partly degenerate a 4 fold band ($\Delta (1929)$): 
\begin{equation}
\vec{n}=(121,211),\Xi (1929)\text{; }\vec{n}=(01\text{-}1,10\text{-}%
1,020,200),\Delta (1929)\text{.}
\end{equation}
or 
\begin{equation*}
\vec{n}=(\text{01-1,10-1}),\Xi (1929)\text{; }\vec{n}=(020,200,121,211),%
\Delta (1929)\text{.}
\end{equation*}

At $E_{P}=\frac{19}{4}$, $\vec{n}=($0-1-1,-10-1,-121,2-11,202,022$)$, $%
J_{P}=2$, the energy fluctuation $\ \Delta \varepsilon =-100(2-1)\Delta S$ $%
=-100\times \Delta S$ from (\ref{flub}). After the first division, we have a
baryon $\Xi (2749)$ and a baryon $\Delta (2549)$. For $\Delta (2549)$, using
(\ref{strangeflu}), we get $S=0+\Delta S==\pm 1$. From $\Delta S=\pm 1,$ we
have $\Delta (2549)\rightarrow $\{[$\Lambda _{C}^{+}(2449)$ ( $\Delta S=+1$ )%
$+\Sigma (2649)$ ($\Delta S=-1$ )] or [$\Sigma _{C}(2449)$ ( $\Delta
C=\Delta S=+1$ )(see (\ref{C&S})) $+\Lambda (2649)$ ($S=-1$)]\}\ to keep the 
$3$ rotary symmetry. It is very important to pay attention to the\textbf{\
baryon }$\Sigma _{C}(2449)$ \textbf{born here, on the 6 fold energy band of
the axis F, after the second division from the fluctuation } $\Delta S=+1$
and $\Delta \varepsilon =-100$ Mev. To sum up, the $6$ fold energy band has
the possibility to represent baryon families: 
\begin{equation}
\text{ }\Xi (2749)\text{, }\Sigma (2649)\text{, }\Lambda _{C}^{+}(2449)\text{%
, }\Sigma _{C}(2449)\text{, }\Lambda (2649)\text{.}  \label{F_B_C}
\end{equation}

At $E_{H}=5$, $\vec{n}=($0-20,-200,-211,1-21,013,103$)$, $J_{H}=1$ the
energy fluctuation$\ \Delta \varepsilon =0$ from (\ref{flub}). We have: 
\begin{equation}
\text{ }\Xi (2739)\text{, }\Delta (2739)\text{.}
\end{equation}

Also at $E_{H}=5$, $\vec{n}=($0-22,-202,-2-11,-1-21,0-13,-103$)$, $J_{H}=2$,
the energy fluctuation$\ \Delta \varepsilon =-100\times \Delta S$ from (\ref
{flub}). We have: 
\begin{equation}
\text{ }\Xi (2839)\text{, }\Sigma (2739)\text{, }\Lambda _{C}^{+}(2539)\text{%
, }\Sigma _{C}(2539)\text{, }\Lambda (2739)\text{.}
\end{equation}

At $E_{P}=27/4$, $\vec{n}=($-12-1,2-1-1,301,031,222,00-2$),J_{P}=3$. Similar
to the case of $E_{P}=19/4$, $\Delta \varepsilon =-200\times \Delta S$ from (%
\ref{flub}) and we have 
\begin{equation}
\text{ }\Xi (3569)\text{, }\Sigma (3369)\text{, }\Lambda _{C}^{+}(2969)\text{%
, }\Sigma _{C}(2969)\text{, }\Lambda (3369)\text{.}  \label{SIGMA(2969)}
\end{equation}

% \UNICODE{0x2026}

\subsection{The Axis $G(M-N)$}

The axis $G(M-N)$ is a $2$ fold symmetry axis. From (\ref{gs}), the strange
number $S=-2$. There are $6$, $4$, and $2$ fold energy bands on the axis
(see Fig. 4(b)).

\subsubsection{The two fold energy bands on the axis $G(M-N)$}

For the $2$ fold energy band (see Fig. 4(b)), $S=-2$, $I=1/2$ from (\ref
{isomax}), and $Q=0,-1$ from (\ref{charge}). From the experimental results
that the family $\Xi $ $(\Xi ^{0},\Xi ^{-})$ has the same intrinsic quantum
numbers, we find that each $2$ fold degenerate energy band with symmetric $%
\vec{n}$ values represents a baryon family $\Xi $ $(\Xi ^{0},\Xi ^{-})$.

At $E_{N}=1/2$, $J_{N}=0,$ $\Delta \varepsilon =0$ from (\ref{fluc}), the
two energy bands with asymmetric $\vec{n}=(000,110)$ are in the first and
second Brillouin zones, respectively. So they will be divided. The energy
band with $\vec{n}=(110)$ (in the second Brillouin zone) represents $\Lambda
(1119)$ with S = -1. Another band with $\vec{n}=(000)$ (in first Brillouin
zone)\textbf{\ represents }$N(939)$ from (\ref{QN of N(939)}).

At $E_{M}=1$, there are two 2 fold energy bands. The 2 fold energy band with
symmetric values $\vec{n}=(101,10-1)$ represents the baryon families $\Xi
(1299).$ Another $2$ fold energy band with asymmetric values $\vec{n}%
=(200,1-10)$ will not be divided (see (\ref{dividing})), it represents the
baryon family $\Xi (1299)$, too.

At $E_{N}=5/2$, $J_{N}=1,$ $\Delta \varepsilon =0$ from (\ref{fluc}). Since $%
\Delta \varepsilon =0$ the 2 fold energy band with $\vec{n}=(020,110)$
should not be divided (see (\ref{dividing})), it represents the baryon
family $\Xi (1839).$

At $E_{M}=5$, the $2$ fold energy band with asymmetric $\vec{n}=($3-10,2-20$%
) $ will not be divided (see (\ref{dividing})), it represents the baryon
family $\Xi (2739)$.

Thus, we have 
\begin{equation}
\begin{array}{llll}
E_{N}=1/2 & \vec{n}=(\text{000,110}) & \varepsilon ^{(0)}=1119 &  \\ 
\text{\ \ \ }J_{N}=0 & \vec{n}=(\text{000}) & \text{S=C=b=0} & N(939) \\ 
& \vec{n}=(\text{110}) & S=-1 & \Lambda (1119) \\ 
E_{M}=1 & \vec{n}=(\text{101,10-1}) & \varepsilon ^{(0)}=1299 & \Xi (1299)
\\ 
& \vec{n}=(\text{200,1-10}) & \varepsilon ^{(0)}=1299 & \Xi (1299) \\ 
E_{N}=3/2 & \vec{n}=(\text{011,01-1}) & \varepsilon ^{(0)}=1479 & \Xi (1479)
\\ 
E_{N}=5/2 & \vec{n}=(\text{020,-110}) & \varepsilon ^{(0)}=1839 & \Xi (1839)
\\ 
\text{ \ \ }J_{N}=1 & \text{ \ \ \ } & \Delta \varepsilon =0 &  \\ 
E_{M}=3 & \vec{n}=(\text{2-11,2-1-1}) & \varepsilon ^{(0)}=2019 & \Xi (2019)
\\ 
E_{M}=5 & \vec{n}=(\text{3-10,2-20}) & \varepsilon ^{(0)}=2739 & \Xi (2739)
\\ 
E_{N}=11/2 & \vec{n}=(\text{-121,-12-1}) & \varepsilon ^{(0)}=2919 & \Xi
(2919) \\ 
\ldots &  &  & 
\end{array}
\label{G_TWO}
\end{equation}

\subsubsection{The four fold energy bands on the axis $G(M-N)$}

According to (\ref{degeneracy}) and (\ref{subdegen}), each $4$ fold energy
band will be divided into two $2$ fold energy bands which represent two
baryons $\Xi $: 
\begin{equation}
\begin{array}{lllll}
E_{M}=3 & \vec{n}=(\text{0-11,0-1-1,} & \text{211,21-1}) & \varepsilon
^{(0)}=2019 &  \\ 
\text{ \ \ }\Delta S=0 & \vec{n}=(\text{0-11,0-1-1}) & \Xi (2019) & \vec{n}=(%
\text{211,21-1}) & \Xi (2019) \\ 
E_{N}=7/2 & \vec{n}=(\text{-101,-10-1,} & \text{121,12-1}) & \varepsilon
^{(0)}=2199 &  \\ 
\text{ \ \ }\Delta S=0 & \vec{n}=(\text{-101,-10-1}) & \Xi (2199) & \vec{n}=(%
\text{121,12-1}) & \Xi (2199) \\ 
E_{M}=5 & \vec{n}=(\text{301,30-1,} & \text{1-21,1-2-1}) & \varepsilon
^{(0)}=2739 &  \\ 
\text{ \ \ }\Delta S=0 & \vec{n}=(\text{301,30-1}) & \Xi (2739) & \vec{n}=(%
\text{1-21,1-2-1}) & \Xi (2739) \\ 
\ldots &  &  &  & 
\end{array}
\end{equation}

\subsubsection{The six fold energy bands on the axis $G(M-N)$}

According to (\ref{degeneracy}) and (\ref{subdegen}), each $6$ fold energy
band will be divided into three $2$ fold energy bands. One of the three $2$
fold sub-degeneracy energy bands has asymmetric $\vec{n}$ values. According
to (\ref{dividing}), the 2 energy bands should be divided further at the
point $N$, but not be at the point $M$. Thus, at point $M$ the each $6$ fold
energy band will represent three baryon families $\Xi $. At the point $N$,
each $6$ fold energy band will represent two $\Xi $ and two single energy
bands .

At $E_{N}=9/2$, $\ \vec{n}=(112,11-2,002,00-2,220,-1-10)$. The $6$ fold
energy band will be divided into: $\vec{n}=(112,11-2)$ ($\Xi (2559)$), $\vec{%
n}=(002,00-2)$ ($\Xi (2559)$), $\vec{n}=(220,-1-10)$. Then the two energy
bands with asymmetric $\vec{n}$ values ($\vec{n}=(220,-1-10)$) will be
further divided. The fluctuation of energy associated with this division is $%
\Delta \varepsilon =$ $200(2-1)\Delta S$ $=200\Delta S$ Mev from (\ref{fluc}%
) ($J_{N}=2$ at $E_{N}=9/2$ for 2-fold asymmetric $\vec{n}$ values, $J_{N}=0$
at $E_{N}=1/2$ and $J_{N}=1$ at $E_{N}=5/2$ (see (\ref{G_TWO})). Since this
is the second division of the $6$ fold energy band, according to \textbf{%
Hypothesis V. 7, (\ref{C&S}), }for $\Delta S=+1$\textbf{\ }we can get a
Charmed number $C=+1$ while keeping the Strange number $S=-2$ unchanged.
Hence the $2$ energy bands will be divided into \ $\vec{n}=$ (-1,-1, 0) $%
\Delta S=+1$ from (\ref{SIGMA_1}) and \ $\vec{n}=$ (2, 2, 0) $\Delta S=-1$%
\begin{equation}
\Delta C=\Delta S=+1\text{, \ }\Omega _{C}(2759)\text{; \ \ \ \ }\Delta S=-1%
\text{, \ }\Omega (2359)\text{.}  \label{G_6_C}
\end{equation}
It is very important to pay attention to \textbf{the baryon} $\Omega _{C}$(%
\textbf{2759}) \cite{OMIGA-C} \textbf{born here}, on the $6$ fold energy
band of the axis G, after the second division with fluctuations $\Delta S=+1$
and $\Delta \varepsilon =+200$ Mev.

At $E_{M}=5$, $\vec{n}=(202,20-2,1-12,1-1-2,310,0-20)$ will be divided into
three $2$ fold energy bands, representing 3 baryon families $\Xi (2739)$.
The 2 fold band with asymmetric $\vec{n}=(310,0-20)$ will not be divided
further at the point $M$ from (\ref{dividing}).

At $E_{N}=13/2$, $\vec{n}=(-112,-11-2,022,02-2,130,-200)$, similar to $%
E_{N}=9/2$, the $6$ fold degeneracy will be divided into two 2 fold
degeneracies, representing two baryon families $\Xi (3279)$. The second
division will result a baryon $\Omega _{C}(3679)$ and a baryon $\Omega
(2879) $ ($J_{N}=3$ and $\Delta \varepsilon =200(3-1)\Delta S=400\Delta S$
Mev).

We have 
\begin{equation}
\begin{array}{llll}
\begin{array}{l}
E_{N}=9/2
\end{array}
& 
\begin{array}{c}
\vec{n}=\text{(112,11-2,002,} \\ 
\text{ \ \ \ \ \ \ 00-2,220,-1-10)}
\end{array}
& 
\begin{array}{l}
\varepsilon ^{(0)}=2559
\end{array}
&  \\ 
\text{ \ \ \ \ }J_{N}=2 & \vec{n}=\text{(112,11-2) \ }\Xi (2559) & \vec{n}=%
\text{(002,00-2)} & \Xi (2559) \\ 
\text{ \ (divided)} & \vec{n}=(220)\text{ \ }\Omega _{C}(2759) & \vec{n}=%
\text{(-1-10)} & \Omega (2359) \\ 
\begin{array}{l}
E_{M}=5
\end{array}
& 
\begin{array}{c}
\vec{n}=\text{(202,20-2,1-12,} \\ 
\text{ \ \ \ \ \ \ 1--1-2,310,0-20)}
\end{array}
& 
\begin{array}{l}
\varepsilon ^{(0)}=2739
\end{array}
& 3\text{ }\Xi (2739) \\ 
\begin{array}{l}
E_{N}=13/2
\end{array}
& 
\begin{array}{c}
\vec{n}=\text{(-112,-11-2,022,} \\ 
\text{ \ \ \ \ \ \ 02-2,130,-200)}
\end{array}
& 
\begin{array}{l}
\varepsilon ^{(0)}=3279
\end{array}
&  \\ 
\text{ \ \ \ \ }J_{N}=3 & \vec{n}=\text{(-112,-11-2) \ \ }\Xi (3279) & \vec{n%
}=\text{(022,02-2)} & \Xi (3279) \\ 
\text{ \ (divided)} & \text{single band \ \ \ \ \ \ \ \ }\Omega _{C}(3679) & 
\text{single band} & \Omega (2879) \\ 
\ldots & \text{ \ \ \ \ \ \ \ \ \ \ \ \ \ \ \ \ \ \ \ \ } & \text{ \ \ \ \ \
\ \ \ \ \ \ \ \ \ \ \ \ } & \text{ \ \ \ \ \ \ \ \ \ \ \ \ \ \ \ \ \ }
\end{array}
\label{G_6_OMEGAC}
\end{equation}

Continuing above procedure, we can use Fig. 2-5 to find the whole baryon
spectrum. Our results are shown in Tables $1$ though $6$ of Section V.

\section{Comparing Results}

We compare the theoretical results of the BCC model to the experimental
results \cite{particle} using Tables 1-6. In the comparison, we will use the
following laws:

(1) We do not take into account the angular momenta of the experimental
results. We assume that the small differences of the masses in the same
group of baryons originate from their different angular momenta. If we
ignore this effect, their masses should be essentially the same.

(2) We use the baryon name to represent the intrinsic quantum numbers as
shown in the second column of Table 1.

(3) For low energy cases, the baryons from different symmetry axes with the
same S, C, b, Q, I, I$_{Z}$, and $\overrightarrow{n}$ value, as well as in
the same Brillouin zone are regarded as the same baryon. The mass of the
baryon is the lowest value of their masses. For example, $\Lambda (1119)$
with $\overrightarrow{n}=(1,1,0)$ on the axis $\Sigma $ (see (\ref{SIGMA_1}%
)), $\Lambda (1119)$ with $\overrightarrow{n}=(1,1,0)$ on the axis $D$ (see (%
\ref{d_two})), $\Lambda (1119)$ with $\overrightarrow{n}=(1,1,0)$ on the
axis $G$ (see (\ref{G_TWO})) and $\Lambda (1209)$ with $\overrightarrow{n}%
=(1,1,0)$ (see (\ref{LAMBDA_3}) and (\ref{F-1}) are all in the second
Brillouin zone (the same energy curved surface in the three dimension phase
space), they are the same baryon $\Lambda (1119)$.

(4) If the same kind of baryons with the same mass but different $%
\overrightarrow{n}$ values, then the possibility of discovery of the baryon
with symmetry $\overrightarrow{n}$ values is much larger than the
possibility of discovery of the baryon\ with asymmetry $\overrightarrow{n}$
values. Thus, we can ignore the baryon with asymmetry $\overrightarrow{n}$
values in the comparison. For example, $\Xi ${\small (1839) }has two
possibility groups of values:$\ $(2, 0, 0; 0, 2, 0) and (0, 2, 0; -1, 1, 0).
We can ignore the baryon with the asymmetric $\overrightarrow{n}$ = (0, 2,
0; -1, 1, 0). However, if there is no same kind of baryons with the same
mass and symmetric $\overrightarrow{n}$ values, then we can not ignore the
baryon with asymmetric $\overrightarrow{n}$ values. For example, $\Delta
(1929)$ has two possible groups of $\overrightarrow{n}$ values $%
\overrightarrow{n}$ = ({\small -101, 0-11, 121,211) and }$\overrightarrow{n}$
= ({\small 01-1, 10-1, 020, 200), }we can not ignore either.

\subsection{The Ground States of Various Kinds of Baryons}

\ The ground states of various kinds of baryons are shown in Table 1. \
These baryons have a relatively long lifetime and are the most important
experimental results of the baryons. The theoretical results are listed
below. From the list, for the long lifetime baryons, we can see some
interesting facts: (1) the unflavored baryon N(939) has $\overrightarrow{n}$
= (0, 0, 0); (2) the strange baryons ($\Lambda (1119)$, $\Sigma (1209)$, $%
\Xi (1299)$, and $\Omega (1659)$) have $\overrightarrow{n}$ values with two
components $\pm 1$ and one component 0; (3) the charmed baryons ($\Lambda
_{c}^{+}$(2279), $\ \Sigma _{c}$(2449), $\Xi _{c}$(2549), and $\Omega _{c}$%
(2759)) have $\overrightarrow{n}$ values with\ at least one component $\pm
2; $ (4) the bottom baryon $\Lambda _{b}$(5639) has a $\overrightarrow{n}$
value with 2 components 3; (5) the family $\Delta (1299)$ is a special case,
they are not long lifetime particles$.$

\begin{tabular}{l}
$
\begin{tabular}{lllllll}
&  &  &  &  &  &  \\ 
{\small Baryon} & {\small N}$(939)$ & $\Lambda (1119)$ & $\ \ \Sigma $%
{\small (1209)} & $\Xi ${\small (1299)} & $\Omega ${\small (1659)} & $%
\Lambda _{c}^{+}${\small (2279)} \\ 
n$_{1}$n$_{2}$n$_{3}$ & {\small 0, 0, 0} & \ {\small 1, 1, 0} & {\small %
110,101,011} & {\small 101,10-1} & {\small -1, -1, 0} & \ {\small 0, 0, -2}
\\ 
({\small Eq. No)} & \ (41) & \ (65) & \ \ (61) & \ (80) & \ (65) & \ (60) \\ 
&  &  &  &  &  & 
\end{tabular}
$ \\ 
$
\begin{tabular}{llllll}
{\small Baryon} & $\ \Sigma _{c}${\small (2449)} & $\Xi _{c}${\small (2549)}
& $\Omega _{c}${\small (2759)} & $\Lambda _{b}${\small (5639)} & $\ \ \
\Delta ${\small (1299)} \\ 
{\small n}$_{1}${\small n}$_{2}${\small n}$_{3}$ & {\small 202,022,2-11.} & 
{\small 202,022.} & \ {\small 2, 2, 0.} & \ {\small 3, 3, 0.} & {\small %
101,-101,011,0-11} \\ 
({\small Eq. No)} & \ \ \ (76) & \ (71) & \ \ (82) & \ (65) & \ \ (51) \\ 
&  &  &  &  & 
\end{tabular}
$%
\end{tabular}
\ \ \ \ \ \ \ \ \ \ \ 

\ From Table 1, we can see that all theoretical intrinsic quantum numbers
(isospin $I$, strange number $S$, charmed number $C$, bottom number $b$, and
electric charge $Q$) are the same as experimental results. Also the
theoretical mass values are in very good agreement with the experimental
values.

\subsection{ The Unflavored Baryons $N$ and $\Delta $}

\ A comparison of the theoretical results with the experimental results of
the unflavored baryons $N$ and $\Delta $ is made in Table 2.

The theoretical results of the low energy baryons $\Delta $ are given by the
following list. In the list, we give all possible $\overrightarrow{n}$
values and the equation numbers inside the``( )'' in whith the baryons $%
\Delta $ are deduced. The $\overrightarrow{n}$ values 1, $\pm 1,0$ represent
two $\overrightarrow{n}$ values (1,+1, 0) and (1, -1, 0)... .

\begin{tabular}{l}
\begin{tabular}{lllllll}
&  &  &  &  &  &  \\ 
Baryon & $\Delta ${\small (1299)} & $\ \Delta ${\small (1659)} & $\ \Delta $%
{\small (1929)} & $\Delta ${\small (2019)} & $\Delta ${\small (2379)} & $\
\Delta ${\small (2649)*} \\ 
$
\begin{tabular}{l}
n$_{1}$n$_{2}$n$_{3}$ \\ 
({\small Eq. No)}
\end{tabular}
$ & 
\begin{tabular}{l}
{\small 1,0,1} \\ 
{\small -1,0,1} \\ 
{\small 0,1,1} \\ 
{\small 0,-1,1} \\ 
$(51)$%
\end{tabular}
& 
\begin{tabular}{l}
{\small 1,}$\pm ${\small 1,0} \\ 
{\small -1,}$\pm ${\small 1,0} \\ 
{\small (}$51)$ \\ 
$\pm ${\small 1,0,-1} \\ 
{\small 0,}$\pm ${\small 1,-1} \\ 
$(51)$%
\end{tabular}
& 
\begin{tabular}{l}
{\small -101,0-11} \\ 
{\small 211,121} \\ 
{\small (66)} \\ 
{\small 01-1,10-1} \\ 
0{\small 20,200} \\ 
{\small (75)}
\end{tabular}
& 
\begin{tabular}{l}
{\small 1,1,2} \\ 
{\small 1,-1,2} \\ 
{\small -1,1,2} \\ 
{\small -1,-1,2} \\ 
{\small (51)}
\end{tabular}
& 
\begin{tabular}{l}
{\small 2,0,0} \\ 
{\small -2,0,0} \\ 
{\small 0,2,0} \\ 
{\small 0,-2,0} \\ 
{\small (51)}
\end{tabular}
& 
\begin{tabular}{l}
{\small -1,1,2} \\ 
{\small 1-12} \\ 
{\small 2, 0, 2} \\ 
{\small 0,2,2} \\ 
{\small (66)}
\end{tabular}
\\ 
Theory & $\Delta ${\small (1299)} & {\small 2}$\Delta ${\small (1659)} & 
{\small 2}$\Delta ${\small (1929)} & $\Delta ${\small (2019)} & $\Delta $%
{\small (2379)} & $\Delta ${\small (2649)} \\ 
&  &  &  &  &  & 
\end{tabular}
\end{tabular}

$\ \ast ${\small next\ }$\Delta ${\small (2739)}$\overrightarrow{n}${\small %
=\ }$\pm ${\small 1}$\pm ${\small 21,}$\pm ${\small 2}$\pm ${\small 11;}$\pm 
${\small 1}$\pm ${\small 21,}$\pm ${\small 2}$\pm ${\small 11;\ }$\pm $%
{\small 202,0}$\pm ${\small 22;0}$\pm ${\small 13,}$\pm ${\small 103\ 5}$%
\Delta ${\small (2739)}

\ \ \ \ \ \ \ \ \ \ \ \ \ \ \ \ \ \ \ \ \ \ \ \ \ \ \ \ \ \ \ \ \ \ \ \ \ \
\ \ \ \ \ \ \ \ \ \ \ \ \ \ \ \ \ \ \ \ \ \ \ \ \ \ \ \ \ \ \ \ \ \ \ \ \ \
\ \ \ \ \ \ \ \ \ \ \ \ \ \ \ \ \ \ \ \ \ \ \ \ \ \ \ \ \ \ \ \ \ \ \ 

The theoretical results of the low energy baryons $N$ are given by the
following list. In the list, we give all possible $\overrightarrow{n}$
values and the equation numbers inside the``( )'' in whith the baryons $N$
are deduced. We also show how to get the final theoretical results of the
baryon N. For example, the baryon N(1209) with $\overrightarrow{n}=($0,1,1;
1,0,1) (see Eq. (67)),\ the baryon N(1209) with $\overrightarrow{n}=($0,1,1;
1,0,1) (see Eq. (69)), and the baryon N(1299) with $\overrightarrow{n}=($%
0,1,1; 1,0,1) (see Eq. (51)) are the same baryon N(1209); the baryon N(1929)
with $\overrightarrow{n}=($0, 0, 2; 1, 1, 2) (67) is ignored since the $%
\overrightarrow{n}$ values are asymmetry valuers.

\begin{tabular}{l}
\begin{tabular}{lllllll}
&  &  &  &  &  &  \\ 
Baryon & {\small N(1209)} & {\small N(1299)} & {\small N(1479)} & {\small %
N(1659)} & {\small N(1839)} & \ \ {\small N(1929)} \\ 
\begin{tabular}{l}
$n_{1}n_{2}n_{3}$ \\ 
({\small Eq. No}.)
\end{tabular}
& 
\begin{tabular}{l}
{\small 0,1,1} \\ 
{\small 1,0,1} \\ 
{\small (67)} \\ 
{\small 0,1,1} \\ 
{\small 1,0,1} \\ 
$(${\small 69)}
\end{tabular}
& 
\begin{tabular}{l}
{\small 0,1,1} \\ 
{\small 1,0,1} \\ 
{\small (51)}
\end{tabular}
& 
\begin{tabular}{l}
{\small 1,0,-1} \\ 
{\small 0,1,-1} \\ 
{\small (67)}
\end{tabular}
& 
\begin{tabular}{l}
{\small 1,1,0} \\ 
{\small 1,-1,0} \\ 
$(51)$ \\ 
{\small 1,0-1} \\ 
{\small -1,0,-1} \\ 
$(51)$%
\end{tabular}
& 
\begin{tabular}{l}
{\small 1,-1,0} \\ 
{\small -1,1,0} \\ 
(66) \\ 
{\small 0,2,0} \\ 
{\small 2,0,0} \\ 
{\small (66)}
\end{tabular}
& 
\begin{tabular}{l}
{\small -101,0-11} \\ 
(66) \\ 
{\small 211,121} \\ 
{\small (66)} \\ 
{\small 002,112} \\ 
\ {\small (67)}
\end{tabular}
\\ 
Theory & {\small N(1209)}$^{\#}$ & $\longleftarrow $ & {\small N(1479)} & 2 
{\small N(1659)} & {\small 2 N(1839)} & \ {\small 2 N(1929)} \\ 
&  &  &  &  &  & 
\end{tabular}
\\ 
\begin{tabular}{lllllll}
Baryon & {\small N(2019)} & {\small N(2199)} & {\small N(2379)} & {\small %
N(2549)} & {\small N(2559)} & \ \ {\small N(2649)*} \\ 
\begin{tabular}{l}
$n_{1}n_{2}n_{3}$ \\ 
({\small Eq. No)}
\end{tabular}
& 
\begin{tabular}{l}
{\small 1,1,2} \\ 
{\small 1,-1,2} \\ 
{\small (51)}
\end{tabular}
& 
\begin{tabular}{l}
{\small 1,2,-1} \\ 
{\small 2,1,-1} \\ 
{\small (66)} \\ 
{\small -1,0,-1} \\ 
{\small 0,-1,-1} \\ 
{\small (66)}
\end{tabular}
& 
\begin{tabular}{l}
{\small 2,0,0} \\ 
{\small -2,0,0} \\ 
$(51)$%
\end{tabular}
& 
\begin{tabular}{l}
{\small 2, 1, -1} \\ 
{\small 1, 2, -1} \\ 
{\small (69)}
\end{tabular}
& 
\begin{tabular}{l}
{\small 2, 2, 0} \\ 
{\small -1,-1,0} \\ 
{\small (67)}
\end{tabular}
& 
\begin{tabular}{l}
{\small -112,1-12} \\ 
\ {\small (66)} \\ 
{\small 202,022} \\ 
{\small \ (66)} \\ 
{\small -121,2-11} \\ 
\ {\small (67)}
\end{tabular}
\\ 
Theory & {\small N(2019)} & {\small 2N(2199)} & {\small N(2379)} & {\small %
N(2549)} & {\small N(2559)} & {\small 3N(2649)}
\end{tabular}
\end{tabular}

\bigskip \# see the following paragraph. * The next baryon\ {\small N(2739) }%
$\overrightarrow{n}$ ={\small \ (1,}$\pm ${\small 2,1), (2,}$\pm ${\small %
1,1), (2,0,}$\pm ${\small 2), (0,}$\pm ${\small 1,3) 4N(2739)\ in the Eq.
(51).}\ 

\ \ \ \ \ \ \ \ \ \ \ \ \ \ \ \ \ \ \ \ 

From Table 2, we can see that the intrinsic quantum numbers of the
theoretical results are the same as the experimental results.\ Also the
theoretical masses of the baryons $N$ and $\Delta $\ are in very good
agreement with the experimental results. The theoretical results $N(1209)$
is not found in experiments. We guess that it is covered up by the
experimental baryon $\Delta (1232)$. The reasons are as follows : (1) they
are unflavored baryons with the same S = C = b =0 and Q (Q$%
_{N^{+}}=Q_{\Delta ^{+}}$ and Q$_{N^{0}}=Q_{\Delta ^{0}})$; (2) they have
the same $\overrightarrow{n}$ values ($\overrightarrow{n}%
_{N(1209)}=(011,101) $, $\overrightarrow{n}_{\Delta (1299)}=($%
101,-101,011,0-11$)$) and they are both in the second \textbf{Brillouin zone}%
; (3) the experimental width (120 Mev) of $\Delta (1232)$ is very large, and
the baryon $N(1209)$ is fall within the width region of $\Delta (1232)$; (4)
the mass ($1209$ Mev) of $N(1209)$ is essentially the same as the mass ($%
1232 $ Mev) of $\Delta (1232)$. The experimental value 1232 is much lower
than the theoretical value 1299 of $\Delta (1299)$ and the experimental
width (120) is much larger than other baryons (with similar masses) support
the explanation.

\subsection{ The Strange Baryons $\Lambda $ and $\Sigma $}

The strange baryons $\Lambda $ and $\Sigma $ are compared in Table 3. The
theoretical results of the baryons $\Lambda $ are shown in the following
list. In the list, we give all low mass possible baryons $\Lambda $ with $%
\overrightarrow{n}$ values and the equation numbers in which the baryons are
deduced. From the list, we can see that\ (1) 3$\times \Lambda (1119)$ with $%
\overrightarrow{n}=$110 and 2$\times \Lambda (1209)$ with $\overrightarrow{n}%
=$110 are the same baryon $\Lambda (1119)$.\ (2) $\Lambda (1929)$ with $%
\overrightarrow{n}=$112 (see Eq. (61)) and $\Lambda (1929)$ with $%
\overrightarrow{n}=$112 (see Eq. (69)) are the same baryon $\Lambda (1929)$.
(3) However, $\Lambda (1299)$ with $\overrightarrow{n}=$ 002 and $\Lambda
(1399)$ with $\overrightarrow{n}=$002 are not the same baryon$,$ because the
baryon $\Lambda (1399)$ has a fluctuation energy $\Delta \varepsilon =100$
Mev (see Eq. (60)) but the baryon $\Lambda (1299)$ does not. (4) $\Lambda
(1299)$ with $\overrightarrow{n}=$002 and $\Lambda (1929)$ with $%
\overrightarrow{n}=$002 (see Eq. (61)) are the same baryon $\Lambda (1299).$

\begin{tabular}{l}
\begin{tabular}{lllllll}
&  &  &  &  &  &  \\ 
Baryon & $\Lambda ${\small (1119)} & $\Lambda ${\small (1209)} & $\Lambda $%
{\small (1299)} & $\Lambda ${\small (1399)} & $\Lambda ${\small (1659)} & $%
\Lambda ${\small (1929)} \\ 
\begin{tabular}{l}
$n_{1}n_{2}n_{3}$ \\ 
({\small Eq. No)}
\end{tabular}
& 
\begin{tabular}{l}
{\small 1,1,0} \\ 
{\small (65)} \\ 
{\small 1,1,0} \\ 
{\small (67)} \\ 
{\small 1,1,0} \\ 
{\small (80)}
\end{tabular}
& 
\begin{tabular}{l}
{\small 1,1,0} \\ 
{\small (61)} \\ 
{\small 1,1,0} \\ 
{\small (68)}
\end{tabular}
& 
\begin{tabular}{l}
{\small 0,0,2} \\ 
{\small \ (69) \ }
\end{tabular}
& {\small 
\begin{tabular}{l}
0,0,2 \\ 
(60)
\end{tabular}
} & 
\begin{tabular}{l}
{\small 0,1,-1} \\ 
{\small (61)} \\ 
{\small 1,0,-1} \\ 
{\small (61)} \\ 
{\small -1,-1,0} \\ 
{\small (61)}
\end{tabular}
& 
\begin{tabular}{l}
{\small 1,1,2 (61)} \\ 
{\small 1,1,2 (69)} \\ 
{\small 1,2,1 (70)} \\ 
{\small 2,1,1 (70)} \\ 
{\small 0,0,2 (61)}
\end{tabular}
\\ 
Theory & $\Lambda ${\small (1119)} &  & $\Lambda ${\small (1299)} & $\Lambda 
${\small (1399)} & 3$\Lambda ${\small (1659)} & 3$\Lambda ${\small (1929)}
\\ 
&  &  &  &  &  & 
\end{tabular}
\\ 
\begin{tabular}{llllll}
Baryon & $\Lambda ${\small (2019)} & $\Lambda ${\small (2359)} & $\Lambda $%
{\small (2379)} & $\Lambda ${\small (2559)} & $\ \ \ \ \Lambda ${\small %
(2649)} \\ 
\begin{tabular}{l}
$n_{1}n_{2}n_{3}$ \\ 
({\small Eq. No)}
\end{tabular}
& 
\begin{tabular}{l}
{\small -1,-1,2} \\ 
{\small (68)} \\ 
{\small -1,-10} \\ 
{\small \ (69)}
\end{tabular}
& 
\begin{tabular}{l}
{\small 1,1,-2} \\ 
{\small (67)}
\end{tabular}
& 
\begin{tabular}{l}
{\small 0,0,-2} \\ 
{\small (61)}
\end{tabular}
& 
\begin{tabular}{l}
{\small 2,2,0} \\ 
{\small (65})
\end{tabular}
& 
\begin{tabular}{l}
{\small 220 (61), 220(69)} \\ 
{\small 121(71), 2-11(71)} \\ 
{\small 1-12 (61)}$,${\small 2-11(61)} \\ 
{\small 0-1-1 (76)}
\end{tabular}
\\ 
Theory & 2$\Lambda ${\small (2019)} & $\Lambda ${\small (2359)} & $\Lambda $%
{\small (2379)} & $\Lambda ${\small (2559)} & \ \ \ 5$\Lambda ${\small (2649)%
} \\ 
&  &  &  &  & 
\end{tabular}
\end{tabular}

All possible theoretical baryons $\Sigma $ which have experimental results
to compare with are shown in following list. From the list, we get that (1)
the 3 asymmetry $\overrightarrow{n}$ values in the 5 possible $\Sigma $%
{\small (1929)} are ignored{\small ;} (2) similarly, the 3 asymmetry $%
\overrightarrow{n}$ values in the possible $\Sigma ${\small (2649)} are
ignored..

\begin{tabular}{l}
\begin{tabular}{lllll}
&  &  &  &  \\ 
Baryon & $\Sigma ${\small (1209)} & $\Sigma ${\small (1299) } & $\ \ \ \ \
\Sigma ${\small (1659)} & $\ \ \ \ \Sigma ${\small (1929)} \\ 
\begin{tabular}{l}
$n_{1}n_{2}n_{3}$ \\ 
({\small Eq. No)}
\end{tabular}
& 
\begin{tabular}{l}
{\small 1,0,1} \\ 
{\small 0,1,1} \\ 
{\small 1,1,0} \\ 
($61)$%
\end{tabular}
& 
\begin{tabular}{l}
{\small 0,0,2} \\ 
{\small -1,0,1} \\ 
{\small 0,-1,1} \\ 
{\small (69)}
\end{tabular}
& 
\begin{tabular}{l}
{\small 1-10,-110,01-1} \\ 
{\small 0-11,10-1,-101} \\ 
{\small -10-1,0-1-1,-1-10} \\ 
{\small \ (}$61)$%
\end{tabular}
& 
\begin{tabular}{l}
{\small 020,002,200 (61)} \\ 
{\small 121,211,112 (61)} \\ 
{\small 112,1-10,-110 (69)} \\ 
{\small 01-1,10-1,121 (70)} \\ 
{\small 020,200,211 (70)}
\end{tabular}
\\ 
Theory & $\Sigma ${\small (1209)} & $\Sigma ${\small (1299)} & \ \ 3 $\Sigma 
${\small (1659)} & \ \ \ 2 $\Sigma ${\small (1929)} \\ 
&  &  &  & 
\end{tabular}
\\ 
$
\begin{tabular}{lllll}
{\small Baryon} & $\Sigma ${\small (2019)} & $\Sigma (2379)$ & {\small \ \ \
\ \ \ \ \ \ \ }$\Sigma ${\small (2649)} & {\small \ \ \ \ \ \ \ }$\Sigma $%
{\small (2739)} \\ 
\begin{tabular}{l}
$n_{1}n_{2}n_{3}$ \\ 
({\small Eq. No)}
\end{tabular}
& 
\begin{tabular}{l}
{\small -1,-1,0} \\ 
{\small 1,1,2} \\ 
{\small 1,-1,2} \\ 
{\small (69)}
\end{tabular}
& 
\begin{tabular}{l}
{\small 0,-2,0} \\ 
{\small -2,0,0} \\ 
{\small 0,0,-2} \\ 
$(61)$%
\end{tabular}
& 
\begin{tabular}{l}
{\small 1-12,-112,21-1 }$(61)$ \\ 
{\small 2-11,12-1,-121 }$(61)$ \\ 
{\small 202,022,220 \ \ \ \ }$(61)$ \\ 
{\small 220,21-1,12-1 \ (69)} \\ 
{\small 220,022,-121 \ \ \ (71)} \\ 
{\small 2-11,0-1-1,-1,0,-1 (71)}
\end{tabular}
& 
\begin{tabular}{l}
{\small 0-20,-200,-211} \\ 
{\small 1-21,013,103 } \\ 
$(72)$ \\ 
{\small 0-22,-202,-2-11 } \\ 
{\small -1-21,0-13,-103 } \\ 
{\small (73)}
\end{tabular}
\\ 
Theory & $\Sigma (2019)$ & $\Sigma (2379)$ & $\ \ \ \ \ 3\Sigma (2649)$ & $\
\ 4\Sigma (2739)$ \\ 
&  &  &  & 
\end{tabular}
$%
\end{tabular}

Using Table 3, we can see that the theoretical masses of the baryons $%
\Lambda $ and $\Sigma $ are in very good agreement with the experimental
results, and their theoretical and experimental intrinsic quantum numbers
are the same.

\subsection{ The Strange Baryons $\Xi $ and $\Omega $}

The theoretical results of the baryons $\Xi $ are deduced in the following
list. In the list, $\overrightarrow{n}=$101,10-1 $\Xi (1299)$ (80) and $%
\overrightarrow{n}=$101,10-1 $\Xi (1479)$ (63) are the same baryon $\Xi
(1299)$. $\overrightarrow{n}=$011,01-1 $\Xi (1479)$\ (63) and $%
\overrightarrow{n}=$011,01-1 $\Xi (1479)$ (80) are the same baryon $\Xi
(1479)$. $\overrightarrow{n}=$0-11,0-1-1 $\Xi (1659)$ (63) and $%
\overrightarrow{n}=$0-11,0-1-1 $\Xi (2019)$ (81) are the same baryon $\Xi
(1659)$. $\overrightarrow{n}=$211,21-1 $\Xi (2019)$ (81) and $%
\overrightarrow{n}=$211,21-1 $\Xi (2199)$ (63) are the same baryon $\Xi
(2019)$. $\overrightarrow{n}=$002,00-2 $\Xi (2379)$ (62) and $%
\overrightarrow{n}=$002,00-2 $\Xi (2559)$ (83) are the same baryon $\Xi
(2379)$. $\overrightarrow{n}=$112,11-2 $\Xi (2559)$ (62) and $%
\overrightarrow{n}=$112,11-2 $\Xi (2559)$ (83) are the same baryon $\Xi
(2559)$. $\overrightarrow{n}=$020,-110 $\Xi (1839)$ (80) is ignored since
the $\overrightarrow{n}$ value is asymmetry. $\overrightarrow{n}=$310,2-20 $%
\Xi (2739)$ (80) and $\overrightarrow{n}=$310,2-20 (83) are ignored also.

\begin{tabular}{l}
\begin{tabular}{lllllll}
Baryon & $\Xi ${\small (1299)} & $\ \Xi ${\small (1479)} & $\Xi ${\small %
(1659)} & $\Xi ${\small (1839)} & $\Xi ${\small (1929)} & $\ \ \Xi ${\small %
(2019)} \\ 
\begin{tabular}{l}
$n_{1}n_{2}n_{3}$ \\ 
({\small Eq. No)}
\end{tabular}
& 
\begin{tabular}{l}
{\small 101} \\ 
{\small 10-1} \\ 
{\small (80)} \\ 
{\small 200} \\ 
{\small 1-10} \\ 
{\small (80)}
\end{tabular}
& 
\begin{tabular}{l}
{\small 101,10-1} \\ 
{\small (}$63)$ \\ 
{\small 011,01-1} \\ 
{\small (63)} \\ 
{\small 011,01-1} \\ 
{\small (}$80)$%
\end{tabular}
& 
\begin{tabular}{l}
{\small 1-10,-110} \\ 
($62)$ \\ 
{\small -101,-10-1} \\ 
{\small (}$63)$ \\ 
{\small 0-11,0-1-1} \\ 
{\small (}$63)$%
\end{tabular}
& 
\begin{tabular}{l}
{\small 200,} \\ 
{\small 020} \\ 
{\small (}$62)$ \\ 
{\small 020} \\ 
{\small -110} \\ 
{\small (80)}
\end{tabular}
& 
\begin{tabular}{l}
{\small 121,} \\ 
{\small 211} \\ 
{\small (or)} \\ 
{\small 01-1} \\ 
{\small 10-1} \\ 
{\small (75)}
\end{tabular}
& 
\begin{tabular}{l}
{\small 2-11,2-1-1} \\ 
({\small 80)} \\ 
{\small 0-11,0-1-1} \\ 
{\small (81)} \\ 
{\small 211,21-1} \\ 
{\small (81)}
\end{tabular}
\\ 
Theory & $\Xi ${\small (1299)} & $\Xi ${\small (1479)} & {\small 3\ }$\Xi $%
{\small (1659)} & $\Xi ${\small (1839)} & $\Xi ${\small (1929)} & \ {\small 2%
}$\Xi ${\small (2019)} \\ 
&  &  &  &  &  & 
\end{tabular}
\\ 
\begin{tabular}{lllll}
Baryon & $\ \ \Xi ${\small (2199)} & $\Xi ${\small (2379)} & $\Xi ${\small %
(2559)} & $\ \ \ \Xi ${\small (2739)} \\ 
\begin{tabular}{l}
$n_{1}n_{2}n_{3}$ \\ 
({\small Eq. No)}
\end{tabular}
& 
\begin{tabular}{l}
{\small -1 0 1,-1 0 -1} \\ 
{\small (81)} \\ 
{\small 1 2 1, 1 2 -1} \\ 
{\small (63)(81)} \\ 
{\small 2 1 1, 2 1 -1} \\ 
{\small (}$63)$%
\end{tabular}
& 
\begin{tabular}{l}
{\small 002,00-2} \\ 
{\small (}$62)$ \\ 
{\small -200,0-20} \\ 
{\small (}$62)$%
\end{tabular}
& 
\begin{tabular}{l}
{\small 1,1,2;1,1,-2} \\ 
{\small (62)} \\ 
{\small 1,1,2;1,1,-2} \\ 
{\small (83)} \\ 
{\small 0,0,2;0,0,-2} \\ 
{\small (83)}
\end{tabular}
& 
\begin{tabular}{l}
{\small 310,2-20 (80)} \\ 
{\small 301,30-1 }({\small 81)} \\ 
{\small 1-21,1-2-1}({\small 81)} \\ 
{\small 202,20-2 (83)} \\ 
{\small 1-12,1-1-2 (83)} \\ 
{\small 310,0-20 (83)} \\ 
{\small 0-20,-200 \ (77)}
\end{tabular}
\\ 
Theory & \ 2$\Xi ${\small (2199)} & $\Xi ${\small (2379)} & $\Xi ${\small %
(2559)} & $\ \ 5\Xi ${\small (2739)} \\ 
&  &  &  & 
\end{tabular}
\end{tabular}

The theoretical results of the baryons $\Omega $ are shown in the following
list. Although $\Omega (1659)$ and $\Omega (2359)$ have the same $%
\overrightarrow{n}=(-1,-1,0),$ they are not the same baryon, because $\Omega
(2359)$ has a fluctuation energy $\Delta \varepsilon =-200$ Mev (see (\ref
{G_6_C})), but the baryon $\Omega (1659)$ does not.

\begin{tabular}{l}
\begin{tabular}{llllll}
&  &  &  &  &  \\ 
Baryon & $\Omega (1659)$ & $\Omega (2359)$ & $\Omega (2879)$ & $\Omega
(3619) $ & $\Omega (7019)$ \\ 
$n_{1}n_{2}n_{3}$ & $-1,-1,0$ & $-1,-1,$ $0$ & $-2,$ $0,$ $0$ & $-2,-2,$ $0$
& -3,-3,0 \\ 
({\small Eq. No)} & ($65)$ & (83) & (83) & ($65)$ & ($65)$ \\ 
&  &  &  &  & 
\end{tabular}
\end{tabular}

The theoretical intrinsic quantum numbers of the baryons $\Xi $ and $\Omega $
are the same as the experimental results (see Table 4). The theoretical
masses of the baryons $\Xi $ and $\Omega $ are in very good agreement with
the experimental results.

\subsection{ The Charmed and Bottom Baryons $\Lambda _{c}^{+}$ and $\Lambda
_{b}^{0}$}

The charmed and bottom baryons $\Lambda _{c}^{+}$ and $\Lambda _{b}^{0}$ can
be found in Table 5. The theoretical results of the baryons $\Lambda
_{c}^{+} $and $\Lambda _{b}$ are listed below: 
\begin{equation*}
\begin{tabular}{llllllll}
&  &  &  &  &  &  &  \\ 
{\small Baryon} & $\Lambda _{c}^{+}${\small (2279)} & $\Lambda _{c}^{+}$%
{\small (2449)} & $\Lambda _{c}^{+}${\small (2539)} & $\Lambda _{c}^{+}$%
{\small (2759)} &  & $\Lambda _{b}${\small (5639)} & $\Lambda _{b}${\small %
(10159)} \\ 
{\small n}$_{1}${\small n}$_{2}${\small n}$_{3}$ & {\small 0, 0, -2} & 
\begin{tabular}{l}
{\small -1,2,1} \\ 
{\small or2-11}
\end{tabular}
& 
\begin{tabular}{l}
{\small 0, -2, 2} \\ 
{\small or-202}
\end{tabular}
& {\small 0,0,-2} &  & 3, 3, 0 & 4, 4, 0 \\ 
{\small (Eq.No)} & {\small (60)} & {\small (76)} & {\small (78)} & {\small %
(67)} &  & (65) & (65) \\ 
Theory & $\Lambda _{c}^{+}${\small (2279)} & $\Lambda _{c}^{+}${\small (2449)%
} & $\Lambda _{c}^{+}${\small (2539)} & $\Lambda _{c}^{+}${\small (2759)} & 
& $\Lambda _{b}${\small (5639)} & $\Lambda _{b}${\small (10159)} \\ 
&  &  &  &  &  &  & 
\end{tabular}
\end{equation*}
From the list, we can see that baryon $\Lambda _{c}^{+}$(2759) and baryon $%
\Lambda _{c}^{+}$(2279) are not the same baryon, since $\Lambda _{c}^{+}$%
(2279) has $\Delta \varepsilon =-100,$ while $\Lambda _{c}^{+}$(2759) has $%
\Delta \varepsilon =200.$

The theoretical intrinsic quantum numbers of the baryons $\Lambda _{c}^{+}$
and $\Lambda _{b}^{0}$ are the same as the experimental results.\ The
experimental masses of the charmed baryons ($\Lambda _{c}^{+}$) and bottom
baryons ($\Lambda _{b}^{0}$) are in very good agreement with the theoretical
results as well.

\subsection{ The Charmed Baryons $\Sigma _{c}$, $\Xi _{c}$ and $\Omega _{C}$}

Finally, we compare the theoretical results with the experimental results
for the charmed strange baryons $\Omega _{C}$, $\Xi _{c}$ and $\Sigma _{c}$
in Table 6.\ Their intrinsic quantum numbers are all matched completely, and
their masses are in very good agreement. The theoretical results of the
baryons $\Sigma _{c},$ $\Xi _{c}$and $\Omega _{c}$ are listed below:

\begin{tabular}{l}
\begin{tabular}{lllllll}
&  &  &  &  &  &  \\ 
Baryon & $\Sigma _{c}(2449)$ & $\Sigma _{c}(2539)$ & $\Sigma _{c}(2969)$ & 
& $\Xi _{c}(2549)$ & $\Xi _{c}(3169)$ \\ 
\begin{tabular}{l}
$n_{1}n_{2}n_{3}$ \\ 
(Eq. No)
\end{tabular}
& 
\begin{tabular}{l}
2, 0, 2 \\ 
0, 2, 2 \\ 
2, -1, 1 \\ 
(76)
\end{tabular}
& 
\begin{tabular}{l}
0, -1, 3 \\ 
-1, 0, 3 \\ 
-2, 0, 2 \\ 
(78)
\end{tabular}
& 
\begin{tabular}{l}
3, 0, 1 \\ 
0, 3, 1 \\ 
2, 2, 2 \\ 
(79)
\end{tabular}
&  & {\small 
\begin{tabular}{l}
2, 0, 2 \\ 
0, 2, 2 \\ 
(71)
\end{tabular}
} & 
\begin{tabular}{l}
3, 0, 1 \\ 
0, 3, 1 \\ 
(74)
\end{tabular}
\\ 
Theory & $\Sigma _{c}(2449)$ & $\Sigma _{c}(2539)$ & $\Sigma _{c}(2969)$ & 
& $\Xi _{c}(2549)$ & $\Xi _{c}(3169)$ \\ 
&  &  &  &  &  & 
\end{tabular}
\end{tabular}

\begin{tabular}{l}
\begin{tabular}{lll}
Baryon & $\Omega _{c}(2759)$ & $\Omega _{c}(3679)$ \\ 
$n_{1}n_{2}n_{3}$ & 2, 2, 0 & 1 3 0 or -200 \\ 
({\small Eq. No)} & (82) & (83) \\ 
Theory & $\Omega _{c}(2759)$ & $\Omega _{c}(3679)$ \\ 
&  & 
\end{tabular}
\end{tabular}

In summary, the BCC model explains all baryon experimental intrinsic quantum
numbers and masses. Virtually no experimentally confirmed baryon is not
included in the model.

It is worth while to pay attention to the \textbf{experimental top limits }%
of the baryons. According to the BCC model, a series of possible baryons
exist. However, when energy goes higher and higher, on one hand, the
theoretical energy bands (baryons) will become denser and denser (such as:
there are {\small 5 }$\Delta ${\small (2739), 3 N(2649), 4 N(2739), }5 $%
\Lambda ${\small (2649), }$3$ $\Sigma (2649)$,$\ 4$ $\Sigma (2739)$, and$\ 5$
$\Xi ${\small (2739)} in mass 2645 Mev-2745 Mev); while on the other hand,
the experimental full widthes of the baryons will become wider and wider (
such as $\Gamma $=650 Mev of N(2600), $\Gamma $=400 Mev of $\Delta (2420)$, $%
\Gamma $=150 Mev of $\Lambda (2350)$, and $\Gamma $=100 Mev of $\Sigma
(2250) $...). making them extremely difficult to be separated. Therefore, \
currently it is very difficult\ to discover higher energy baryons predicted
by the BCC model. We believe that many new baryons will be discovered in the
future with the development of more sensitive experimental techniques.

\section{Predictions and Discussion}

\subsection{Some New Baryons}

The following new baryons predicted by the model seem to have a better
chance to be discovered in a not too distant future 
\begin{equation}
\begin{array}{cccccc}
I=\text{ }0\text{ \ } & S=-3 & Q=-1 & \Omega ^{-}(3619) &  & \text{(\ref
{SIGMA_1})} \\ 
I=\text{ }0\text{ \ } & b=-1 & Q=\text{ \ }0 & \Lambda _{b}^{0}(10159) &  & 
\text{(\ref{SIGMA_1})} \\ 
I=\text{ }0\text{\ \ } & C=+1 & Q=+1 & \Lambda _{C}^{+}(6599) &  & \text{(%
\ref{DELTA-ONE})} \\ 
I=\text{ }0\text{ \ } & S=-1 & Q=\text{ \ }0 & \Lambda ^{0}(2559) &  & \text{%
(\ref{SIGMA_1})} \\ 
I=1/2 & S=-1 & C=+1 & Q=1,0\text{ \ \ } & \Xi _{C}(3169) & \text{(\ref
{F_6_KESIC})} \\ 
I=\text{ }1\text{ \ } & S=-1 & C=+1 & Q=2,1,0 & \Sigma _{C}(2969) & \text{(%
\ref{SIGMA(2969)})}
\end{array}
\end{equation}
In the last column, we give the equation number where the baryons are first
deduced in this paper.

\subsection{The Super Heavy Electron Spectrum}

Similar to the energy bands of the Lee Particle, we can also deduce the
energy bands for the Thomson \cite{Thomson} particle (a vacuum state
electron). When a Thomson particle is in the vacuum state the lepton number $%
L=0$, but when it is excited from vacuum, the lepton number $L=1$. The
energy $\varepsilon ^{(0)}(\vec{k},\vec{n})$ of the excited Thomson particle
are determined (similarly to (\ref{mass})) by 
\begin{equation}
\varepsilon ^{(0)}(\vec{k},\vec{n})=V_{Th}+C_{Th}E(\vec{k},\vec{n}),
\label{Thommass}
\end{equation}
where 
\begin{equation}
C_{Th}=h^{2}/(2m_{Th}a_{x}^{2}).  \label{Thomc}
\end{equation}

It is easy to determine the constant $V_{Th}$. From Eq.(\ref{Thommass}) we
can see that the lowest energy is $\varepsilon ^{(0)}(\vec{k},\vec{n}%
)=V_{Th} $. Similar to (\ref{vzero}), we know that the lowest energy is the
mass of the electron ($0.511$ Mev), so 
\begin{equation}
V_{Th}=m_{e}=0.511\text{ Mev.}  \label{ve}
\end{equation}
In order to determine the parameter $C_{Th}$, we assume that the masses of
the Thomson particle and the Lee Particle are proportional to the masses of
the ground states of the excited Thomson particle (the electron) and the
excited Lee Particle (the proton and the neutron): 
\begin{equation}
m_{Lee}=cm_{N}\text{ \ ,}
\end{equation}
\begin{equation}
m_{Th}=cm_{e}\text{ \ .}
\end{equation}
Thus 
\begin{equation}
m_{Th}=cm_{e}=(m_{Lee}/m_{N})m_{e}\text{ \ . }  \label{const}
\end{equation}
Using (\ref{alpha}), (\ref{Thomc}), (\ref{ve}), and (\ref{const}), we get 
\begin{equation}
C_{Th}=\frac{h^{2}}{2m_{Th}a_{x}^{2}}=\frac{m_{N}}{m_{e}}\frac{h^{2}}{%
2m_{Lee}a_{x}^{2}}=\frac{939}{0.511}360\text{Mev}=662\text{Gev.}
\end{equation}
Hence 
\begin{equation}
\varepsilon ^{(0)}(\vec{k},\vec{n})=V_{Th}+C_{Th}E(\vec{k},\vec{n}%
)=[0.511\times 10^{-3}+662\times E(\vec{k},\vec{n})]\text{Gev.}
\end{equation}

Fig. 2-4 show the energy bands of the body center cubic periodic field. The
periodic field is a strong interaction field which originates from the Lee
Particles. Because the Thomson particles (the vacuum state electrons) do not
have any strong interactions, they can not see the body center cubic
periodic field. They can see, however, a simple cubic electromagnetic
periodic field which originates from the charged Lee Particles (the point
aproximation of the colorless group uud). Similar to Fig. 2-4, we can find
the energy bands of the simple cubic periodic field. Since the isospin,
strange number, charmed number, and bottom number of the leptons have not
been observed in the experiments, we do have left them out. We provide the
electric charge and mass for a few lower energy excited states of the
Thomson particle, as the touchstone of the BCC model. The electric charges
are all the same with the Thomson particle. If they are really discovered,
the conjecture of the BCC model of the vacuum material is somehow confirmed.

\ \ \ \ \ \ \ \ \ \ \ \ \ \ \ \ \ \ \ \ \ \ \ \ \ \ \ \ \ \ \ \ \ \ \ \ \ \
\ \ \ \ \ \ \ \ \ \ \ \ \ 

$
\begin{tabular}{lllllll}
\hline
$E(\vec{k},\vec{n})$ & 0 & 1/4 & 1/2 & 3/4 & 1 & 2 \\ 
Theory(Gev) & e$^{\text{-}}$(0.511 Mev) & E(166) & E(331) & E(497) & E(662)
& E(1324) \\ 
Expt. \ \ (Gev) & e$^{\text{-}}$(0.511 Mev) & E(167) &  &  &  &  \\ 
$\Delta $M/M & 0.000 & 0.006 &  &  &  &  \\ \hline
\end{tabular}
$

\ \ \ \ \ \ \ \ \ \ \ \ \ \ \ \ \ \ \ \ \ \ \ \ \ \ \ \ \ \ \ \ \ \ \ \ \ \
\ \ \ \ \ \ \ \ \ \ \ \ \ \ \ \ \ \ \ \ \ \ \ \ \ \ \ \ \ \ \ 

We would like to call the excited Thomson particle super heavy electrons
with symbole $E$ according to T. D. Lee in his book \cite{Leebook}. Among
them, E($331$ Gev), E($497$ Gev), E($662$ Gev) and E($1324$ Gev) have not
been discovered yet. The reason is that their energies are too high.
However, a particle with a mass of $167$ Gev has been found by the
experimentt recently\cite{E167}. It might be the super heavy electron
E(166), which is about $177$ times heavier than a proton.

\subsection{ Experimental Verification of the BCC Model}

\begin{enumerate}
\item  From (\ref{SIGMA_1}) and Fig. 5 (c), we see three ``brother''
baryons: at E$_{N}=1/2\ \overrightarrow{n}=(1,1,0)$ $\Lambda (1119)$, at E$%
_{N}=9/2$ $\overrightarrow{n}=(2,2,0)$ $\Lambda (2559)$, at E$_{N}=25/2$ $%
\overrightarrow{n}=(3,3,0)$ $\Lambda _{b}(5639)$. They are born on the same
symmetry axis $\Sigma $ and at the same symmetry point $N$. The three
``brothers'' have the same isospin I = 0, the same electric charge Q = 0,
and the same generalized strange number (see (\ref{strangegen})) S$_{G}$ = S
+ C + b = -1. Among the three ``brothers'', the light one ($\Lambda (1119)$)
and the heavy one ($\Lambda _{b}(5639))$ both have long lifetimes ($\tau =$
2.6$\times $10$^{-10}s$ for $\Lambda (1119)$, $\tau =$ 1.1$\times $10$%
^{-12}s $ for $\Lambda _{b}(5639)$), but the middle one ($\Lambda (2559)$)
has not been discovered. Thus, we propose that one search for the long
lifetime baryon $\Lambda (2559)$ (I = 0, S = -1, Q = 0, M = 2559 Mev, and
lifetime 2.6$\times $10$^{-10}s>\tau >$1.1$\times $10$^{-12}s$ ). The
discovery of the baryon $\Lambda (2559)$ will provide a strong support for
the BCC model.

\item  According to the BCC model, the excited Lee particles which are in
different energy band states form a baryon spectrum. Similarly, the excited
Thomson \cite{Thomson} particles shall form a super heavy electron spectrum.
if we real experimentally observe the spectrum, we will confirm the BCC
model, and confirm the body center cubic symmetry of the vacuum material. 
\textbf{It will be a great discovery\ in the history of physics}. Therefore,
we suggest that one search experimentally for the super heavy electron
spectrum. The lowest super heavy electron of the spectrum may have already
been discovered \cite{TOPQUARK}\ \cite{E167}$.$\ \ \ 
\end{enumerate}

\subsection{Discussions}

\begin{enumerate}
\item  From (\ref{alpha}), we have 
\begin{equation}
m_{b}a_{x}^{2}=h^{2}/720\text{ Mev.}
\end{equation}
Although we do not know the values of $m_{b}$ and $a_{x}$, we find that $%
m_{b}a_{x}^{2}$ is a constant. According to the renormalization theory \cite
{renormal}, the bare mass of the Lee Particle should be infinite, so that $%
a_{x}$ will be zero. Of course, the infinite and the zero are physical
concepts in this case. We understand that the ``infinite'' means $m_{b}$ is
huge and the ``zero'' means $a_{x}$ is much smaller than the nuclear radius.
``$m_{b}$ is huge'' guarantees that we can use the Schr\"{o}dinger equation (%
\ref{Schrodinger}) instead of the Dirac equation, and ``$a_{x}$ is much
smaller than the nuclear radius'' makes the structure of the vacuum material
very difficult to be discovered.

\item  In a sense, the vacuum material with the body center cubic structure
works like an ultra-superconductor.\ There is no electric and mechanical
resistance to any particle and any physical body (with or without electric
charge) moving inside the vacuum material. Since the energy gaps are so
large (for electron the energy gap is about 0.5 Mev; for proton and neutron
the energy gap is about 939 Mev), the vacuum remains unchanged when a
physical body is moving with a constant velocity. At the same time, the
motion of the physical body remains unchanged also.

\item  The BCC model presents not only a baryon spectrum, but also a natural
and reasonable explanation for the experimental fact that all baryons
automatically decay to nucleons ($p$ or $n$) in a very short time ($<10^{-9}$
second). Although the decay forms are various, all baryons must decay to the
same kind of particles - nucleons. There is no baryon that does not decay,
and there is no baryon that decays into other particle. The reason is very
simple: the baryons are energy band excited states of the Lee Particle,
while the nucleons are the ground band states of the Lee Particles. It is a
well known law in physics that the excited states will decay into the ground
state. Thus, the long life baryons $\Lambda (1116),$ $\Sigma (1193),$ $\Xi
(1318),$ $\Omega (1672),$ $\Lambda _{c}(2285),$ $\Lambda _{b}(5641)...$may
be the metastable states of the Lee Particles.

\item  Since the theoretical baryon mass spectrum of the free particle
approximation (V($\vec{r}$) = V$_{o}$ and the wave functions satisfy the
body center cubic periodic symmetries) is very close to the experimental
mass spectrum, the amplitude (A) of the strong interaction periodic field
should be much smaller than the average of the periodic field (V$_{0}$).
According the BCC model, Dirac's sea concept\cite{diarcsea} is a complete
free approximation for the vacuum periodic field (V($\vec{r}$) = V$_{o}$ and
the wave functions do not have to satisfy the body center cubic periodic
symmetries). Because the amplitude of the vacuum periodic field is very
small and the periodic constant $a_{x}$ is very small also, Dirac's sea
concept is a very good approximation. In fact, Dirac's sea concept has
deeply embedded in quantum field theory and the Standard Model.
\end{enumerate}

\section{Conclusions\ \ \ }

1. Although baryons ($\Delta ,$ $N,$ $\Lambda ,$ $\Sigma ,$ $\Xi ,$ $\Omega
, $ $\Lambda _{C,}$ $\Xi _{C},$ $\Sigma _{C},$ and $\Lambda _{b}...$) are so
different from one another in I, S, C, b, Q, and M, they may be the same
kind of particles (the Lee Particles), which are in different energy band
states. The long life baryons $\Lambda (1116),$ $\Sigma (1193),$ $\Omega
(1672),$ $...$ may be the metastable states. The conclusion is supported not
only by the results of the BCC model, but also by many experiments which
demonstrate that all baryons (their lives are very short, $\tau <10^{-9}\sec 
$.) decay to the same kind of particles - nucleons (the ground energy bands).

2. The BCC model has deduced the strange number S, the charmed number C, and
the bottom number b from the symmetry of body center cubic periodic field.
It means that the quantum number S, C, and b may be not from the quarks (s,
c, b), they may be from new symmetries. Frank Wilczek point out in Review of
Modern Physics \cite{Wilczek2}: Some ``appropriate symmetry principles and
degrees of freedom, in terms of which the theory should be formulated, have
not yet been identified.'' \ We believe that the body center cubic periodic
symmetry of the vacuum material may be the appropriate symmetry.

3. The BCC model has shown that there may be only 2 kinds of quarks (u and
d), each of them has three colored members, in the quark family. The
super-strong attractive forces (color) make the colorless Lee Particle (uud
and udd). The Lee Particles constitute a body center cubic lattice in the
vacuum. In other words, the vacuum materials has body center cubic symmetry.

4. Due to the existence of the vacuum material, all observable particles are
constantly affected by the vacuum material (the Lee Particle lattice). Thus,
some laws of statistics (such as fluctuation) can not be ignored.

5. Although the BCC model successfully explain the baryon spectrum, the
baryon spectrum is deduced from 5 phenomenological Hypotheses and 3
phenomenological formulas in the BCC model. Thus, the BCC model is only a
phenomenological model.

\begin{center}
\bigskip \textbf{Acknowledgment}
\end{center}

I would like to express my heartfelt gratitude to Dr. Xin Yu for checking
the calculations of the energy bands and helping write the paper. I
sincerely thank Professor Robert L. Anderson for his valuable advice. I also
acknowledge\textbf{\ }my indebtedness to Professor D. P. Landau for his
help. I specially thank Professor T. D. Lee, for his \textit{particle physics%
} class in Beijing and his CUSPEA program which gave me an opportunity for
getting my Ph. D. I thank Professor W. K. Ge very much for his valuable help
and for recommending Wilczek's paper \cite{wilczek}. I thank my friend Z. Y.
Wu very much for his help in preparing the paper. I thank my classmate J. S.
Xie very much for checking the calculations of the energy bands. I thank
Professor Y. S. Wu, H. Y. Guo, and S. Chen \cite{XUarticle} very much for
very useful discussions.

\begin{center}
\bigskip {\LARGE FIGURES}
\end{center}

Fig. 1. \ The first Brillouin zone of the body center cubic lattice. The the
symmetry points and axes are indicated. The center of the first Brillouin
zone is at the point $\Gamma $. The axis $\Delta $ (the axis $\Gamma $-H) is
a $4$ fold rotation axis, the strange number S = 0, the baryon family $%
\Delta $ ($\Delta ^{++},$ $\Delta ^{+},$ $\Delta ^{0},$ $\Delta ^{-}$) will
appear on the axis. The axes $\Lambda $ (the axis $\Gamma $-P) and F (the
axis P-H) are $3$ fold rotation axes, the strange number S =\ -1, the baryon
family $\Sigma $ ($\Sigma ^{+},$ $\Sigma ^{0},$ $\Sigma ^{-})$ will appear
on the axes. The axes $\Sigma $ (the axis $\Gamma $-N) and G (the axis M-N)
are $2$ fold rotation axes, the strange number S = -2, the baryon family $%
\Xi $ ($\Xi ^{0},$ $\Xi ^{-}$) will appear on the axes. The axis D (the axis
P-N) is parallel to the axis $\Delta $, S = 0. And the axis is a $2$ fold
rotation axis, the baryon family N (N$^{+}$, N$^{0}$) will be on the axis.

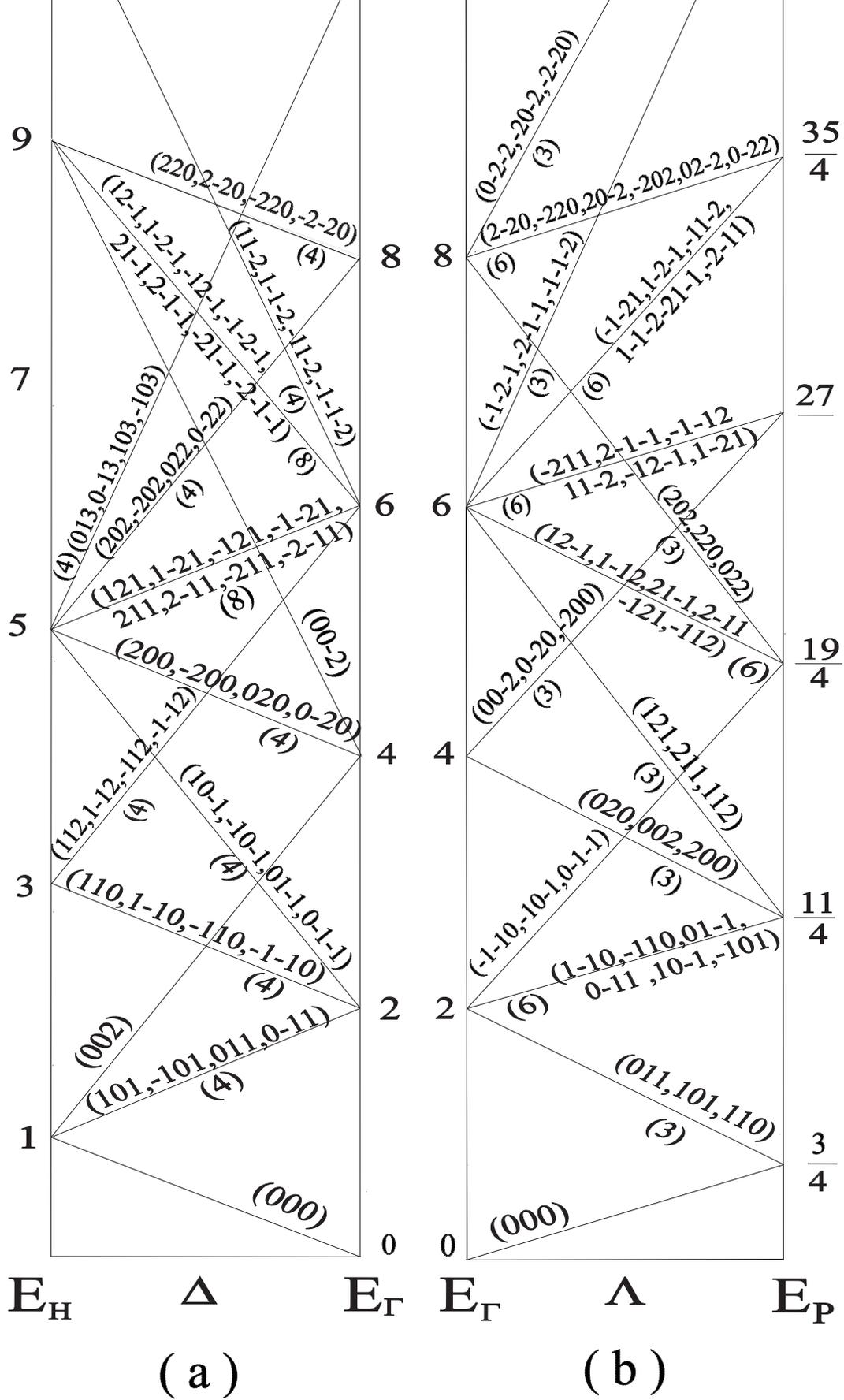
Fig. 2. \ \ (a) The energy bands on the axis $\Delta $ (the axis $\Gamma -$%
H). The numbers above the lines are the values of $\vec{n}$ = ($n_{1}$, $%
n_{2}$, $n_{3}$). The numbers under the lines are the fold numbers of the
degeneracy. E$_{\Gamma }$ is the value of E($\vec{k}$, $\vec{n}$) (see Eq. (%
\ref{energy})) at the end point $\Gamma ,$ while E$_{H}$ is the value of E($%
\vec{k}$, $\vec{n}$) at other end point H. \ \ (b) The energy bands on the
axis $\Lambda $ (the axis $\Gamma $-P). E$_{\Gamma }$ is the value of E($%
\vec{k}$, $\vec{n}$) (see Eq. (\ref{energy})) at the end point $\Gamma ,$
while E$_{P}$ is the value of E($\vec{k}$, $\vec{n}$) at other end point P.
The numbers above the lines are the values of $\vec{n}$ = ($n_{1}$, $n_{2}$, 
$n_{3}$). The numbers under the lines are the fold numbers of the degeneracy.

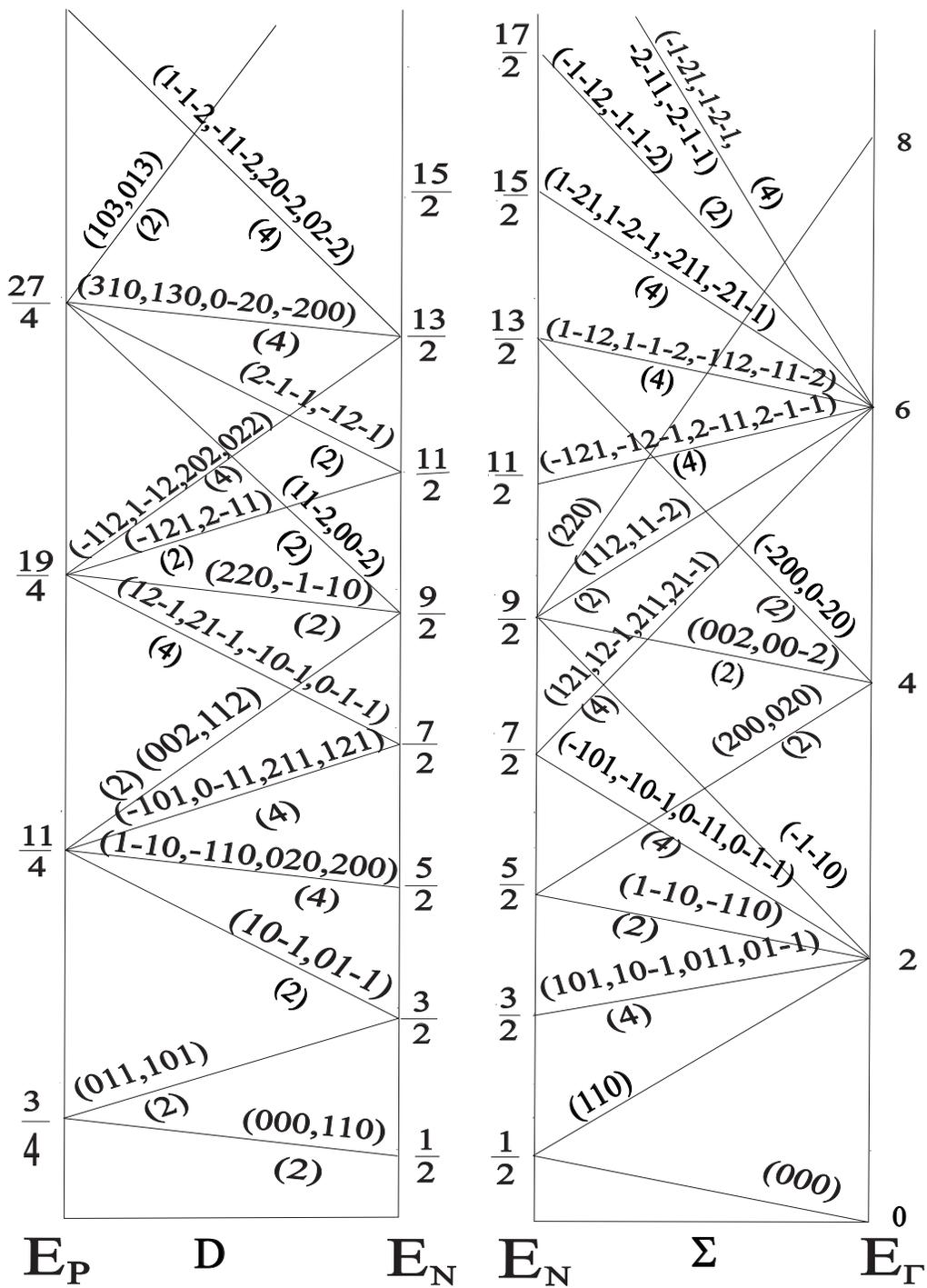
Fig. 3. \ \ (a) The energy bands on the axis $\Sigma $ (the axis $\Gamma $%
-N). The numbers above the lines are the values of $\vec{n}$ = ($n_{1}$, $%
n_{2}$, $n_{3}$). The numbers under the lines are the fold numbers of the
degeneracy. E$_{\Gamma }$ is the value of E($\vec{k}$, $\vec{n}$) (see Eq. (%
\ref{energy})) at the end point $\Gamma ,$ while E$_{N}$ is the value of E($%
\vec{k}$, $\vec{n}$) at other end point N. \ (b) The energy bands on the
axis $D$ (the axis P-N). E$_{P}$ is the value of E($\vec{k}$, $\vec{n}$)
(see Eq. (\ref{energy})) at the end point P$,$ while E$_{N}$ is the value of
E($\vec{k}$, $\vec{n}$) at other end point N. The numbers above the lines
are the values of $\vec{n}$ = ($n_{1}$, $n_{2}$, $n_{3}$). The numbers under
the lines are the fold numbers of the degeneracy.

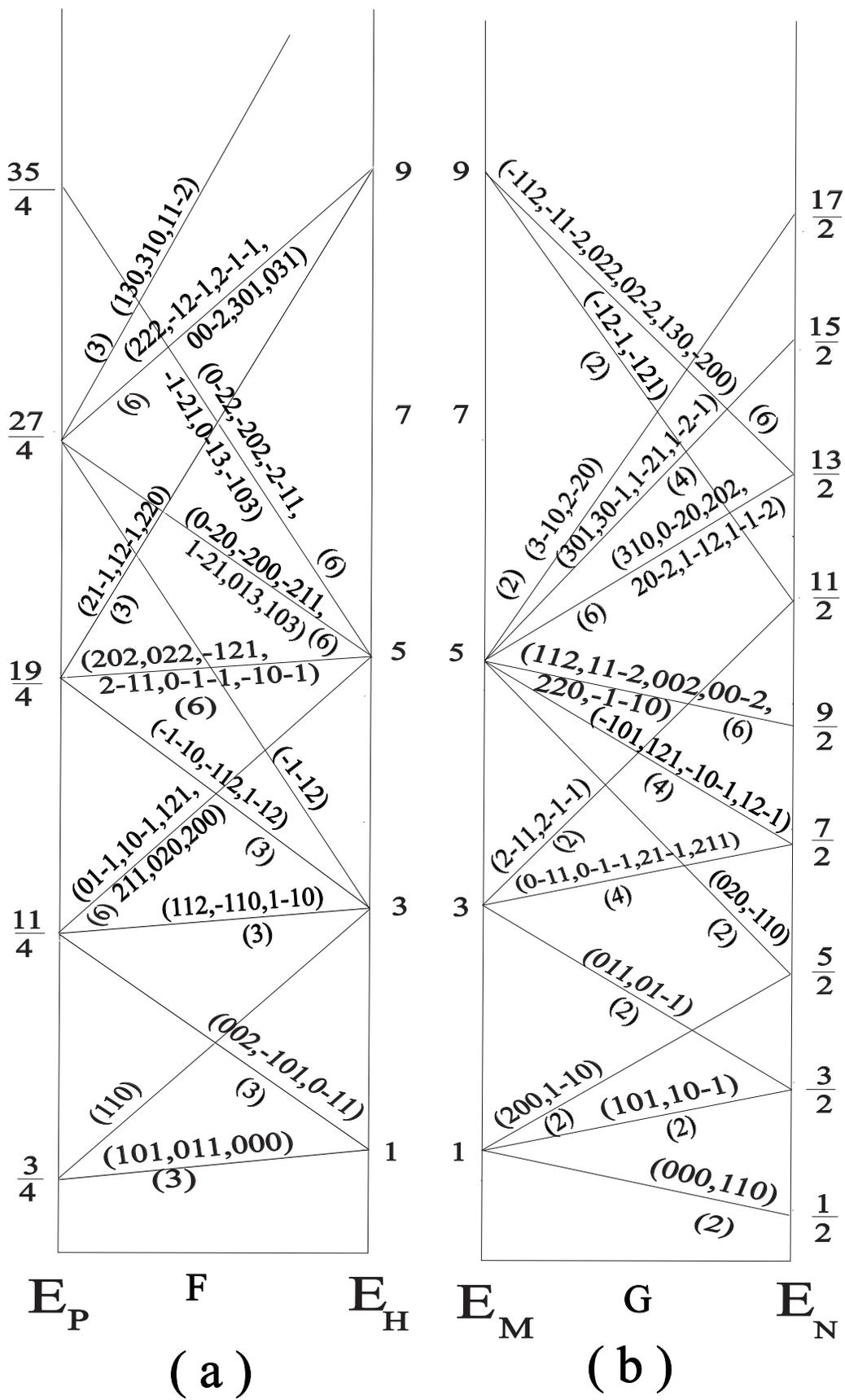
Fig. 4. \ \ (a) The energy bands on the axis $F$ (the axis P-H). The numbers
above the lines are the values of $\vec{n}$ = ($n_{1}$, $n_{2}$, $n_{3}$).
The numbers under the lines are the fold numbers of the degeneracy. E$_{P}$
is the value of E($\vec{k}$, $\vec{n}$) (see Eq. (\ref{energy})) at the end
point $P,$ while E$_{H}$ is the value of E($\vec{k}$, $\vec{n}$) at other
end point H.\ \ (b) The energy bands on the axis $G$ (the axis M-N).\ E$_{M}$
is the value of E($\vec{k}$, $\vec{n}$) (see Eq. (\ref{energy})) at the end
point M$,$ while E$_{N}$ is the value of E($\vec{k}$, $\vec{n}$) at other
end point N. The numbers above the lines are the values of $\vec{n}$ = ($%
n_{1}$, $n_{2}$, $n_{3}$). The numbers under the lines are the fold numbers
of the degeneracy.\ \ 

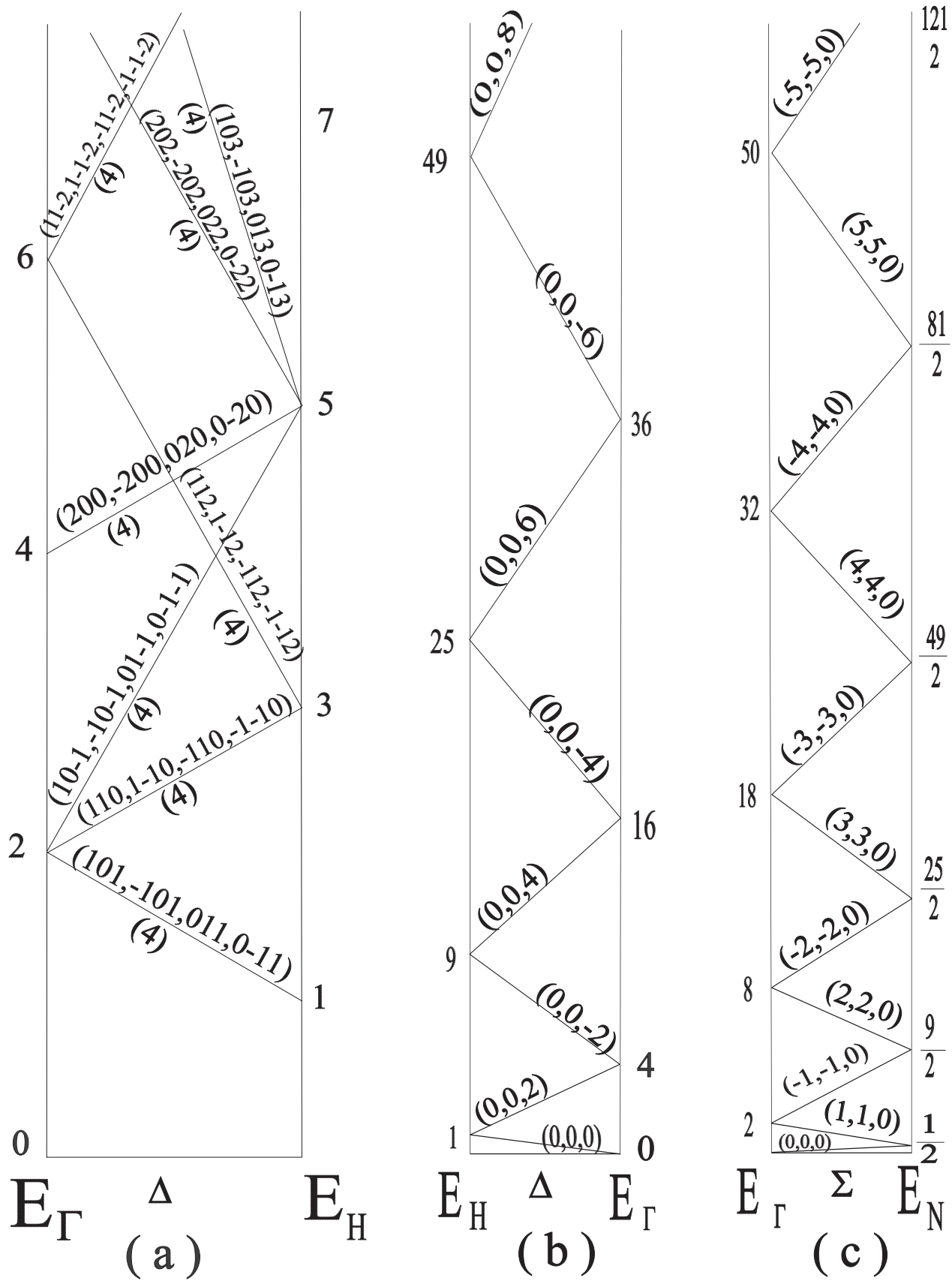
Fig. 5. \ \ (a) The $4$ fold degenerate energy bands (selected from Fig.
2(a)) on the axis $\Delta $ (the axis $\Gamma $-H). The numbers above the
lines are the values of $\vec{n}$ (n$_{1}$, n$_{2}$, n$_{3}$). The numbers
under the lines are the numbers of the degeneracy of the energy bands. \ \
(b) The single energy bands (selected from Fig. 2(a)) on the axis $\Delta $
(the axis $\Gamma $-H). The numbers above the lines are the values of $\vec{n%
}$ (n$_{1}$, n$_{2}$, n$_{3}$). \ \ (c) The single energy band (selected
from Fig. 3(a)) on the axis $\Sigma $ (the axis $\Gamma $-N). The numbers
above the lines are the values of $\vec{n}$ (n$_{1}$, n$_{2}$, n$_{3}$)

\newpage \pagebreak

\bigskip

\bigskip

\bigskip

\bigskip

\bigskip

\bigskip

\bigskip

\bigskip

\bigskip

\bigskip

\bigskip

\bigskip

\newpage

\begin{center}
\bigskip\ \ \ \ \ \ 

{\LARGE TABLE}
\end{center}

\qquad \qquad \qquad \qquad Table 1. \ The Ground States of the Various
Baryons.

\begin{tabular}{|l|l|l|l|l|l|}
\hline
{\small Theory} & Quantum. No &  & {\small Experiment} & R & Life Time \\ 
\hline
Name({\small M}) & S, C, \ b, \ \ I, \ Q & Eq. No & Name({\small M}) &  & 
\\ \hline
p(939) & 0,\ \ 0, \ 0, {\small 1/2, \ \ }1 & (\ref{QN of N(939)}) & p(938) & 
0.1 & $ >10^{31}years$ \\ \hline
n(939) & 0, \ 0, \ 0, {\small 1/2, \ \ }0 & (\ref{QN of N(939)}) & n(940) & 
0.1 & 1.0$\times 10^{8}$ s \\ \hline
$\Lambda (1119)$ & -1, \ 0, \ 0, \ \ 0, \ 0 & (\ref{SIGMA_1}) & $\Lambda
(1116)$ & 0.3 & 2.6$\times $10$^{-10}s$ \\ \hline
$\Sigma (1209)^{+}$ & -1, \ 0, \ 0, \ \ 1, \ 1 & (\ref{LAMBDA_3}) & $\Sigma
(1189)^{+}$ & 1.7 & .80$\times $10$^{-10}s$ \\ \hline
$\Sigma (1209)^{0}$ & -1, \ 0, \ 0, \ \ 1, \ 0 & (\ref{LAMBDA_3}) & $\Sigma
(1193)^{0}$ & 1.4 & 7.4$\times $10$^{-20}s$ \\ \hline
$\Sigma (1209)^{-}$ & -1, \ 0, \ 0, \ \ 1, -1 & (\ref{LAMBDA_3}) & $\Sigma
(1197)^{-}$ & 1.0 & 1.5$\times $10$^{-10}s$ \\ \hline
$\Xi (1299)^{0}$ & -2, \ 0, \ 0, {\small 1/2}, 0 & (\ref{G_TWO}) & $\Xi
(1315)^{0}$ & 1.2 & 2.9$\times $10$^{-10}s$ \\ \hline
$\Xi (1299)^{-}$ & -2, \ 0, \ 0, {\small 1/2}, -1 & (\ref{G_TWO}) & $\Xi
(1321)^{-}$ & 1.7 & 1.6$\times $10$^{-10}s$ \\ \hline
$\Omega (1659)^{-}$ & -3, \ 0, \ 0. \ \ 0, -1 & (\ref{SIGMA_1}) & $\Omega
(1672)^{-}$ & 0.8 & .82$\times $10$^{-10}s$ \\ \hline
$\Lambda _{c}^{+}(2279)$ & 0, \ 1, \ 0, \ \ 0, \ 1 & (\ref{DELTA-ONE}) & $%
\Lambda _{c}^{+}(2285)$ & 0.3 & .21$\times $10$^{-12}s$ \\ \hline
$\Xi _{c}^{+}(2549)$ & -1, \ 1, \ 0, {\small 1/2}, 1 & (\ref{KERSA-C}) & $%
\Xi _{c}^{+}(2466)$ & 3.4 & .35$\times $10$^{-12}$ \\ \hline
$\Xi _{c}^{0}(2549)$ & -1, \ 1, \ 0, {\small 1/2}, 1 & (\ref{KERSA-C}) & $%
\Xi _{c}^{0}(2470)$ & 3.2 & .10$\times $10$^{-12}s$ \\ \hline
$\Sigma _{c}^{++}(2449)$ & 0, \ 1, \ 0, \ 1, \ \ 2 & (\ref{F_B_C}) & $\Sigma
_{c}^{++}(2453)$ & 0.2 &  \\ \hline
$\Sigma _{c}^{+}(2449)$ & 0, \ 1, \ 0, \ 1, \ \ 1 & (\ref{F_B_C}) & $\Sigma
_{c}^{+}(2454)$ & 0.2 &  \\ \hline
$\Sigma _{c}^{0}(2449)$ & 0, \ 1, \ 0, \ 1, \ \ 0 & (\ref{F_B_C}) & $\Sigma
_{c}^{0}(2452)$ & 0.1 &  \\ \hline
$\Omega _{c}(2759)$ & 0, \ 0, -1, \ 0, \ 0 & (\ref{G_6_C}) & $\Omega
_{c}(2704)$ & 2.0 & .64$\times $10$^{-13}s$ \\ \hline
$\Lambda _{b}(5639)$ & 0, \ 0, -1, \ 0, \ 0 & (\ref{SIGMA_1}) & $\Lambda
_{b}(5641)$ & .04 & 1.1$\times $10$^{-12}s$ \\ \hline
$\Delta (1299)^{++}$ & 0, \ 0, \ 0, {\small 3/2}, \ 2 & (\ref{DELTA_4}) & $%
\Delta (1232)^{++}$ & 5.2 & $\Gamma $=120 Mev \\ \hline
$\Delta (1299)^{+}$ & 0, \ 0, \ 0, {\small 3/2}, \ 1 & (\ref{DELTA_4}) & $%
\Delta (1232)^{+}$ & 5.4 & $\Gamma $=120 Mev \\ \hline
$\Delta (1299)^{0}$ & 0, \ 0, \ 0, {\small 3/2}, \ 0 & (\ref{DELTA_4}) & $%
\Delta (1232)^{0}$ & 5.4 & $\Gamma $=120 Mev \\ \hline
$\Delta (1299)^{-}$ & 0, \ 0, \ 0, {\small 3/2}, -1 & (\ref{DELTA_4}) & $%
\Delta (1232)^{-}$ & 5.4 & $\Gamma $=120 Mev \\ \hline
\end{tabular}

{\small * }In the third column, we give the equation number where the baryon

can be found in this paper. In the fifth column, R =($\frac{\Delta \text{M}}{%
\text{M}}$){\small \%.}

\bigskip

\bigskip

\bigskip

\bigskip

\bigskip

\bigskip

\bigskip

\bigskip

\bigskip

\bigskip

\bigskip

\bigskip

\bigskip

\bigskip

\bigskip

\bigskip \newpage \qquad\ \ \ \ \ \ \ \ \ \ Table 2. The Unflavored Baryons $%
N$ and $\Delta $ ($S$= $C$=$b$ = 0) \ 

\begin{tabular}{|l|l|l||l|l|l|}
\hline
Theory & Experiment & $\frac{\Delta \text{M}}{\text{M}}\%$ & Theory & 
Experiment & $\frac{\Delta \text{M}}{\text{M}}\%$ \\ \hline
$\mathbf{\bar{N}(939)}$ & $\mathbf{\bar{N}(939)}$ & \textbf{0.0} & $\mathbf{%
\bar{\Delta}(1254)}^{\#}$ & $\mathbf{\bar{\Delta}(1232)}$ & \textbf{1.8} \\ 
\hline
$N(1479)$ & 
\begin{tabular}{l}
$N(1440)$ \\ 
$N(1520)$ \\ 
$N(1535)$%
\end{tabular}
&  &  &  &  \\ \hline
$\mathbf{\bar{N}(1479)}$ & $\mathbf{\bar{N}(1498)}$ & \textbf{1.2} &  &  & 
\\ \hline
\begin{tabular}{l}
$N(1659)$ \\ 
$N(1659)$%
\end{tabular}
& 
\begin{tabular}{l}
$N(1650)$ \\ 
$N(1675)$ \\ 
$N(1680)$ \\ 
$N(1700)$ \\ 
$N(1710)$ \\ 
$N(1720)$%
\end{tabular}
&  & 
\begin{tabular}{l}
$\Delta (1659)$ \\ 
$\Delta (1659)$%
\end{tabular}
& 
\begin{tabular}{l}
$\Delta (1600)$ \\ 
$\Delta (1620)$ \\ 
$\Delta (1700)$%
\end{tabular}
&  \\ \hline
$\mathbf{\bar{N}(1659)}$ & $\mathbf{\bar{N}(1689)\ \ }$ & \textbf{1.7} & $%
\mathbf{\bar{\Delta}(1659)}$ & $\mathbf{\bar{\Delta}(1640)}$ & \textbf{1.2}
\\ \hline
\begin{tabular}{l}
$N(1839)$ \\ 
$N(1839)$ \\ 
$N(1929)$ \\ 
$N(1929)$ \\ 
$N(2019)$%
\end{tabular}
& 
\begin{tabular}{l}
$N(1900)\ast $ \\ 
$N(1990)\ast $ \\ 
$N(2000)\ast $ \\ 
$N(2080)\ast $%
\end{tabular}
&  & 
\begin{tabular}{l}
$\Delta (1929)$ \\ 
$\Delta (1929)$ \\ 
$\Delta (2019)$%
\end{tabular}
& 
\begin{tabular}{l}
$\Delta (1900)$ \\ 
$\Delta (1905)$ \\ 
$\Delta (1910)$ \\ 
$\Delta (1920)$ \\ 
$\Delta (1930)$ \\ 
$\Delta (1950)$%
\end{tabular}
&  \\ \hline
$\mathbf{\bar{N}(1914)}$ & $\mathbf{\bar{N}(1923)}$ & \textbf{0.5} & $%
\mathbf{\bar{\Delta}(1959)}$ & $\mathbf{\bar{\Delta}(1919)}$ & \textbf{2.1}
\\ \hline
\begin{tabular}{l}
$N(2199)$ \\ 
$N(2199)$%
\end{tabular}
& 
\begin{tabular}{l}
$N(2190)$ \\ 
$N(2220)$ \\ 
$N(2250)$%
\end{tabular}
&  &  &  &  \\ \hline
$\mathbf{\bar{N}(2199)}$ & $\mathbf{\bar{N}(2220)}$ & \textbf{0.9} &  &  & 
\\ \hline
\begin{tabular}{l}
$N(2379)$ \\ 
$N(2549)$%
\end{tabular}
&  &  & $\Delta (2379)$ & $\Delta (2420)$ &  \\ \hline
$\ 
\begin{tabular}{l}
$N(2549)$ \\ 
$N(2559)$%
\end{tabular}
$ &  &  & $\mathbf{\bar{\Delta}(2379)}$ & $\mathbf{\bar{\Delta}(2420)}$ & 
\textbf{1.6} \\ \hline
$\mathbf{3N(2649)}$ & $\mathbf{N(2600)}$ & \textbf{1.9} & $\Delta (2649)$ & 
&  \\ \hline
4$N(2739)$ &  &  & 5$\Delta (2739)$ &  &  \\ \hline
\end{tabular}

\#{\small The average of N(1209) and }$\Delta (1299).$

*{\small Evidences are fair, they are not listed in the Baryon Summary Table 
\cite{particle}}.

\bigskip

\bigskip

\bigskip

\bigskip

\bigskip

\bigskip

\bigskip

\bigskip

\bigskip

\bigskip

\bigskip

\bigskip

\bigskip

\bigskip

\bigskip

\bigskip

\bigskip

\bigskip \newpage\ \ \ \ \ \qquad\ \ \ Table 3. Two Kinds of Strange Baryons 
$\Lambda $ and $\Sigma $ ($S=-1$)

\begin{tabular}{|l|l|l||l|l|l|}
\hline
Theory & Experiment & $\frac{\Delta \text{M}}{\text{M}}\%$ & Theory & 
Experiment & $\frac{\Delta \text{M}}{\text{M}}\%$ \\ \hline
$\mathbf{\Lambda (1119)}$ & $\mathbf{\Lambda (1116)}$ & \textbf{0.36} & $%
\mathbf{\Sigma (1209)}$ & $\mathbf{\Sigma (1193)}$ & \textbf{1.4} \\ \hline
\begin{tabular}{l}
$\Lambda (1299)$ \\ 
$\Lambda (1399)$%
\end{tabular}
&  &  &  &  &  \\ \hline
$\overline{\mathbf{\Lambda }}\mathbf{(1349)}$ & $\overline{\mathbf{\Lambda }}%
\mathbf{(1405)}$ & \textbf{4.0} & $\mathbf{\Sigma (1299)}$ & $\mathbf{\Sigma
(1385)}$ & \textbf{6.2} \\ \hline
\begin{tabular}{l}
$\Lambda (1659)$ \\ 
$\Lambda (1659)$ \\ 
$\Lambda (1659)$%
\end{tabular}
& 
\begin{tabular}{l}
$\Lambda (1520)$ \\ 
$\Lambda (1600)$ \\ 
$\Lambda (1670)$ \\ 
$\Lambda (1690)$%
\end{tabular}
&  & 
\begin{tabular}{l}
$\Sigma (1659)$ \\ 
$\Sigma (1659)$ \\ 
$\Sigma (1659)$%
\end{tabular}
& 
\begin{tabular}{l}
$\Sigma (1660)$ \\ 
$\Sigma (1670)$ \\ 
$\Sigma (1750)$ \\ 
$\Sigma (1775)$%
\end{tabular}
&  \\ \hline
$\mathbf{\bar{\Lambda}(1659)}$ & $\mathbf{\bar{\Lambda}(1620)}$ & \textbf{2.4%
} & $\mathbf{\bar{\Sigma}(1659)}$ & $\mathbf{\bar{\Sigma}(1714)}$ & \textbf{%
3.2} \\ \hline
\begin{tabular}{l}
$\Lambda (1929)$ \\ 
$\Lambda (1929)$ \\ 
$\Lambda (1929)$%
\end{tabular}
& 
\begin{tabular}{l}
$\Lambda (1800)$ \\ 
$\Lambda (1810)$ \\ 
$\Lambda (1820)$ \\ 
$\Lambda (1830)$ \\ 
$\Lambda (1890)$%
\end{tabular}
&  & 
\begin{tabular}{l}
$\Sigma (1929)$ \\ 
$\Sigma (1929)$%
\end{tabular}
& 
\begin{tabular}{l}
$\Sigma (1915)$ \\ 
$\Sigma (1939)$%
\end{tabular}
&  \\ \hline
$\mathbf{\bar{\Lambda}(1929)}$ & $\mathbf{\bar{\Lambda}(1830)}$ & \textbf{5.4%
} & $\mathbf{\bar{\Sigma}(1929)}$ & $\mathbf{\bar{\Sigma}(1928)}$ & \textbf{%
.05} \\ \hline
\begin{tabular}{l}
$\Lambda (2019)$ \\ 
$\Lambda (2019)$%
\end{tabular}
& 
\begin{tabular}{l}
$\Lambda (2100)$ \\ 
$\Lambda (2110)$%
\end{tabular}
&  & $\mathbf{\Sigma (2019)}$ & $\mathbf{\Sigma (2030)}$ & \textbf{.54} \\ 
\hline
$\mathbf{\bar{\Lambda}(2019)}$ & $\mathbf{\Lambda (2105)}$ & \textbf{4.1} & 
&  &  \\ \hline
$
\begin{tabular}{l}
$\Lambda (2359)$ \\ 
$\Lambda (2379)$%
\end{tabular}
$ & $\Lambda (2350)$ &  & $\Sigma (2379)$ & 
\begin{tabular}{l}
$\Sigma (2250)$ \\ 
$\Sigma (2455)^{\ast }$%
\end{tabular}
&  \\ \hline
$\mathbf{\bar{\Lambda}(2369)}$ & $\mathbf{\bar{\Lambda}(2350)}$ & \textbf{0.8%
} & $\mathbf{\bar{\Sigma}(2379)}$ & $\mathbf{\bar{\Sigma}(2353)}$ & \textbf{%
1.1} \\ \hline
$\mathbf{\Lambda (2559)}$ & $\mathbf{\Lambda (2585)}^{\ast }$ & \textbf{0.9}
&  &  &  \\ \hline
5$\Lambda (2649)$ &  &  & \textbf{3} $\mathbf{\Sigma (2649)}$ & $\mathbf{%
\Sigma (2620)}$ & \textbf{1.1} \\ \hline
&  &  & 4 $\mathbf{\Sigma (2739)}$ &  &  \\ \hline
\end{tabular}
\ \ \ \ \ \ \ \ \ \ \ \ \ \ \ \ \ \ \ \ \ \ \ \ \ \ \ \ \ \ \ 

{\small \ }*Evidences of existence for these baryons are only fair, they are
not

listed in the Baryon Summary Table \cite{particle}.

\bigskip

\bigskip \newpage

\qquad\ \ \ \ \ \ \ Table 4. The Baryons $\Xi $ and the Baryons $\Omega $

\begin{tabular}{|l|l|l||l|l|l|}
\hline
Theory & Experiment & $\frac{\Delta \text{M}}{\text{M}}\%$ & Theory & 
Experiment & $\frac{\Delta \text{M}}{\text{M}}\%$ \\ \hline
$\ \mathbf{\Xi (1299)}$ & $\mathbf{\Xi (1318)}$ & \textbf{1.5} & $\mathbf{%
\Omega (1659)}$ & $\mathbf{\Omega (1672)}$ & 0\textbf{.8} \\ \hline
$\ \mathbf{\Xi (1479)}$ & $\mathbf{\Xi (1530)}$ & \textbf{3.3} & $\Omega
(2359)$ & 
\begin{tabular}{l}
$\Omega (2250)$ \\ 
$\Omega (2380)$ \\ 
$\Omega (2470)$%
\end{tabular}
&  \\ \hline
\textbf{3}$\mathbf{\Xi (1659)}$ & $\mathbf{\Xi (1690)}$ & \textbf{1.8} & $%
\mathbf{\bar{\Omega}(2359)}$ & $\mathbf{\bar{\Omega}(2367)}$ & \textbf{0.4}
\\ \hline
$\ \mathbf{\Xi (1839)}$ & $\mathbf{\Xi (1820)}$ & \textbf{1.1} & $\Omega
(2879)$ &  &  \\ \hline
\textbf{\ }$\mathbf{\Xi (1929)}$ & $\mathbf{\Xi (1950)}$ & \textbf{1.1} & $%
\Omega (3619)$ &  &  \\ \hline
\textbf{2}$\mathbf{\Xi (2019)}$ & $\mathbf{\Xi (2030)}$ & \textbf{1.0} & $%
\Omega (7019)$ &  &  \\ \hline
\textbf{2}$\mathbf{\Xi (2199)}$ & $\mathbf{\Xi (2250)}^{\ast }$ & \textbf{2.3%
} &  &  &  \\ \hline
$\ \ \mathbf{\Xi (2379)}$ & $\mathbf{\Xi (2370)}^{\ast }$ & \textbf{0.4} & 
&  &  \\ \hline
$\ \Xi (2559)$ &  &  &  &  &  \\ \hline
5$\Xi (2739)$ &  &  &  &  &  \\ \hline
\end{tabular}

\bigskip *Evidences of existence for these baryons are only fair, they are
not

listed in the Baryon Summary Table \cite{particle}.

{\small \bigskip }\newpage

\qquad \qquad Table 5. Charmed \ $\Lambda _{c}^{+}$\ and Bottom $\Lambda _{b%
\text{ }}^{0}$ Baryons \ 

\begin{tabular}{|l|l|l||l|l|l|}
\hline
Theory & Experiment & $\frac{\Delta \text{M}}{\text{M}}\%$ & Theory & 
Experiment & $\frac{\Delta \text{M}}{\text{M}}\%$ \\ \hline
$\mathbf{\Lambda }_{c}^{+}\mathbf{(2279)}$ & $\mathbf{\Lambda }_{c}^{+}%
\mathbf{(2285)}$ & \textbf{0.22} & $\mathbf{\Lambda }_{b}^{0}\mathbf{(5639)}$
& $\mathbf{\Lambda }_{b}^{0}\mathbf{(5641)}$ & 0.035 \\ \hline
$
\begin{tabular}{l}
$\Lambda _{c}^{+}(2449)$ \\ 
$\Lambda _{c}^{+}(2539)$%
\end{tabular}
$ & 
\begin{tabular}{l}
$\Lambda _{c}^{+}(2593)$ \\ 
$\Lambda _{c}^{+}(2625)$%
\end{tabular}
&  & $\Lambda _{b}^{0}(10159)$ &  &  \\ \hline
$\mathbf{\bar{\Lambda}}_{c}^{+}\mathbf{(2494)}$ & $\mathbf{\bar{\Lambda}}%
_{c}^{+}\mathbf{(2609)}$ & \textbf{4.4} &  &  &  \\ \hline
$\Lambda _{c}^{+}(2759)$ &  &  &  &  &  \\ \hline
$\Lambda _{c}^{+}(2969)$ &  &  &  &  &  \\ \hline
$\Lambda _{c}^{+}(6599)$ &  &  &  &  &  \\ \hline
\end{tabular}

\bigskip *Evidences of existence for these baryons are only fair, they are
not

listed in the Baryon Summary Table \cite{particle}.

{\small \bigskip }

\bigskip

\newpage

\qquad \qquad Table 6. Charmed Strange Baryon $\Xi _{c}$, $\Sigma _{c}$ and $%
\Omega _{C}$ \ 

\begin{tabular}{|l|l|l||l|l|l|}
\hline
Theory & Experiment & $\frac{\Delta \text{M}}{\text{M}}\%$ & Theory & 
Experiment & $\frac{\Delta \text{M}}{\text{M}}\%$ \\ \hline
$\Xi _{c}(2549)$ & 
\begin{tabular}{l}
$\Xi _{c}(2468)$ \\ 
$\Xi _{c}(2645)$%
\end{tabular}
&  & 
\begin{tabular}{l}
$\Sigma _{c}(2449)$ \\ 
$\Sigma _{c}(2539)$%
\end{tabular}
& $
\begin{tabular}{l}
$\Sigma _{c}(2455)$ \\ 
$\Sigma _{c}(2530)\ast $%
\end{tabular}
$ &  \\ \hline
$\mathbf{\bar{\Xi}}_{c}\mathbf{(2549)}$ & $\mathbf{\bar{\Xi}}_{c}\mathbf{%
(2557)}$ & \textbf{0.3} & $\mathbf{\bar{\Sigma}}_{c}\mathbf{(2495)}$ & $%
\mathbf{\bar{\Sigma}}_{c}\mathbf{(2493)}$ & \textbf{0.08} \\ \hline
$\Xi _{C}(3169)$ &  &  & $\Sigma _{c}(2969)$ &  &  \\ \hline
&  &  &  &  &  \\ \hline
&  &  & $\mathbf{\Omega }_{C}\mathbf{(2759)}$ & $\mathbf{\Omega }_{C}\mathbf{%
(2704)}$ & \textbf{2.0} \\ \hline
&  &  & $\Omega _{C}(3679)$ &  &  \\ \hline
\end{tabular}
\ \ \ 

*Evidences of existence for these baryons are only fair, they are not

listed in the Baryon Summary Table \cite{particle}.

\bigskip

\bigskip

\bigskip

\bigskip

\bigskip

\bigskip

\bigskip


\begin{thebibliography}{99}
\bibitem{QuarkModel}  M. Gell-Mann, Phys. Lett.\textbf{\ 8}, 214 (1964); G.
Zweig, CERN Preprint Th 401, 402 (1964); A. J. G. Hey and R. L. Kelly, Phys.
Reports, \textbf{96}, 71 (1983); N.Isgur, Int. Mod. Phys.\textbf{\ E1}, 465
(1992).

\bibitem{RELATION of MASS}  S. Okubo, Phys. Lett. \textbf{4}. 14 (1963); G.
Morpurgo, Phys. Rev. Lett., \textbf{68}, 139 (1992).

\bibitem{QUARKS}  R. M. Barrett \textit{et al.}, Rev. Mod. Phys., \textbf{%
68, }303 and 658 (1996).

\bibitem{Three Color}  O. W. Greenberg, Phys. Rev. Lett. \textbf{13} (1964)
598; M. Y. Han and Y. Nanbu,Three-Triplet Model with Double SU(3) Symmetry,
Phys. Rev. \textbf{139B}, (1965)1006-1010.

\bibitem{Vacuum engineering}  T. D. Lee, \textit{Particle Physics and
Introduction to Field Theory }(harwood academic, New York, 1981) 824.

\bibitem{Hand in}  The Oecd Frum, \textit{Particle Physics (Head of
Publications Service, OECD) 55 (1995). }A. Pais, Rev. Mod. Phys., \textbf{71
No.2,} S16 (1999).

\bibitem{QUARK DECAY}  S. L. Glashow, J. Iliopoulos, and L. Maiani, Phy.
Rev. \textbf{D2} (1970) 1285; S. Weinberg Phys. Rev. Lett. \textbf{19 }%
(1967) 1254; R M. Barrett \textit{et al.}, Rev. Mod. Phys., \textbf{68,} 642
(1996).

\bibitem{Free QUARK}  R. M. Barrett \textit{et al.}, Rev. Mod. Phys., 
\textbf{68,} 315 (1996).

\bibitem{CONFINEMENT}  T. D. Lee, \textit{Particle Physics and Introduction
to Field Theory }( Harwood Academic, New York, 1981) 391; M. Creutz, Phy.
Rev. D \textbf{21}, 2308 (1980); K. G. Wilson, Phys. Rev. D \textbf{10, 2445
(1974}); A. Pais, Rev. Mod. Phys., \textbf{71 No.2,} S16 (1999).

\bibitem{particle}  R. M. Barrett \textit{et al.}, Rev. Mod. Phys., \textbf{%
68,} 642 (1996).

\bibitem{Leeparticle}  \ T. D. Lee, Rev. Mod. Phys. \textbf{47, }267 (1975).

\bibitem{asymmetry}  T. D. Lee, \textit{Particle Physics and Introduction to
Field Theory }( Harwood Academic, New York, 1981) 378.

\bibitem{li pARTICLE}  (1) T. D. Lee has pointed out that vacuum is the
source of asymmetry in his book (\textit{Particle Physics and Introduction
to Field Theory, 1981, page }378); furthermore, in the section ``the
possibility of vacuum engineering'' of his book , T. D. Lee pointed out:
``we believe our vacuum, ..., to be quite complicated. Like any other
physical medium, it can carry long-range-order parameters and it may also
undergo phase transitions... ''. His ideas inspired us to research the
structure of the vacuum. (2) T. D. Lee has pointed out: ``We are still in
the transition period. In order to apply the present theories, we need about
seventeen ad hoc parameters. All these theories are based on symmetry
considerations, yet most of the symmetry quantum numbers do not appear to be
conserved. All hadrons are made of quarks and yet no single quark can be
individually observed. ... We are in a serious dilemma about how to make the
next giant step. Because the challenge is related to the very foundation of
the totality of physics, a breakthrough is bound to bring us a profound
change in basic science.'' His statements$\ $strengthen our confidence in
the research. (3) We thank T. D. Lee for his \textit{particle physics} class
in Beijing and his CUSPEA program which gave me an opportunity to get my Ph.
D. in the U. S. A. (4) Lee is both a Chinese and an American name. Lee is
the largest chinese family name and it may be the largest family name in the
world. We use it to name the particles in order to represent ordinary people
also. The vacuum particle in the particle world can not been seen directly
just like ordinary people can not been seen in human history. Ordinary
people (represented by the name Lee) form the social base. Similarly, the
Lee Particles form the base of the perfect material universe.

\bibitem{wilczek}  F. Wilczek, Phys. Tod. \textbf{Jan.} 11(1998) .

\bibitem{diarcsea}  P. A. M. Dirac, \textit{The Principles of Quantum
Mechanics} (Oxford at the Clarendon Press, Fourth Edition,1981) 273; L. H.
Ryder, \textit{Quantum Field} \textit{Theory }(Great Britain at the
University Press, Cambridge, 1966) 45.heory

\bibitem{Chromodynamics}  H. Fritzsch, M. Gell-Mann, and H. Leutwyler, Phys.
Lett. 47B (1973) 365; D. J. Gross and F. Wilczek, Phys. Rev. \textbf{D 8 }%
(1973) 3633\textbf{; }S. Weinberg, Phys. Rev. Lett. 31\textbf{\ }(1973) 494%
\textbf{;} F. J. Ydurain, \textit{Quntum Chromodynamics, An Introduction to
the Theory of Quarks and Flavors}. Springer, Berlin, Heidelberg 1983.

\bibitem{Solidstates}  R. C. Evans, \textit{An Introduction to Crystal
Chemistry} (Cambridge, 1964) 41; X. D. Xie and J. X. Fang, \textit{Solid
State Physics} (in Chinese,1964) 46.\ \ \ \ 

\bibitem{bodycenter}  R. A. Levy, \textit{Principles of Solid State Physics}
(Academic, New York, \ 1972) 41.

\bibitem{eband}  J. Callaway, \textit{Energy Band\ Theory} (Academic Press,
New York and London, 1964) ; H. Jones, \textit{The Theory of Brillouin Zones
and Electronic States in Crystals} ( 2nd revised edition, Noth-Holand,
Amsterdam,1975) Chapter 3.

\bibitem{TOPQUARK}  F. Abe \textit{et al}., Phys. Rew. Lett., \textbf{74, }%
2626 (1995)\textbf{; }S. Abachi \textit{et al}., Phys. Rew. Lett.,\textbf{\
74}, 2632 (1995).

\bibitem{freeparticle}  H. Jones, \textit{The Theory of Brillouin Zones and
Electronic States in Crystals} ( 2nd revised edition, Noth-Holand,
Amsterdam,1975) 117; P. T. Landsberg, \textit{Solid State Theory Methods and
Applications} ( Wiley-Interscience, New York, 1969) 222.

\bibitem{quantumfield}  L. S. Brown, \textit{Quantum Field Theory}
(Cambridge University Press, 1992) 61 and 283.

\bibitem{TDLEE}  T. D. Lee, \textit{Particle Physics and Introduction to
Field Theory }(harwood academic, New York, 1981) 1.\ \ 

\bibitem{real value of S}  We were enlightened by the Monte Carlo methods:
K. Binder, \textit{Monte Carlo Methods in Statistical Physics}
(Springer-Verlag, Berlin 1979); D. P. Landau, Phys. Rev. \textbf{B} \textbf{%
14}, 4054 (1979).

\bibitem{GellMann}  M. Gell-Mann, Phys. Rev. \textbf{92, }833\textbf{\ }%
(1953); K. Nishijima and T. Nakano, Prog. Theor. Phys. \textbf{10}, 581
(1953).

\bibitem{renormal}  J. D. Bjorken and S. D. Drell, \textit{Relativistic
Quantum Fields }(McGraw-Hill, Ney York, 1965) Section 19.10 and 19.11; S.
Weinberg, \textit{The Quantum Theory of Frields} ({\small CAMBRIDGE}, New
York, 1995) 499.

\bibitem{Brillouin}  A. W. Joshi, \textit{Elements of Group Theory for
Physicists} ({\small JOHN WILEY \& SONS}, New York, 1977) 284.

\bibitem{DoubleGroup}  H. Jones, \textit{The Theory of Brillouin Zones and
Electronic States in Crystals} ( 2nd revised edition, Noth-Holand,
Amsterdam,1975) Chapter 7.

\bibitem{SymmetryOp}  H. Jones, \textit{The Theory of Brillouin Zones and
Electronic States in Crystals} ( 2nd revised edition, Noth-Holand,
Amsterdam,1975) Chapter 3.

\bibitem{charmed}  J. J. Aubert, et al., Phys. Rev. Lett. \textbf{33} (1974)
1404; J. E. Augustin, et al., Phys. Rev. Lett. \textbf{33 }(1974) 1406; E.
G. Gazzoli \textit{et al}., Phys. Rev. Lett. \textbf{34}, 1125 (1975).

\bibitem{omiga}  V. E. Barnes \textit{et al}., Phys. Rev. Lett., \textbf{12}%
, 204 (1964).

\bibitem{bottom}  S. W. Herb, et al., Phys. Rev. Lett. 39 (1977) 252; C.
Albajar, Phys. Lett. \textbf{B} \textbf{273}, 540 (1991).

\bibitem{KESI_C}  P. Coteus \textit{et al., } Phys. Rev. Lett. \textbf{59},
1530 (1987); P. Avery \textit{et al., }Phys. Rev. Lett, \textbf{62}, 863
(1989); P. Avery \textit{et al.,} Phys. Rev. Lett. \textbf{75}, 4364 (1995);
L. Gibbons \textit{et al}.,\ Phys. Rev. Lett. \textbf{77}, 810 (1996).\ 

\bibitem{SEGMA_C}  M. Procario \textit{et al.}, Phys. Rev. Lett., \textbf{73}%
, 1472 (1994); G. Brandenburg \textit{et al., }Phys. Rev. Lett., \textbf{78}%
, 2304 (1997).

\bibitem{OMIGA-C}  R. M. Barrett \textit{et al.}, Rev. Mod. Phys., \textbf{%
68,} 651 (1996).

\bibitem{Thomson}  J. J.Thomson, Recollection and Reflections, G. Bell,
London, 1936.

\bibitem{Leebook}  T. D. Lee, \textit{Particle Physics and Introduction to
Field Theory }( harwood academic, New York, 1981) 825.

\bibitem{E167}  B. Abbott \textit{et al}, Phys. Rev. Lett. \textbf{80, }2063
(1998); F. Abe \textit{et al}, Phys. Rev. Lett., \textbf{80, }2767 (1998).

\bibitem{Wilczek2}  F. Wilczek, Rev. Mod. Phys., \textbf{71}, S85 (1999).

\bibitem{XUarticle}  J. L. Xu, Y. S. Wu, H. Y. Guo, and S. Chen, H. Energy
and Nucl. Phys., \textbf{2}, 251 (Beijing, 1980).

\newpage
\end{thebibliography}
\end{document}